\documentclass{LMCS}

\usepackage{array}
\usepackage{latexsym}
\usepackage{amssymb}
\usepackage{pstricks}
\usepackage{pst-node}
\usepackage{stmaryrd}
\usepackage{mathtools}
\usepackage{enumerate,hyperref}

\theoremstyle{plain}\newtheorem{property}[thm]{Property}

\newcommand{\lang}{\mathcal{L}}

\newcommand{\pdsreach}{\stackrel{*}{\hookrightarrow}}
\newcommand{\idG}{\widetilde{G}}
\newcommand{\idallG}{\widetilde{\mathcal{G}}}
\newcommand{\newautG}{\hat{\mathcal{G}}}
\newcommand{\newsingleautG}{\hat{G}}

\newcommand{\finitesingleautG}{G}
\newcommand{\etal}{~\textit{et~al.}}

\def\doi{4 (4:14) 2008}
\lmcsheading%
{\doi}
{1--45}
{}
{}
{Sep.~26, 2007}
{Dec.~\phantom{0}5, 2008}
{}

\begin{document}

\title{Symbolic Backwards-Reachability Analysis for Higher-Order Pushdown
Systems}
\author[M.~Hague]{Matthew Hague}
\address{Oxford University Computing Laboratory, Wolfson Building,
Parks Road, Oxford, UK, OX1 3QD}
\email{\{Matthew.Hague,Luke.Ong\}@comlab.ox.ac.uk}

\author[C.-H.~L.~Ong]{Luke Ong}


\begin{abstract}
Higher-order pushdown systems (PDSs) generalise pushdown systems through the
use of higher-order stacks, that is, a nested ``stack of stacks'' structure.
These systems may be used to model higher-order programs and are
closely related to the Caucal hierarchy of infinite graphs and safe higher-order
recursion schemes.

We  consider the backwards-reachability problem over higher-order Alternating
PDSs (APDSs), a generalisation of higher-order PDSs.  This builds on and extends
previous work on pushdown systems and context-free higher-order processes in a
non-trivial manner.  In particular, we show that the set of configurations from
which a regular set of higher-order APDS configurations is reachable is regular
and computable in $n$-EXPTIME.  In fact, the problem is $n$-EXPTIME-complete.

We show that this work has several applications in the verification of
higher-order PDSs, such as linear-time model-checking, alternation-free
$\mu$-calculus model-checking and the  computation of winning regions of reachability
games.
\end{abstract}

\maketitle


\section{Introduction}

\subsection{Pushdown Automata and Pushdown Systems}

Pushdown automata are an extension of finite state automata.  In
addition to a finite set of control states, a pushdown automaton has a stack which can
be manipulated with the usual push and pop operations.  Transitions of
the automaton depend on both the current control state and the
top item of the stack.  During the execution of a transition, a push or
pop operation is applied to the stack.  Since there is no bound on the size of the
stack, the resulting automaton has an infinite number of ``states'' or
configurations, which consist of the current control state and the
contents of the stack.  This allows the definition of such non-regular
languages as the well known $\{\ a^nb^n\ |\ n \geq 0\ \}$.

Higher-order pushdown automata (PDA) generalise pushdown automata through
the use of higher-order stacks.  Whereas a stack in the sense of a
pushdown automaton is an order-one stack --- that is, a stack of
characters --- an order-two stack is a
stack of order-one stacks.  Similarly, an order-three stack is a
stack of order-two stacks, and so on.  An order-$n$ PDA has push and pop
commands for every $1 \leq l \leq n$.  When $l > 1$ a pop command
removes the topmost order-$l$ stack.  Conversely, the push command
duplicates the topmost order-$l$ stack.

Higher-order PDA were originally introduced by Maslov~\cite{M76} in
the 1970s as generators of (a hierarchy of) finite word
languages. \emph{Higher-order pushdown systems} (PDSs) are
higher-order PDA viewed as generators of infinite trees or
graphs. These systems provide a natural infinite-state model for
higher-order programs with recursive function calls and are therefore
useful in software verification. Several notable advances in recent
years have sparked off a resurgence of interest in higher-order
PDA/PDSs in the Verification community. E.g.~Knapik \emph{et
al.}~\cite{KNU02} have shown that the ranked trees
generated by deterministic order-$n$ PDSs are exactly those that are
generated by order-$n$ recursion schemes satisfying the \emph{safety}
constraint; Carayol and W\"ohrle~\cite{CW03} have shown that the
$\epsilon$-closure of the configuration graphs of higher-order PDSs
exactly constitute Caucal's graph hierarchy \cite{Cau02}. Remarkably
these infinite trees and graphs have decidable monadic second-order
(MSO) theories~\cite{MS85,CW03,KNU02}.

\subsection{Backwards Reachability}

The decidability results discussed above only allow us to check that a
property holds from a given configuration.  Alternatively, we may wish
to compute the set of configurations that satisfy a given property,
especially since there may be an infinite number of such configurations.
An important step in solving this problem is the backwards reachability
problem.  That is, given a set of configurations $C_{Init}$, compute the set
of configurations that can, via any number of transitions, reach a
configuration in $C_{Init}$.
This is an important verification problem in its own right:
many properties required in industry are safety properties --- that is,
an undesirable program state (such as deadlock) is never reached.

This problem was solved for order-one pushdown systems by
Bouajjani\etal~\cite{BEM97}.  In particular, they gave a
method for computing the regular set of configurations $Pre^*(C_{Init})$ that could
reach a given regular set of configurations $C_{Init}$.  A regular set of
configurations is represented in the form of a finite
multi-automaton.  That is, a finite automaton that accepts finite words
(representing stacks) with an initial state for each control state of
the PDS.  A configuration is accepted if the stack (viewed as a word)
is accepted
from the appropriate initial state.
  $Pre^*(C_{Init})$ is computed through
the addition of a number of transitions, determined by the transition relation of
the PDS, to the automaton accepting $C_{Init}$, until a fixed point is
reached.  A fixed point is guaranteed since no states are added and
the alphabet is finite:
eventually the automaton will become \textit{saturated}.

This idea was generalised by Bouajjani and Meyer to the case of
higher-order context-free systems~\cite{BM04}, which are
higher-order PDSs with a single control state.
A key innovation in their work was the introduction of a new class of
(finite-state) automata called \emph{nested store automata}, which
captures an intuitive notion of regular sets of $n$-stores.
An order-$n$ nested store automaton is a finite automaton whose
transitions are labelled by order-$(n-1)$ nested store automata.  In
this way the structure of a higher-order store is reflected.  The
procedure is similar to the algorithm for the order-one case:
transitions are added until a fixed point is reached.  Termination in
this case is more subtle.  Since products are formed when processing
higher-order push commands, the state space increases.  However, it can
be shown that only a finite number of products will be constructed and
that termination follows.

Bouajjani and Meyer also show that forward reachability analysis does
not result in regular sets of configurations.

\subsection{Our Contribution}

Our paper is concerned with the non-trivial problem\footnote{``This
  does not seem to be technically trivial, and na\"ive extensions of
  our construction lead to procedures which are not guaranteed to
  terminate.''~\cite[p.~145]{BM04}} of extending the backwards
  reachability result of Bouajjani and Meyer to the general case of
  higher-order PDSs (by taking into account a set of control
  states). In fact, we consider (and solve) the backwards reachability
  problem for the more general case of higher-order
  \emph{alternating} pushdown systems (APDSs). Though slightly
  unwieldy, an advantage of the alternating framework is that it
  conveniently lends itself to a number of higher-order PDS
  verification problems.  Following the work of
  Cachat~\cite{C03}, we show that the winning region of a
  reachability game played over a higher-order PDS can be computed by
  a reduction to the backwards reachability problem of an appropriate
  APDS. We also generalise results due to Bouajjani~\etal~\cite{BEM97}
  to give a method for computing the set of
  configurations of a higher-order PDS that satisfy a given formula of
  the alternation-free $\mu$-calculus or a linear-time temporal
logic.

The algorithm uses a similar form of nested automata to represent
configurations and uses a similar routine of adding transitions
determined by the transition relation of the higher-order APDS.  However, na\"ive
combinations of the multi-automaton and nested-store automaton
techniques do not lead to satisfactory solutions.
During our own efforts with simple combined techniques, it was unclear how to form the product of two automata and
maintain a distinction between the different control states as required.  To perform such
an operation safely it seemed that additional states were required
on top of those added by the basic product operation, invalidating the
termination arguments.  We overcome this problem by using alternating automata and by modifying the termination
argument.  Additionally, we reduce the complexity of Bouajjani and
Meyer from a tower of exponentials twice the size of $n$, to a tower
of exponentials as large as $n$.  In fact, the problem is $n$-EXPTIME-complete.

Termination is reached through a cascading of fixed points.  Given
a (nested) store-automaton, we fix the order-$n$ state-set.  During a
number of iterations, we add a
finitely bounded number of
new transitions to order $n$ of the automaton.  We also update the
automata labelling the previously added transitions to reflect the new
transition structure.
Eventually we reach a stage where no new transitions are being added
at order $n$, although the automata labelling their edges will
continue to be
replaced.  At this point the updates become
repetitive and we are able to freeze the state-set at the
second highest order.  This is done by adding possibly cyclical
transitions between the existing states, instead of
chains of transitions between an infinite set of new states.
Because the state-set does not change, we reach another fixed point
similar to that at order $n$.  In this
way the fixed points cascade to order-one, where the finite alphabet
ensures that the automaton eventually becomes saturated.  We are left
with an automaton representing the set $Pre^*(C_{Init})$.

\subsection{Related Work}

In this section we discuss several areas of related work.  These are
higher-order pushdown games, alternative notions of regularity, and higher-order
recursion schemes.

\subsubsection{Higher-Order Pushdown Games}

The definition of higher-order PDSs may be extended to higher-order pushdown
games.  In this scenario, control states are partitioned into to sets
$\exists$ and $\forall$.  When the current configuration contains a
control state in $\exists$, the player Eloise chooses the next
configuration with respect to the transition relation.  Conversely,
Abelard chooses the next transition from a control state in $\forall$.
The winner of the game depends on the winning condition.  A
configuration is winning for Eloise if she can satisfy the winning
condition regardless of the choices made by Abelard.  A winning region
for Eloise is the set of all configurations from which Eloise can
force a win.  Two particular
problems for these games are calculating whether a given configuration
is winning for Eloise and computing the winning region for Eloise.

In the order-one case, the problem of determining whether a configuration is winning for
Eloise with a parity winning condition was
solved by Walukiewicz in 1996~\cite{W96}.  The order-one backwards
reachability algorithm of Bouajjani\etal\ was adapted by
Cachat to compute the winning regions of order-one reachability and
B\"uchi games~\cite{C03}.  Techniques for computing winning regions in
the order-one case when the winning
condition is a parity condition have been discovered independently by
both Cachat~\cite{C03} and Serre~\cite{S02}.  These results for
pushdown games have been extended to a number of winning
conditions~\cite{CDT02,BSW03,G04,S04,LMS04}.  In the higher-order
case with a parity winning condition, a method for deciding whether a
configuration is winning has been provided by Cachat~\cite{C03}.

\subsubsection{C-Regularity}

Prompted by the fact that the set of
configurations reachable from a given configuration of a higher-order
PDS is not regular in the sense of Bouajjani and Meyer (the stack
contents cannot be represented by a finite automaton over words),
Carayol~\cite{C05} has proposed an alternative definition of
regularity for higher-order stacks, which we shall call
\emph{C-regularity}.  Our notion of regularity coincides with that of
Bouajjani and Meyer, which, when confusion may arise, we shall call \emph{BM-regularity}.

A set of order-$n$ stacks is C-regular if it is obtained by a
regular sequence of order-$n$ stack operations.  This notion of
regularity is not equivalent to BM-regularity.  For example, the set
of order-$2$ stacks defined by the expression $(push_a)^*; push_2$ are all
stacks of the form $[[a^n][a^n]]$.  This set is clearly
unrecognisable by any finite state automaton, and thus, it is not BM-regular.

Carayol shows that C-regularity coincides
with MSO definability over the canonical structure $\Delta^n_2$
associated with order-$n$ stacks.  This implies, for instance, that
the winning region of a parity game over an order-$n$ pushdown graph
is also C-regular, as it can be defined as an MSO formula~\cite{C03}.

In this paper we solve the backwards reachability problem for higher-order PDSs
and apply the solution to reachability games and model-checking.  In this sense
we give a weaker kind of result that uses a different notion of regularity.
Because C-regularity does not imply BM-regularity, our result is not subsumed by
the work of Carayol.  However, a detailed comparison of the two approaches may
provide a fruitful direction for further research.

\subsubsection{Higher-Order Recursion Schemes}

Higher-order recursion schemes (HORSs) represent a further area of related work.
A long standing open problem is whether a condition called \emph{safety} is a
genuine restriction on the expressiveness of a HORS.  If not, then HORSs are
equivalent to higher-order PDSs.  It is known that safety is not a restriction
at order-two for word languages~\cite{AdMO05}.  This is conjectured not to be
the case at higher orders.

MSO decidability for trees generated by arbitrary (i.e.~not necessarily safe)
HORSs has been shown by Ong~\cite{O06}.  A variant kind of higher-order PDSs
called \textit{collapsible pushdown automata} (extending \textit{panic automata}
\cite{KNUW05} or \textit{pushdown automata with links} \cite{AdMO05} to all
finite orders) has recently been shown to be equi-expressive with HORSs for
generating ranked trees \cite{HMOS06}. These new automata are conjectured to
enrich the class of higher-order systems and provide many new avenues of
research.

\subsection{Document Structure}

In Section~\ref{prelim} we give the definitions of higher-order (A)PDS
and $n$-store multi-automata.  We describe the backwards-reachability
algorithm in the order-two case in three stages in
Section~\ref{thealgorithmorder2}: firstly we use an example to give an
intuitive explanation of the algorithm.  We then give a description of
its framework and explain how we can generate an infinite sequence of
$2$-store multi-automata capturing the set $Pre^*(C_{Init})$.  Finally,
we show how this sequence can be finitely represented (and
constructed).  The section finishes with a brief discussion of the order-$n$
case, and the complexity of the algorithm.  Section~\ref{applications}
discusses the applications of the main result to LTL model-checking,
reachability games and alternation-free $\mu$-calculus model-checking
over higher-order PDSs.   Finally, we conclude in
Section~\ref{conclusion}.  Additional proofs and algorithms are given
in the appendix.

\section{Preliminaries}
\label{prelim}

\subsection{Alternation}

In the sequel we will introduce several kinds of alternating
automata.  For convenience, we will use a non-standard definition of alternating
automata that is equivalent to the standard definitions of Brzozowski
and Leiss~\cite{BL80} and Chandra, Kozen and Stockmeyer~\cite{CKS81}.
Similar definitions have been used for the analysis of pushdown
systems by Bouajjani\etal~\cite{BEM97} and Cachat~\cite{C03}.
The alternating transition relation $\Delta \subseteq \mathcal{Q} \times \Gamma \times
2^{\mathcal{Q}}$ --- where $\Gamma$ is an alphabet and
$\mathcal{Q}$ is a state-set --- is given in
disjunctive normal form.  That is, the image $\Delta(q, \gamma)$ of $q \in \mathcal{Q}$
and $\gamma \in \Gamma$ is a set $\{Q_1,\ldots,Q_m\}$ with $Q_i
\in 2^{\mathcal{Q}}$ for $i \in \{1,\ldots,m\}$.  When the automaton is viewed as a game, Eloise
--- the existential player --- chooses a set $Q \in \Delta(q,
\gamma)$;  Abelard --- the universal player --- then chooses a state
$q \in Q$.  The existential component of the automaton is
reflected in Eloise's  selection of an element $(q, \gamma, Q)$ from
$\Delta$ for a given $q$ and $\gamma$.  Abelard's choice of a state $q$
from $Q$ represents the universal aspect of the automaton.

\subsection{(Alternating) Higher-Order Pushdown Systems}

A higher-order pushdown system comprises a finite set of control states
and a higher-order store.  Transitions of the higher-order PDS depend on both the
current control state and the top symbol of the higher-order store.  Each
transition changes the control state and manipulates the store.

The main result of this paper is presented over \emph{alternating}
higher-order pushdown systems.  This is because, although we apply our
results to higher-order PDSs, the power of alternation is required to provide
solutions to reachability games and alternation-free
$mu$-calculus model-checking over higher-order PDSs.

We begin by defining higher-order stores and their operations.  We
will then define higher-order PDSs in full.

\begin{defi}[$n$-Stores]
\label{nstoredef}
The set $C^\Sigma_1$ of $1$-stores over an alphabet $\Sigma$ is the
set of words of the form
$[a_1,\ldots,a_m]$ with $m \geq 0$ and $a_i \in \Sigma$ for all $i \in
\{1,\ldots,m\}$, $[ \notin \Sigma$ and $] \notin \Sigma$.  For $n >
1$, $C^\Sigma_n = [w_1,\ldots,w_m]$ with $m \geq 1$ and $w_i \in
C^{\Sigma}_{n-1}$ for all $i \in \{1,\ldots,m\}$.
\end{defi}

There are three types of operations applicable to $n$-stores: $push$,
$pop$ and $top$.  These are defined inductively.  Over a $1$-store, we
have (for all $w \in \Sigma^*$),
\[ \begin{array}{rcl}
   push_w[a_1\ldots a_m] &=& [wa_2\ldots a_m] \\
   top_1[a_1\ldots a_m] &=& a_1 \\
   \end{array}
 \]
We may define the abbreviation $pop_1 = push_\varepsilon$.  When $n > 1$, we have,
\[ \begin{array}{rcll}
      push_w[\gamma_1\ldots \gamma_m] &=& [push_w(\gamma_1)\gamma_2\ldots\gamma_m] &\\

      push_l[\gamma_1\ldots\gamma_m] &=&
[push_l(\gamma_1)\gamma_2\ldots\gamma_m] & \mathrm{if\ } 2 \leq l < n
\\

      push_n[\gamma_1\ldots\gamma_m] &=&
[\gamma_1\gamma_1\gamma_2\ldots\gamma_m] & \\

      pop_l[\gamma_1\ldots\gamma_m] &=&
[pop_l(\gamma_1)\gamma_2\ldots\gamma_m] & \mathrm{if\ } 1 \leq l < n
\\

      pop_n[\gamma_1\ldots\gamma_m] &=& [\gamma_2\ldots\gamma_m] &
\mathrm{if\ } m > 1
\\

      top_l[\gamma_1\ldots\gamma_m] &=& top_l(\gamma_1) &
\mathrm{if\ } 1 \leq l < n \\

      top_n[\gamma_1\ldots\gamma_m] &=& \gamma_1 & \\
   \end{array}
\]
Note that we assume without loss of generality $\Sigma \cap \mathcal{N} = \emptyset$, where
$\mathcal{N}$ is the set of natural numbers.  Furthermore, observe
that when $m = 1$, $pop_n$ is undefined.  We define
$\mathcal{O}_n = \{\ push_w\ |\ w \in \Sigma^*\ \} \cup \{\ push_l, pop_l\
|\ 1 < l \leq n\ \}$.
The definition of higher-order PDSs follows,
\begin{defi}
An \emph{order-$n$ PDS} is a tuple $(\mathcal{P}, \mathcal{D}, \Sigma)$ where $\mathcal{P}$ is a finite set
of control states $p$, $\mathcal{D} \subseteq
\mathcal{P}\times\Sigma\times\mathcal{O}_n\times\mathcal{P}$ is a finite set of commands
$d$, and $\Sigma$ is a finite alphabet.

A configuration of a higher-order PDS is a pair $\langle p, \gamma\rangle$
where $p \in \mathcal{P}$ and $\gamma$ is an $n$-store.  We have a
transition $\langle p, \gamma\rangle \hookrightarrow \langle p',
\gamma'\rangle$ iff we have $(p, a, o, p') \in \mathcal{D}$,
$top_1(\gamma) = a$ and $\gamma' = o(\gamma)$.

We define $\pdsreach$ to be the transitive closure of
$\hookrightarrow$.  For a set of configurations $C_{Init}$ we define
$Pre^*(C_{Init})$ as the set of configurations $\langle p, \gamma\rangle$
such that, for some configuration $\langle p', \gamma'\rangle \in C_{Init}$,
we have $\langle p, \gamma\rangle \pdsreach \langle p', \gamma'\rangle$.
\end{defi}

We may generalise this definition to the case of Alternating
higher-order PDSs.

\begin{defi}
An \emph{order-$n$  APDS} is a tuple $(\mathcal{P}, \mathcal{D}, \Sigma)$ where $\mathcal{P}$ is a finite set
of control states $p$, $\mathcal{D} \subseteq
\mathcal{P}\times\Sigma\times2^{\mathcal{O}_n\times\mathcal{P}}$ is a finite set of commands
$d$, and $\Sigma$ is a finite alphabet.

A configuration of a higher-order APDS is a pair $\langle p, \gamma\rangle$
where $p \in \mathcal{P}$ and $\gamma$ is an $n$-store.  We have a
transition $\langle p, \gamma\rangle \hookrightarrow C$ iff we have $(p, a, OP) \in \mathcal{D}$,
$top_1(\gamma) = a$, and
\[C = \{\ \langle p',
\gamma'\rangle\ |\ (o, p') \in OP\ \land\ \gamma' = o(\gamma)\
\} \cup \{\ \langle p, \triangledown\rangle\ |\ \mathrm{if\ } (o, p') \in
OP\ \mathrm{and\ } o(\gamma)
\mathrm{\ is\ not\ defined}\ \}\]
The transition relation generalises to sets of
configurations via the following  rule:
\[
{
         \frac{\langle p, \gamma \rangle \; \hookrightarrow \; C}
         {C' \cup \langle p, \gamma\rangle \; \hookrightarrow \; C' \cup C}
}      \quad
      \langle p, \gamma \rangle \notin C'
\]

%


We define $\pdsreach$ to be the transitive closure of
$\hookrightarrow$.  For a set of configurations $C_{Init}$ we define
$Pre^*(C_{Init})$ as the set of configurations $\langle p, \gamma\rangle$
such that
we have $\langle p, \gamma\rangle \pdsreach C$ and $C \subseteq C_{Init}$.
\end{defi}

\begin{exa}
We present an example to illustrate the definition of
$Pre^*(C_{Init})$ for higher-order APDSs.
\begin{figure}
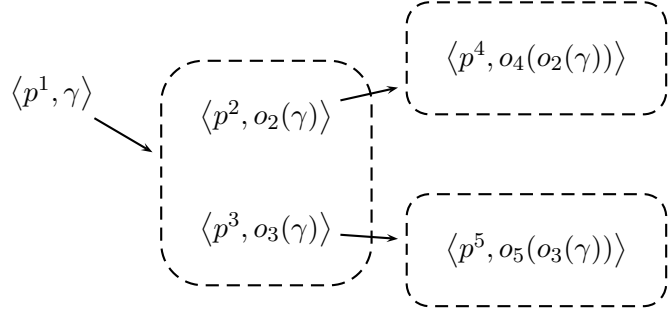

\begin{center}
\psset{framearc=.4,linestyle=dashed,framesep=.5}
\begin{tabular}{m{2cm}m{3cm}m{3cm}}
  \begin{center}
  \begin{psmatrix}
     \rnode{P1}{$\left\langle p^1, \gamma\right\rangle$} \\
     $\quad$
  \end{psmatrix}
  \end{center}

  &

  \begin{center}
    \Rnode{P1next}{\psframebox{\begin{psmatrix}[rowsep=1]
                    \rnode{P2}{$\left\langle p^2, o_2(\gamma)\right\rangle$} \\
                    \rnode{P3}{$\left\langle p^3, o_3(\gamma)\right\rangle$}
                \end{psmatrix}}}
  \end{center}

  &

  \begin{center}
  \begin{psmatrix}[rowsep=1]
    \Rnode{P4}{\psframebox{$\left\langle p^4, o_4(o_2(\gamma))\right\rangle$}} \\
    \Rnode{P5}{\psframebox{$\left\langle p^5, o_5(o_3(\gamma))\right\rangle$}}
  \end{psmatrix}
  \end{center}

  \psset{arrows=->,linestyle=solid,nodesep=1mm}

  \ncline{P1}{P1next}
  \ncline{P2}{P4}
  \ncline{P3}{P5}

\end{tabular}
\end{center}
\caption{The configuration graph (excerpt) of an example higher-order APDS.}
\label{preeg}
\end{figure}
Figure~\ref{preeg} shows an excerpt of the configuration graph of a
higher-order APDS with the commands,
\[
  \begin{array}{c}
    (p^1, \_, \{(o_2, p^2), (o_3, p^3)\}) \\
    (p^2, \_, \{(o_4, p^4)\}) \\
    (p^3, \_, \{(o_5, p^5)\})
  \end{array}
\]
We consider a number of different values of $C_{Init}$.
\begin{enumerate}[(1)]
\item
Let $C_{Init} = \{\langle p^2, o_2(\gamma)\rangle\}$.  In this case
$Pre^*(C_{Init}) = C_{Init}$.  The configuration $\langle
p^1,\gamma\rangle$ is not in $Pre^*(C_{Init})$ since the configuration
$\langle p^3, o_3(\gamma)\rangle$ cannot be in $Pre^*(C_{Init})$.

\item
Let $C_{Init} = \{\langle p^2, o_2(\gamma)\rangle, \langle p^3, o_3(\gamma)\rangle\}$.
In this case $Pre^*(C_{Init}) = C_{Init} \cup \{\langle
p^1,\gamma\rangle\}$.  This is because the transition from $\langle p^1,
\gamma\rangle$ reaches a set that is a subset of $C_{Init}$.

\item
Let $C_{Init} = \{\langle p^4, o_4(o_2(\gamma))\rangle\}$.  In this case
$Pre^*(C_{Init}) = C_{Init} \cup \{\langle p^2, o_2(\gamma)\}$.  The
configuration $\langle p^2,
o_2(\gamma)\rangle$ is in the set because its transition moves to a set which
is a subset of $C_{Init}$. The pair $\langle p^1, \gamma\rangle$ is not in the set
because, although $\langle p^2, o_2(\gamma)\rangle$ is in $Pre^*(C_{Init})$,
the configuration $\langle p^3, o_3(\gamma)\rangle$ is not.

\item
Let $C_{Init} = \{\langle p^4, o_4(o_2(\gamma))\rangle, \langle p^3, o_3(\gamma)\rangle\}$.
In this case $Pre^*(C_{Init})$ is the set $C_{Init} \cup \{\langle p^2,
o_2(\gamma)\rangle, \langle p^1, \gamma\rangle\}$.  We have $\langle p^2, o_2(\gamma)\rangle
\in Pre^*(C_{Init})$ as before.  Furthermore, we have the following
run from $\langle p^1, \gamma\rangle$,
\[ \langle p^1, \gamma\rangle \hookrightarrow \{\langle p^2, o_2(\gamma)\rangle, \langle
p^3, o_3(\gamma)\rangle\} \hookrightarrow \{\langle p^4, o_4(o_2(\gamma))\rangle, \langle
p^3, o_3(\gamma)\rangle \} \]
Hence, $\langle p^1, \gamma\rangle \in Pre^*(C_{Init})$.
\end{enumerate}
Finally, suppose the higher-order APDS also has a command of the form,
\[ (p^5, \_, \{(push_l, p^4)\}) \]
And it is the case that (only) $push_l(o_5(o_3(\gamma)))$ is undefined.  If
$C_{Init} = \{\langle p^5, \triangledown\rangle\}$, then $Pre^*(C_{Init}) =
C_{Init} \cup \{\langle p^5, o_5(o_3(\gamma))\rangle, \langle p^3, o_3(\gamma)\rangle\}$.
\end{exa}

Observe that since no transitions are possible from an ``undefined''
configuration $\langle p, \triangledown\rangle$ we can reduce the
reachability problem for higher-order PDSs to the reachability problem
over higher-order APDSs
in a straightforward manner.

In the sequel, to ease the presentation, we assume $n > 1$.  The case
$n=1$ was investigated by Bouajjani\etal~\cite{BEM97}.

\subsection{$n$-Store Multi-Automata}

To represent sets of configurations symbolically we will use \emph{$n$-store multi-automata}.
These are alternating automata whose transitions are labelled by
\emph{$(n-1)$-store automata}, which are also alternating.  A set of configurations is
\textit{regular} iff it can be represented using an $n$-store
multi-automaton.  This notion of regularity coincides with the
definition of Bouajjani and Meyer (see
Appendix~\ref{ourregularityis}).  In Appendix~\ref{storeautalg} we
give algorithms for enumerating runs of $n$-store automata, testing
membership and performing boolean operations on the automata.

\begin{defi} \
\begin{enumerate}[(1)]
\item
A \emph{1-store automaton} is a tuple $(\mathcal{Q}, \Sigma, \Delta, q_0,
\mathcal{Q}_f)$ where $\mathcal{Q}$ is a finite set of states, $\Sigma$ is a
finite alphabet, $q_0$ is the initial state and $\mathcal{Q}_f
\subseteq \mathcal{Q}$ is a
set of final states.  It is the case that $\Delta \subseteq \mathcal{Q} \times \Sigma \times
2^{\mathcal{Q}}$ is a finite transition relation.

\item
Let $\mathfrak{B}^\Sigma_{n-1}$ be the (infinite) set of all $(n-1)$-store
automata over the alphabet $\Sigma$.
An \emph{$n$-store automaton} over the alphabet $\Sigma$ is a tuple
$(\mathcal{Q}, \Sigma, \Delta, q_0, \mathcal{Q}_f)$ where $\mathcal{Q}$ is a
finite set of states,
$q_0 \notin \mathcal{Q}_f$ is the initial state, $\mathcal{Q}_f\subseteq \mathcal{Q}$ is a set of
final states, and  $\Delta \subseteq \mathcal{Q} \times
\mathfrak{B}^\Sigma_{n-1} \times
2^{\mathcal{Q}}$ is a \emph{finite} transition relation.  Furthermore, let
$\mathfrak{B}^\Sigma_0 = \Sigma$.

\item
An \emph{$n$-store multi-automaton} over the alphabet $\Sigma$ is a tuple
\[(\mathcal{Q}, \Sigma, \Delta, \{q^1,\ldots,q^z\}, \mathcal{Q}_f)\]
where $\mathcal{Q}$ is a finite set of states, $\Sigma$ is a finite alphabet,
$q^i$ for $i \in \{1,\ldots,z\}$ are pairwise distinct initial states with $q^i
\notin \mathcal{Q}_f$ and $q^i \in \mathcal{Q}$; $\mathcal{Q}_f\subseteq
\mathcal{Q}$ is a set of final states, and,
\[  \Delta \subseteq (\mathcal{Q} \times
\mathfrak{B}^\Sigma_{n-1} \times
2^\mathcal{Q}) \cup (\{q^1,\ldots,q^z\} \times \{\triangledown\} \times
\{q^\varepsilon_f\}) \]
is a \emph{finite} transition relation where $q^\varepsilon_f \in
\mathcal{Q}_f$ has no outgoing transitions.
\end{enumerate}
\end{defi}
To indicate a transition $(q, B,
\{q_1,\ldots, q_m\}) \in \Delta$ we write,
\[ q \stackrel{B}{\longrightarrow} \{q_1,\ldots, q_m\} \]
A transition of the form $q^j \stackrel{\triangledown}{\longrightarrow}
\{q^\varepsilon_f\}$ indicates that the undefined configuration $\langle p^j,
\triangledown\rangle$ is accepted.
Runs of the automata from a state $q$ take the form,
\[ q \stackrel{\widetilde{B}_0}{\longrightarrow} \{q^1_1,\ldots, q^1_{m_1}\}
\stackrel{\widetilde{B}_1}{\longrightarrow} \ldots \stackrel{\widetilde{B}_m}{\longrightarrow}
\{q^{m+1}_1, \ldots, q^{m+1}_{m_l}\} \]
where transitions between configurations
$\{q^x_1,\ldots,q^x_{m_{x}}\} \stackrel{\widetilde{B}_x}{\longrightarrow} \{q^{x+1}_1,\ldots,q^{x+1}_{m_{x+1}}\}$ are
such that we have $q^x_y \stackrel{B_y}{\longrightarrow} Q_y$ for all $y \in
\{1,\ldots,m_x\}$ and $\bigcup_{y \in \{1,\ldots,m_x\}} Q_y =
\{q^{x+1}_1,\ldots,q^{x+1}_{m_{x+1}}\}$ and additionally $\bigcup_{y \in
\{1,\ldots,m_x\}}\{B_y\} = \widetilde{B}_x$.  Observe that $\widetilde{B}_0$
is necessarily a singleton set.  A run over a word $\gamma_1\ldots\gamma_m$,
denoted $q \xrightarrow{\gamma_1\ldots\gamma_m} Q$, exists whenever,
\[ q
\stackrel{\widetilde{B}_0}{\longrightarrow} \ldots
\stackrel{\widetilde{B}_m}{\longrightarrow} Q \]
and for all $0 \leq i \leq m. \gamma_i \in \lang(\widetilde{B}_i)$,  where
$\gamma \in \lang(\widetilde{B})$ iff $\gamma \in \lang(B)$ (defined below) for
all $B \in \widetilde{B}$.  If a run occurs in an automaton forming part of a
sequence of automata $A_0, A_1,\ldots$, we may write $\longrightarrow_i$ to indicate
which automaton $A_i$ the run belongs to.

 We define $\lang(a) = a$ for all $a \in \Sigma = \mathfrak{B}^\Sigma_0$.
An $n$-store $[\gamma_1\ldots\gamma_m]$ is accepted by an $n$-store automaton $A$
(that is $[\gamma_1\ldots\gamma_m] \in \lang(A)$) iff we have a run $q_0
\xrightarrow{\gamma_1\ldots\gamma_m} Q$ in $A$ with $Q \subseteq \mathcal{Q}_f$.
For a given $n$-store multi-automaton $A = (Q, \Sigma, \Delta,
\{q^1,\ldots,q^z\}, \mathcal{Q}_f)$ we define,
\[ \begin{array}{rcl}
    \lang(A^{q^j}) &=& \{\ [\gamma_1\ldots\gamma_m]\ |\ q^j
\xrightarrow{\gamma_1\ldots\gamma_m} Q\
\land\ Q \subseteq \mathcal{Q}_f\  \} \\
     & & \cup\ \{\ \triangledown\ |\ q^j \stackrel{\triangledown}{\longrightarrow}
\{q^\varepsilon_f\}\ \}
   \end{array} \]
and
\[ \lang(A) = \{\ \langle p^j, \gamma\rangle\ |\ j \in \{1,\ldots,z\} \land \gamma \in
\lang(A^{q^j})\ \} \]

Finally, we define the automata $B^a_l$ for all $1 \leq l \leq n$ and
$a \in \Sigma$ and the notation $q^\theta$.  The $l$-store automaton $B^a_l$ accepts any $l$-store
$\gamma$ such that $top_1(\gamma) = a$.  If $\theta$ represents a store automaton, the state $q^\theta$
refers to the initial state of the automaton represented by $\theta$.

\section{Backwards Reachability: The Order-Two Case}
\label{thealgorithmorder2}

Since the backwards reachability problem for higher-order PDSs permits a direct reduction to
the same problem for higher-order APDSs, we solve the backwards reachability
problem for higher-order APDSs.  Due to space constraints we present the
order-$2$ case.  The general case is addressed briefly at the end of this
section and is due to appear in Hague's Ph.D. thesis~\cite{HThesis}.

\begin{thm}
  Given an $2$-store multi-automaton $A_0$ accepting the set
  of configurations $C_{Init}$ of an order-$2$ APDS, we can construct
  in $2$-EXPTIME (in the size of $A_0$) an $2$-store multi-automaton $A_*$ accepting the set
  $Pre^*(C_{Init})$.  Thus, $Pre^*(C_{Init})$ is regular.\qed
\end{thm}

Fix an order-$2$ APDS. We begin by showing how to generate an infinite
sequence of automata $A_0,A_1,\ldots$, where $A_0$ is such that
$\lang(A_0) = C_{Init}$.  This sequence is increasing in the sense
that $\lang(A_{i}) \subseteq \lang(A_{i+1})$ for all $i$, and sound
and complete with respect to $Pre^*(C_{Init})$; that is $\bigcup_{i
\geq 0} \lang(A_i) = Pre^*(C_{Init})$.  To conclude the algorithm, we
construct a single automaton $A_*$ such that $\mathcal{L}(A_*) =
\bigcup_{i \geq 0} \lang(A_i)$.

We assume, without loss of generality, that all initial
states in $A_0$ have no incoming transitions and there exists in $A_0$
a state $q^*_f$ from which all valid $2$-stores are accepted and a
state $q^\varepsilon_f \in \mathcal{Q}_f$ that has no outgoing
transitions.

\subsection{Example}

\begin{figure}
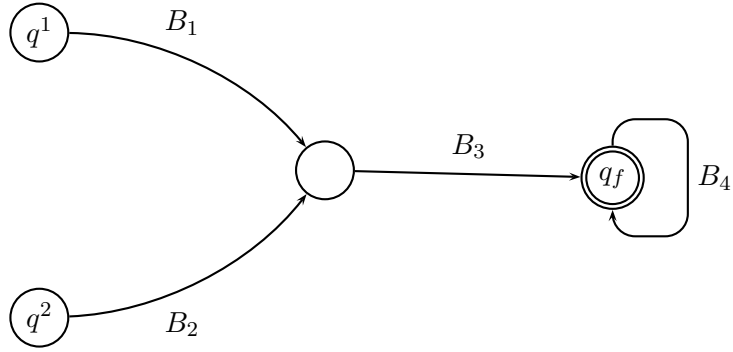

\begin{center}
\begin{psmatrix}[colsep=3,rowsep=1]
  \circlenode{Q1}{$q^1$} \\
  & \circlenode{Q3}{$\phantom{q^2}$} &  \circlenode[doubleline=true]{Q4}{$q_f$}
\\
  \circlenode{Q2}{$q^2$}

  \psset{arrows=->}
  \ncarc[arcangle=25]{Q1}{Q3}^{$B_1$}
  \ncarc[arcangle=-25]{Q2}{Q3}_{$B_2$}
  \ncline{Q3}{Q4}^{$B_3$}
  \ncloop[angleA=90,angleB=-90,loopsize=1,linearc=.3]{Q4}{Q4}\Aput*[0.1]{$B_4$}
\end{psmatrix}
\end{center}
\caption{The initial 2-store multi-automaton}
\label{fig1}
\end{figure}

We give an intuitive explanation of the algorithm by means of an example.
Fix the following two-state order-two PDS:
\[ \begin{array}{l}
        d_1 = (p^1, a, push_2, p^1) \\
        d_2 = (p^1, a, push_\varepsilon, p^1) \\
        d_3 = (p^2, a, push_w, p^1) \\
        d_4 = (p^2, a, pop_2, p^1) \\
   \end{array}
\]
And a 2-store multi-automaton $A_0$ shown in Figure~\ref{fig1} with some $B_1, B_2, B_3$ and $B_4$.

\begin{figure}
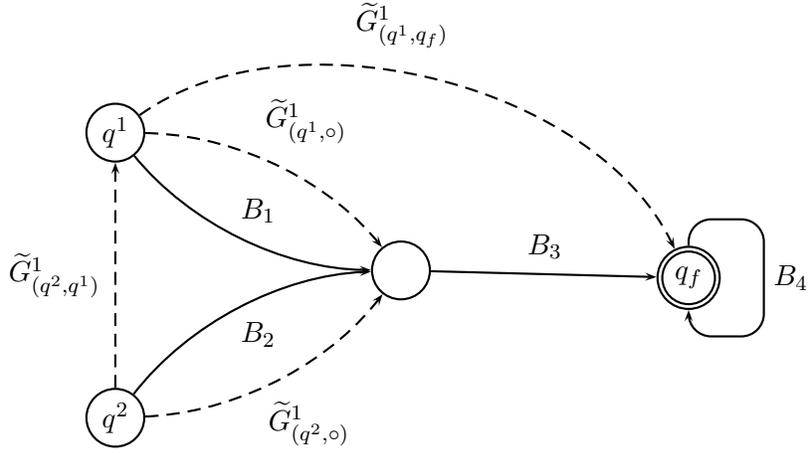

\begin{center}
\begin{psmatrix}[colsep=3,rowsep=1]
  $\quad$ \\
  \circlenode{Q1}{$q^1$} \\
  & \circlenode{Q3}{$\phantom{q^2}$} &  \circlenode[doubleline=true]{Q4}{$q_f$}
\\
  \circlenode{Q2}{$q^2$}

  \psset{arrows=->}
  \ncarc[arcangle=-25]{Q1}{Q3}^{$B_1$}
  \ncarc[arcangle=25]{Q2}{Q3}_{$B_2$}
  \ncline{Q3}{Q4}^{$B_3$}
  \ncloop[angleA=90,angleB=-90,loopsize=1,linearc=.3]{Q4}{Q4}\Aput[0.1]{$B_4$}
  \ncarc[arcangle=50,linestyle=dashed]{Q1}{Q4}^{$\idG^1_{(q^1,
q_f)}$}
  \ncarc[arcangle=25,linestyle=dashed]{Q1}{Q3}\Aput{$\idG^1_{(q^1, \circ)}$}
  \ncarc[arcangle=-25,linestyle=dashed]{Q2}{Q3}\Bput{$\idG^1_{(q^2, \circ)}$}
  \ncline[linestyle=dashed]{Q2}{Q1}\Aput{$\idG^1_{(q^2, q^1)}$}
\end{psmatrix}
\end{center}
\caption{The automaton $A_1$}
\label{fig2}
\end{figure}

We proceed via a number of iterations, generating the automata $A_0,
A_1, \ldots$.  We construct $A_{i+1}$ from $A_i$ to reflect an
additional inverse application of the commands $d_1,\ldots,d_4$.
Rather than manipulating the order-1 store automata labelling the
edges of $A_0$ directly, we introduce new transitions (at most
one between each pair of states $q_1$ and $q_2$) and label these edges
with the set $\idG^1_{(q_1, q_2)}$.  This set is a recipe for the
construction of an order-$1$ store automaton that will ultimately
label the edge.  The set $\mathcal{G}^1$ is the set of all sets
$\idG^1_{(q_1,q_2)}$ introduced.  The resulting $A_1$ is given in Figure~\ref{fig2}
where the contents of
\[ \mathcal{G}^1 =
\{\idG^1_{(q^1,\circ)},\idG^1_{(q^1,q_f)},\idG^1_{(q^2,\circ)},\idG^1_{(q^2,q^1)}\}
\]
are given in Table~\ref{tab1}.  The columns indicate which command
introduced each element to the set.

\begin{table}
\begin{center}
\psframebox[framearc=.1,framesep=-.005]{%
\begin{tabular}{l|l|l|l|l}
 &  $d_1$ &
    $d_2$ &
    $d_3$ &
    $d_4$  \\
\hline
 & & & &  \\
 $\idG^1_{(q^1, \circ)}$ & & $\{(a, push_\varepsilon, B_1)\}$ & & \\
 & & & &  \\
 $\idG^1_{(q^1, q_f)}$ & $\{B^a_1, B_1, B_3\}$ & & & \\
 & & & &  \\
 $\idG^1_{(q^2, \circ)}$ & & & $\{(a, push_w, B_1)\}$ & \\
 & & & &  \\
 $\idG^1_{(q^2, q^1)}$ & & & & $\{B^a_1\}$ \\
 & & & &  \\
\end{tabular}}
\end{center}
\caption{The contents of the sets in $\idallG^1$.}
\label{tab1}
\end{table}

To process the command $d_1$ we need to add to the set of configurations accepted by $A_1$
all configurations of the
form $\langle p_1, [\gamma_1\ldots\gamma_m]\rangle$ with
$top_1(\gamma_1) = a$ for each configuration $\langle p_1,
[\gamma_1\gamma_1\ldots\gamma_m]\rangle$ accepted by $A_0$.  This is because
$push_2[\gamma_1\ldots\gamma_m] = [\gamma_1\gamma_1\ldots\gamma_m]$.
Hence we add the transition from $q^1$ to $q_f$.  The contents of
$\idG^1_{(q^1,q_f)}$ indicate that this edge must accept the product
of $B^a_1$, $B_1$ and $B_3$.

The commands $d_2$ and $d_3$ update the $top_2$ stack of any
configuration accepted from $q^1$ or $q^2$ respectively.  In both
cases this updated stack must be accepted from $q^1$ in $A_0$.  Hence,
the contents of $\idG^1_{(q^1,\circ)}$ and $\idG^1_{(q^2, \circ)}$
specify that the automaton $B_1$ must be manipulated to produce the
automaton that will label these new transitions.
Finally, since $pop_2[\gamma_1\ldots\gamma_m] = [\gamma_2\ldots\gamma_m]$,
$d_4$ requires an additional $top_2$ stack with $a$ as
its $top_1$ element to be added to any stack accepted from $q^1$.
Thus, we introduce the transition from $q^2$ to $q^1$.

\begin{figure}
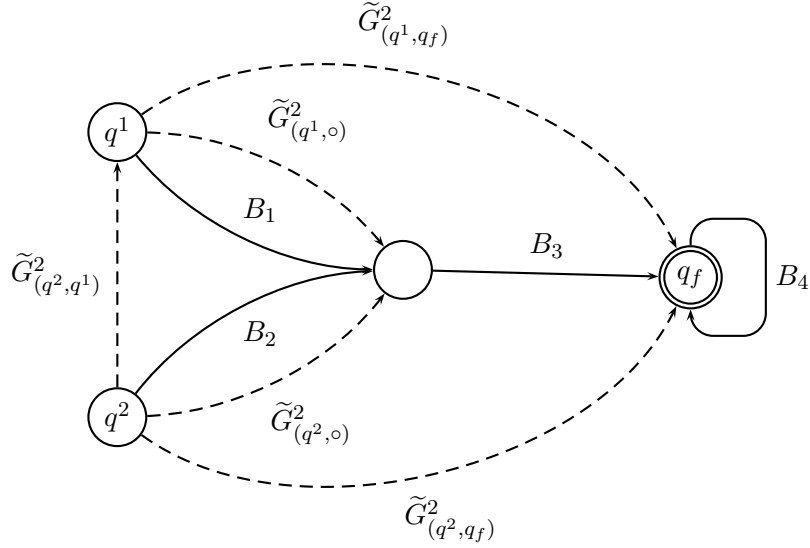

\begin{center}
\begin{psmatrix}[colsep=3,rowsep=1]
  $\quad$ \\
  \circlenode{Q1}{$q^1$} \\
  & \circlenode{Q3}{$\phantom{q^2}$} &  \circlenode[doubleline=true]{Q4}{$q_f$}
\\
  \circlenode{Q2}{$q^2$}
\\
  \psset{arrows=->}
  \ncarc[arcangle=-25]{Q1}{Q3}^{$B_1$}
  \ncarc[arcangle=25]{Q2}{Q3}_{$B_2$}
  \ncline{Q3}{Q4}^{$B_3$}
  \ncloop[angleA=90,angleB=-90,loopsize=1,linearc=.3]{Q4}{Q4}\Aput[0.1]{$B_4$}
  \ncarc[arcangle=50,linestyle=dashed]{Q1}{Q4}^{$\idG^2_{(q^1,
q_f)}$}
  \ncarc[arcangle=25,linestyle=dashed]{Q1}{Q3}\Aput{$\idG^2_{(q^1, \circ)}$}
  \ncarc[arcangle=-25,linestyle=dashed]{Q2}{Q3}\Bput{$\idG^2_{(q^2, \circ)}$}
  \ncline[linestyle=dashed]{Q2}{Q1}\Aput{$\idG^2_{(q^2, q^1)}$}
  \ncarc[arcangle=-50,linestyle=dashed]{Q2}{Q4}\Bput{$\idG^2_{(q^2, q_f)}$}
\end{psmatrix}
\end{center}
\caption{The automaton $A_2$.}
\label{fig4}
\end{figure}

To construct $A_2$ from $A_1$ we repeat the above procedure, taking
into account the additional transitions in $A_1$.  Observe that we do
not add additional transitions between pairs of states that already
have a transition labelled by a set.  Instead, each labelling set may
contain several element sets.  The resulting $A_2$ is given in
Figure~\ref{fig4} where the contents of
\[ \mathcal{G}^2 =
\{\idG^2_{(q^1,\circ)},\idG^2_{(q^1,q_f)},\idG^2_{(q^2,\circ)},\idG^2_{(q^2,q^1)},\idG^2_{(q^2,q_f)}\}
\]
are given in Table~\ref{tab2}.  The columns indicate which command
introduced each element to the set.

\begin{table}
\begin{center}
\psframebox[framearc=.1,framesep=-.005]{%
\begin{tabular}{l|l|l|l|l}
 &
    $d_1$ &
    $d_2$ &
    $d_3$ &
    $d_4$  \\
\hline
 & & &  & \\
 $\idG^2_{(q^1, \circ)}$ & & $\{(a, push_\varepsilon,B_1)\}$ & & \\
                         & & $\{(a, push_\varepsilon,\idG^1_{(q^1,\circ)})\}$ & & \\
 & & & & \\
 $\idG^2_{(q^1, q^f)}$  & $\{B^a_1, B_1,B_3\}$ & $\{(a, push_\varepsilon, \idG^1_{(q^1,q_f)})\}$ & & \\
                        & $\{B^a_1, \idG^1_{(q^1, q_f)}, B_4\}$ & & & \\
                        & $\{B^a_1, \idG^1_{(q^1, \circ)}, B_3\}$ & & & \\
 & & & & \\
 $\idG^2_{(q^2, \circ)}$ & & & $\{(a, push_w, B_1)\}$ & \\
                         & & & $\{(a, push_w,\idG^1_{(q^1, \circ)})\}$ & \\
 & & & & \\
 $\idG^2_{(q^2, q^1)}$ & & & & $\{B^a_1\}$ \\
 & & & & \\
 $\idG^2_{(q^2, q_f)}$ & & & $\{(a, push_w, \idG^1_{(q^1,q_f)})\}$  \\
 & & &  & \\
\end{tabular}}
\end{center}
\caption{The contents of the sets in $\idallG^2$.}
\label{tab2}
\end{table}

If we were to repeat this procedure to construct $A_3$ we would notice
that a kind of fixed point has been reached.  In particular, the
transition structure of $A_3$ will match that of $A_2$ and each
$\idG^3_{(q, q')}$ will match $\idG^2_{(q, q')}$ in everything but the
indices of the labels $\idG^1_{(\_,\_)}$ appearing in the element sets.  We
may write $\idG^3_{(q, q')} = \idG^2_{(q, q')}[2/1]$ where the
notation $[2/1]$ indicates a substitution of the element indices.

So far we have just constructed sets to label the transitions of $A_1$ and
$A_2$.  To complete the construction of $A_1$ we need to construct
the automata $G^1_{(q, q')}$ represented by the
labels $\idG^1_{(q, q')}$ for the
appropriate $q, q'$.  Because each of these new automata will be constructed
from $B_1,\ldots,B_4,B^a_1$, we
build them simultaneously, constructing a single ($1$-store
multi-)automaton $\mathcal{G}^1$ with an
initial state $g^1_{(q, q')}$ for each $\idG^1_{(q, q')}$.  The automaton
$\mathcal{G}^1$ is constructed through the addition of states and
transitions to the disjoint union of $B_1,\ldots,B_4,B^a_1$.  Creating the
automaton $A_2$ is analagous and
$\mathcal{G}^2$ is built through the addition of states and
transitions to $\mathcal{G}^1$.

\begin{figure}
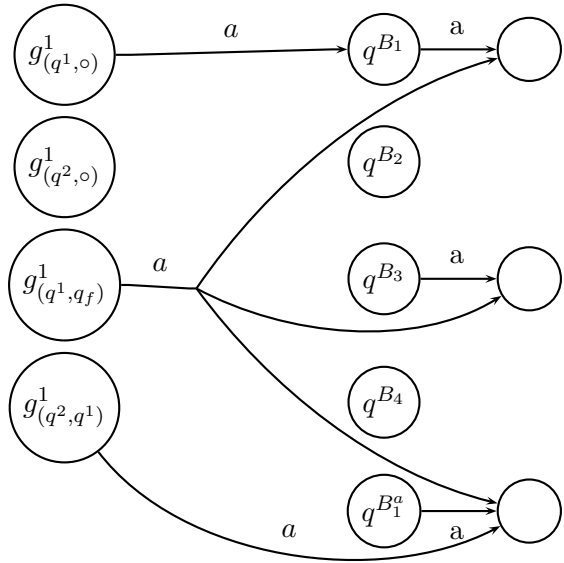

\begin{center}
\begin{psmatrix}[colsep=1,rowsep=0.1]
  \circlenode{G1}{$g^1_{(q^1,\circ)}$} &  & &
\circlenode{B1}{$q^{B_1}$} & \circlenode{B1next}{$\phantom{q^B}$} \\
  \circlenode{G3}{$g^1_{(q^2, \circ)}$} & & &
  \circlenode{B2}{$q^{B_2}$} \\
  \circlenode{G2}{$g^1_{(q^1, q_f)}$}  & \rnode{K1}{}  & &
\circlenode{B3}{$q^{B_3}$} &  \circlenode{B3next}{$\phantom{q^B}$} \\
  \circlenode{G4}{$g^1_{(q^2, q^1)}$} & & & \circlenode{B4}{$q^{B_4}$}\\
    & & & \circlenode{Ba}{$q^{B^a_1}$} &  \circlenode{Banext}{$\phantom{q^B}$}\\

  \psset{angleB=180,arm=0.2,linearc=.1}
  \ncdiag[arm=0.1]{G2}{K1}\Aput{$a$}
  \ncarc[arcangle=20]{->}{K1}{B1next}
  \ncarc[arcangle=-30]{->}{K1}{B3next}
  \ncarc[arcangle=-20]{->}{K1}{Banext}
  \ncdiag{->}{G1}{B1}\Aput{$a$}
  \ncarc[arcangle=-40]{->}{G4}{Banext}\Aput{$a$}
  \ncdiag{->}{B1}{B1next}^a
  \ncdiag{->}{B3}{B3next}^a
  \ncdiag{->}{Ba}{Banext}_a

\end{psmatrix}
\end{center}
\caption{A selective view of $\mathcal{G}^1$.}
\label{fig3}
\end{figure}

The automaton $\mathcal{G}^1$ is given in Figure~\ref{fig3}.
We do not display this automaton in full since the number of
alternating transitions entails a diagram too
complicated to be illuminating.  Instead we will give the basic
structure of the automaton with many transitions omitted.  In
particular we show a transition derived from $\{B^a_1, B_1,
B_3\}$ (from state $g^1_{(q^1, q_f)}$), a transition derived from $\{(a,
push_\varepsilon, B_1)\}$ (from state $g^1_{(q^1, \circ)}$)
 and a transition derived from $\{B^a_1\}$ (from state $g^1_{(q^2, q^1)}$).
Notably, we have omitted any transitions derived from the $push_w$
command.  This is simply for convenience since we do not wish to
further explicate $B_1, B_2, B_3$ or $B_4$.
From this automaton we derive $G^1_{(q^1, \circ)},
G^1_{(q^1, q_f)}, G^1_{(q^2, \circ)}$ and $G^1_{(q^2, q^1)}$ by setting the
initial state to $g^1_{(q^1, \circ)},
g^1_{(q^1, q_f)}, g^1_{(q^2, \circ)}$ and $g^1_{(q^2, q^1)}$ respectively.

\begin{figure}
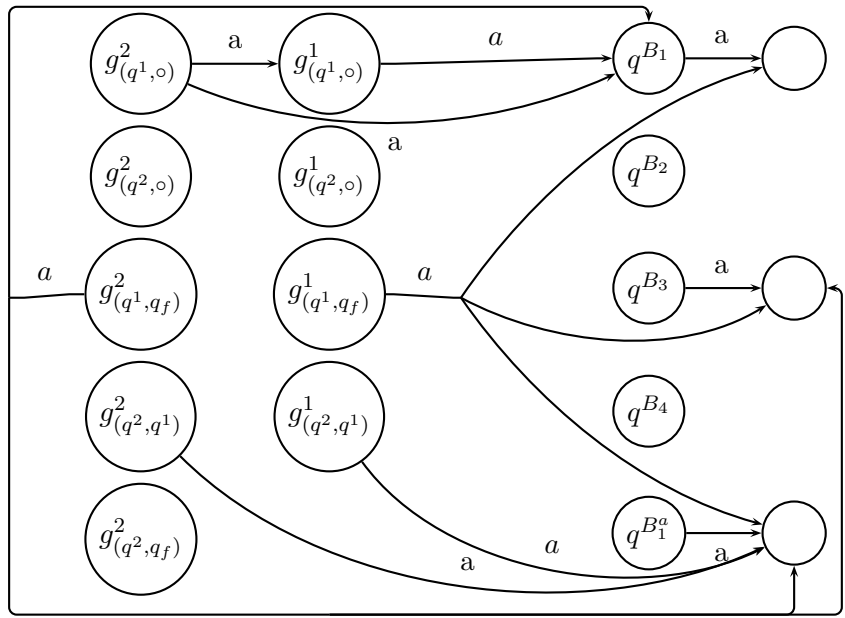
 \begin{center}
\begin{psmatrix}[colsep=1,rowsep=0.1]

  &   \circlenode{G12}{$g^2_{(q^1,\circ)}$} &
  \circlenode{G1}{$g^1_{(q^1,\circ)}$} &  & &
\circlenode{B1}{$q^{B_1}$} & \circlenode{B1next}{$\phantom{q^B}$} \\

  &  \circlenode{G32}{$g^2_{(q^2, \circ)}$} &
  \circlenode{G3}{$g^1_{(q^2, \circ)}$} & & &
  \circlenode{B2}{$q^{B_2}$} \\

  \rnode{K2}{} &  \circlenode{G22}{$g^2_{(q^1, q_f)}$}  &
 \circlenode{G2}{$g^1_{(q^1, q_f)}$}  & \rnode{K1}{}  & &
\circlenode{B3}{$q^{B_3}$} &  \circlenode{B3next}{$\phantom{q^B}$} \\

 &  \circlenode{G42}{$g^2_{(q^2, q^1)}$}  &
  \circlenode{G4}{$g^1_{(q^2, q^1)}$} & & & \circlenode{B4}{$q^{B_4}$}\\

&   \circlenode{G52}{$g^2_{(q^2,q_f)}$}  &
  & & & \circlenode{Ba}{$q^{B^a_1}$} &  \circlenode{Banext}{$\phantom{q^B}$}\\
  & & \rnode{K3}{} & &  & \\

  \psset{angleB=180,arm=0.2,linearc=.1}
  \ncdiag[arm=0.1]{G2}{K1}\Aput{$a$}
  \ncarc[arcangle=20]{->}{K1}{B1next}
  \ncarc[arcangle=-30]{->}{K1}{B3next}
  \ncarc[arcangle=-20]{->}{K1}{Banext}
  \ncdiag{->}{G1}{B1}\Aput{$a$}
  \ncarc[arcangle=-40]{->}{G4}{Banext}\Aput{$a$}
  \ncdiag{->}{B1}{B1next}^a
  \ncdiag{->}{B3}{B3next}^a
  \ncdiag{->}{Ba}{Banext}_a

  \ncdiag[border=2mm]{-}{K2}{G22}\Aput{$a$}
  \ncangle[angleA=-90, armB=0]{-}{K2}{K3}
  \ncangle[angleA=-90, angleB=0]{<-}{Banext}{K3}
  \ncbar{<-}{B3next}{K3}
  \ncangles[angleA=90, angleB=90]{<-}{B1}{K2}

  \ncdiag{->}{G12}{G1}^a
  \ncarc[arcangle=-24]{->}{G12}{B1}_a
  \ncarc[arcangle=-35]{->}{G42}{Banext}^a

\end{psmatrix}
\end{center}
\caption{A selective view of $\mathcal{G}^2$.}
\label{fig5}
\end{figure}

The automaton $\mathcal{G}^2$ is shown in Figure~\ref{fig5}.  Again, due to the
illegibility of a complete diagram, we omit many of the transitions.  The new
transition from $g^2_{(q^1, q_f)}$ is derived from the set $\{B^a_1, B_3,
\idG^1_{(q^1, \circ)}\}$.  One of the transitions from $g^2_{(q^1,\circ)}$ and
the only transition from $g^2_{(q^2, q^1)}$ are inherited from their
corresponding states in the previous automaton.  This inheritance ensures that
we do not lose information from the previous iteration. The uppermost transition
from $g^2_{(q^1, \circ)}$ derives from $\{(a, push_\varepsilon, \idG^1_{(q^1,
\circ)})\}$.  
From this automaton we derive $G^1_{(q^1, \circ)},
G^1_{(q^1, q_f)}, G^1_{(q^2, \circ)}$ and $G^1_{(q^2, q^1)}$.

We have now constructed the automata $A_1$ and $A_2$.  We could then
repeat this procedure to generate $A_3, A_4,\ldots$, resulting in an
infinite sequence of automata that is sound and complete with respect
to $Pre^*(\lang(A_0))$.

\begin{figure}
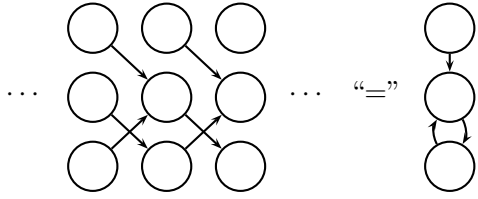

\begin{center}
\begin{psmatrix}[rowsep=0.2,colsep=0.3]
  &
  \circlenode{L1}{$\phantom{q_1}$} &
  \circlenode{L2}{$\phantom{q_1}$} &
  \circlenode{L3}{$\phantom{q_1}$} &
  & &
  \circlenode{L4}{$\phantom{q_1}$} &
  \\
  $\cdots$ &
  \circlenode{N1}{$\phantom{q_1}$} &
  \circlenode{N2}{$\phantom{q_1}$} &
  \circlenode{N3}{$\phantom{q_1}$} &
  $\cdots$ & ``='' &
  \circlenode{N4}{$\phantom{q_1}$} &

  \\
   &
  \circlenode{M1}{$\phantom{q_1}$} &
  \circlenode{M2}{$\phantom{q_1}$} &
  \circlenode{M3}{$\phantom{q_1}$} &
   &   &
  \circlenode{M4}{$\phantom{q_1}$} &

  \ncline{->}{N1}{M2}
  \ncline{->}{N2}{M3}
  \ncline{->}{M1}{N2}
  \ncline{->}{M2}{N3}

  \ncarc[arcangle=30]{->}{N4}{M4}
  \ncarc[arcangle=30]{->}{M4}{N4}

  \ncline{->}{L1}{N2}
  \ncline{->}{L2}{N3}
  \ncline{->}{L4}{N4}
\end{psmatrix}
\caption{Collapsing a repetitive chain of new states.}
\label{selfloops}
\end{center}
\end{figure}

To construct $A_*$ such that $\lang(A_*) = \bigcup_{i \geq 0} \lang(A_i)$ we observe that since a fixed point was reached at
$A_2$, the update to each $\mathcal{G}^i$ to create
$\mathcal{G}^{i+1}$ will use similar recipes and hence become repetitive.
  This will lead to an
infinite chain with an unvarying pattern of edges.  This chain can be
collapsed as shown in Figure~\ref{selfloops}.

In particular, we are no longer required to add new states to
$\mathcal{G}^2$ to construct $\mathcal{G}^i$ for $i > 2$.
Instead, we fix the update instructions $\idG^2_{(q, q')}[2/1]$ for
all $q, q'$ and manipulate $\mathcal{G}^2$ as we manipulated the
order-$2$ structure of $A_0$ to create $A_1$ and $A_2$.  We write
$\newautG^i$ to distinguish these automata from the automata
$\mathcal{G}^i$ generated without fixing the state-set.

Because $\Sigma$ and the state-set are finite (and remain unchanged), this procedure will
reach another fixed point $\newautG^*$ when the transition relation is \emph{saturated}
and $\newautG^i = \newautG^{i+1}$.  The automaton $A_*$ has the
transition structure that became fixed at $A_2$ labelled with automata
derived from $\newautG^*$.  This automaton will be
sound and complete with respect to $Pre^*(\lang(A_0))$.

An abbreviated diagram of $\newautG^*$ is given in
Figure~\ref{fig8}.  We have hidden, for clarity, the transition derived from
$\{B^a_l,B_3, \idG^1_{(q^1,\circ)}\}$ in Figure~\ref{fig5}.
Instead, we show the transition introduced
for the set $\{B^a_1,B_3, \idG^1_{(q^1,\circ)}\}[2/1] = \{B^a_1,B_3, \idG^2_{(q^1,\circ)}\}$  during the construction of $\newautG^*$.
We have also added the self-loop added by
$\{(a, push_\varepsilon, G^1_{(q^1, \circ)})\}[2/1] = \{(a, push_\varepsilon, G^2_{(q^1, \circ)})\}$ that enabled the
introduction of this transition.
\begin{figure}
\begin{center}
\begin{psmatrix}[colsep=1,rowsep=0.1]
 \\
 \\
  &   \circlenode{G12}{$g^2_{(q^1,\circ)}$} &
  \circlenode{G1}{$g^1_{(q^1,\circ)}$} &  & &
\circlenode{B1}{$q^{B_1}$} & \circlenode{B1next}{$\phantom{q^B}$} \\

  &  \circlenode{G32}{$g^2_{(q^2, \circ)}$} &
  \circlenode{G3}{$g^1_{(q^2, \circ)}$} & & &
  \circlenode{B2}{$q^{B_2}$} \\

  \rnode{K2}{} &  \circlenode{G22}{$g^2_{(q^1, q_f)}$}  &
 \circlenode{G2}{$g^1_{(q^1, q_f)}$}  & \rnode{K1}{}  & &
\circlenode{B3}{$q^{B_3}$} &  \circlenode{B3next}{$\phantom{q^B}$} \\

 &  \circlenode{G42}{$g^2_{(q^2, q^1)}$}  &
  \circlenode{G4}{$g^1_{(q^2, q^1)}$} & & & \circlenode{B4}{$q^{B_4}$}\\

&   \circlenode{G52}{$g^2_{(q^2,q_f)}$}  &
  & & & \circlenode{Ba}{$q^{B^a_1}$} &  \circlenode{Banext}{$\phantom{q^B}$}\\
  & & \rnode{K3}{} & &  & \\

  \psset{angleB=180,arm=0.2,linearc=.1}
  \ncdiag[arm=0.1]{G2}{K1}\Aput{$a$}
  \ncarc[arcangle=20]{->}{K1}{B1next}
  \ncarc[arcangle=-30]{->}{K1}{B3next}
  \ncarc[arcangle=-20]{->}{K1}{Banext}
  \ncdiag{->}{G1}{B1}\Aput{$a$}
  \ncarc[arcangle=-40]{->}{G4}{Banext}\Aput{$a$}
  \ncdiag{->}{B1}{B1next}^a
  \ncdiag{->}{B3}{B3next}^a
  \ncdiag{->}{Ba}{Banext}_a

  \ncdiag[border=2mm]{-}{K2}{G22}\Aput{$a$}
  \ncangle[angleA=-90, armB=0]{-}{K2}{K3}
  \ncangle[angleA=-90, angleB=0]{<-}{Banext}{K3}
  \ncbar{<-}{B3next}{K3}
  \ncangle[angleA=180, angleB=90]{<-}{G12}{K2}
  \nccircle{->}{G12}{0.75}^a

  \ncdiag{->}{G12}{G1}^a
  \ncarc[arcangle=-24]{->}{G12}{B1}_a
  \ncarc[arcangle=-35]{->}{G42}{Banext}^a

\end{psmatrix}
\end{center}
\caption{A selective view of $\newautG^*$.}
\label{fig8}
\end{figure}

\subsection{Preliminaries}

We now discuss the algorithm more formally.  We begin by describing the
transitions labelled by $G^i_{(q_1, Q_2)}$ before discussing the construction of
the sequence $A_0, A_1, \ldots$ and the automaton $A^*$.

To aid in the construction of an automaton representing
$Pre^*(C_{Init})$ we introduce a new kind of transition to the
$2$-store automata.  These new transitions are introduced during the processing of the
APDS commands.  They are labelled with
place-holders that will eventually be converted into $1$-store
automata.

Between any state $q_1$ and set of states $Q_2$ we add
at most one transition.  We associate this transition with an
identifier $\idG_{(q_1, Q_2)}$.  To describe our algorithm we will
define sequences of automata, indexed by $i$.  We superscript the identifier to
indicate to which automaton in the sequence it belongs.  The identifier
$\idG^i_{(q_1, Q_2)}$ is associated with a set that acts as a recipe
for updating the $1$-store automaton described by
$\idG^{i-1}_{(q_1, Q_2)}$ or creating a new automaton if
$\idG^{i-1}_{(q_1, Q_2)}$ does not exist.  Ultimately, the constructed
$1$-store automaton will label the new transition.  In the sequel, we will
confuse the notion of an identifier and its associated set.  The intended usage
should be clear from the context.

The sets are in a kind of disjunctive normal form.  A set
$\{S_1,\ldots,S_m\}$ represents an automaton that accepts the
union of the languages accepted by the automata described by $S_1,\ldots,S_m$.  Each
set $S \in \{S_1,\ldots,S_m\}$ corresponds to a possible effect of a
command $d$ at order-$1$ of the automaton.  The automaton
described by $S = \{\alpha_1,\ldots,\alpha_m\}$ accepts the intersection of languages described by
its elements $\alpha_t$ ($t \in \{1,\ldots,m\}$).  An element that is an automaton $B$ refers directly to
the automaton $B$.  Similarly, an identifier $\idG^i_{(q_1, Q_2)}$ refers to its
corresponding automaton.  Finally, an element of the form $(a, push_w,
\theta)$ refers to an automaton capturing the effect of applying the
inverse of the $push_w$ command to the stacks accepted by the automaton
represented by $\theta$; moreover, the $top_1$ character of the stacks
accepted by the new automaton will be $a$.  It is a consequence of
the construction that for any $S$ added during the algorithm, if $(a, push_w,
\theta) \in S$ and $(a', push_{w'}, \theta') \in S$ then $a = a'$.

Formally, to each $\idG^i_{(q_1, Q_2)}$ we attach a
subset of
\[ 2^{\hbox{$ \mathcal{B} \cup \idallG^{i-1} \cup (\Sigma \times \mathcal{\mathcal{O}}_1 \times (\mathcal{B} \cup
\idallG^{i-1}))$}} \] where $\mathcal{B}$ is
the set of all $1$-store automata occurring in $A_0$ and all
automata of the form $B^a_{1}$.  Further, we denote
the set of all identifiers $\idG^{i}_{(q, Q)}$ in
$A_i$ as $\idallG^i$.  The sets $\mathcal{B}$ and
$\mathcal{O}_1$ are finite by definition.  The size of the set
$\idallG^i$ for any $i$ is finitely bound by the (fixed) state-set of $A_i$.

We build the automata for all
$\idG^{i}_{(q_1, Q_2)} \in \idallG^i$ simultaneously.  That is, we
create a single automaton $\mathcal{G}^i$ associated with the set
$\idallG^i$.  This automaton has
a state $g^{i}_{(q_1,Q_2)}$ for each $\idG^{i}_{(q_1,Q_2)} \in \idallG^i$.  The automaton
$G^{i}_{(q_1, Q_2)}$
labelling the transition  $q_1 \longrightarrow_i Q_2$ is the automaton
$\mathcal{G}^i$ with $g^i_{(q_1,Q_2)}$ as its initial state.

The automaton $\mathcal{G}^i$ is built inductively.  We set
$\mathcal{G}^0$ to be the disjoint union of all automata in
$\mathcal{B}$.
We define $\mathcal{G}^{i+1} =
T_{\idallG^{i+1}}(\mathcal{G}^i)$ where
$T_{\idallG^j}(\mathcal{G}^i)$ is given in Definition~\ref{TGorder2}.
In Section~\ref{astarorder2} it will be seen that $j$ is not always $(i + 1)$.
\begin{defi}\rm
\label{TGorder2}
Given an automaton $\mathcal{G}^i = (\mathcal{Q}^i, \Sigma,
\Delta^i,\_,\mathcal{Q}_f)$ and a set of identifiers (and associated sets)
$\idallG^j_1$, we define,
\[ \mathcal{G}^{i+1} =
   T_{\idallG^j}(\mathcal{G}^i) =
   (\mathcal{Q}^{i+1}, \Sigma, \Delta^{i+1}, \_, \mathcal{Q}_f) \]
where $\mathcal{Q}^{i+1} = \mathcal{Q}^i \cup \{\ g^{j}_{(q_1,
Q_2)}\ |\ \idG^j_{(q_1, Q_2)} \in \idallG^j\ \}$, $\Delta^{i+1} =
\Delta^{inherited} \cup
\Delta^{new} \cup \Delta^i$, and,
\[ \begin{array}{rcl}
     \Delta^{inherited} &=& \{\ g^j_{(q_1, Q_2)}
\stackrel{a}{\longrightarrow} Q\ |\ (g^{j-1}_{(q_1, Q_2)}
\stackrel{a}{\longrightarrow} Q) \in \Delta^i\ \} \\
     \Delta^{new} &=& \left\{\ g^j_{(q_1, Q_2)}
\stackrel{b}{\longrightarrow} Q\ |\
                      \begin{array}{l}
                        \idG^j_{(q_1, Q_2)} \in \idallG^j\mathrm{\ and\ }
                        b \in \Sigma \mathrm{\ and\ } (1)
                      \end{array}\ \right\}
   \end{array}
\]
where $(1)$ requires $\{\alpha_1,\ldots,\alpha_r\} \in
\idG^j_{(q_1, Q_2)}$, $Q = Q_1 \cup
\ldots \cup Q_r$ and for each $t \in \{1,\ldots,r\}$ we have,
\begin{enumerate}[$\bullet$]
\item
If $\alpha_t = \theta$, then $(q^{\theta}
\stackrel{b}{\longrightarrow} Q_t) \in \Delta^i$.

\item
If $\alpha_t = (a, push_w, \theta)$, then $b=a$
and $q^{\theta} \stackrel{w}{\longrightarrow} Q_t$ is a run of $\mathcal{G}^i$.
\end{enumerate}
\end{defi}

  There are two key
parts to Definition~\ref{TGorder2}.  During the first stage we
add a new initial state for each automaton forming a part of
$\mathcal{G}^{i+1}$.  By adding new initial states, rather than
using the previous set of initial states, we guarantee that no
unwanted cycles are introduced, which may lead to the erroneous acceptance of
certain stores.  We ensure that each $1$-store accepted by
$\mathcal{G}^i$ is accepted by $\mathcal{G}^{i+1}$ --- and the
set of accepted stores is increasing --- by inheriting transitions
from the previous set of initial states.

During the second stage we add transitions between the set of new initial
states and the state-set of $\mathcal{G}^i$ to capture the effect of a backwards
application of
the
APDS commands to $\lang(A_i)$.  Intuitively, we only add new transitions to
the initial states
because all stack operations affect the top of the stack, leaving the
remainder unchanged.

There are two different forms for the elements $\alpha_t \in
\{\alpha_1,\ldots,\alpha_r\}$.  If $\alpha_t$ refers directly to an
automaton, then we require that the new store is also accepted by
the automaton referred to by $\alpha_t$.
We simply inherit the initial transitions of that automaton in a similar
manner to the first stage of $T_{\idallG^j}(\mathcal{G}^i)$.  If $\alpha_t$ is of the
form $(a, push_w, \theta)$, then it corresponds to the effects of a command
$(p, a, \{\ldots,(push_w, p'),\ldots\})$. The new
store must have the character $a$ as its  $top_1$ character, and the
store resulting from the application of the operation $push_w$ must be
accepted by the automaton represented by $\theta$.
That is, the new state must accept
all stores of the form $a w'$ when the store $w w'$ is accepted by $\theta$.

\subsection{Constructing the Sequence $A_0, A_1, \ldots$}

For a given order-$2$ APDS with commands $\mathcal{D}$ we define
$A_{i+1} = T_\mathcal{D}(A_i)$ where the operation $T_{\mathcal{D}}$
follows.  We assume $A_0$ has a state $q^\varepsilon_f$ with no outgoing
transitions and a state $q^*_f$ from which all stores are accepted.
\begin{defi}
\label{gena0a1order2}
Given an automaton $A_i = (\mathcal{Q}, \Sigma, \Delta^i,\{q^1,\ldots,q^z\},\mathcal{Q}_f)$ and a set of commands $\mathcal{D}$,
we define,
\[ A_{i+1} =
   T_{\mathcal{D}}(A_i) =
   (\mathcal{Q}, \Sigma, \Delta^{i+1}, \{q^1,\ldots,q^z\}, \mathcal{Q}_f) \]
where $\Delta^{i+1}$ is given below.

\medskip

We begin by defining the set of labels
$\idallG^{i+1}$.  This set contains labels on transitions
present in $A_i$, and
labels on transitions derived from $\mathcal{D}$.  That is,
\[ \begin{array}{rcl}
     \idallG^{i+1} &=&
              \begin{array}{l}
                \left\{\ \idG^{i+1}_{(q^j, Q)}\ |\ (q^j
\xrightarrow{\idG^i_{(q^j,
Q)}} Q) \in \Delta^i \mathrm{\ and\ } j \in \{1,\ldots,z\}\ \right\}\ \cup\
                \left\{\ \idG^{i+1}_{(q^j, Q)}\ |\ (2)\ \right\}
              \end{array}
    \end{array}
\]
The contents of the associated sets $\idG^{i+1}_{(q, Q)} \in
\idallG^{i+1}$ are defined $\idG^{i+1}_{(q^j, Q)} = \left\{\ S\ |\
 (2)\ \right\}$ where $(2)$ requires $(p^j, a, \{(o_1, p^{k_1}),\ldots,(o_m, p^{k_m})\}) \in \mathcal{D}$,
$Q = Q_1 \cup \ldots \cup Q_m$, $S = S_1 \cup \ldots \cup S_m$ and for each $t \in \{1,\ldots,m\}$ we have,
\begin{enumerate}[$\bullet$]
\item
If $o_t = push_2$, then $S_t = \{B^a_{1}\} \cup
\widetilde{\theta}_1\cup\widetilde{\theta}_2$ and there
exists a path $q^{k_t} \stackrel{\widetilde{\theta}_1}{\longrightarrow}_i Q'
\stackrel{\widetilde{\theta}_2}{\longrightarrow}_i Q_t$ in $A_i$.

\item
If $o_t = pop_2$, then $S_t = \{B^a_{1}\}$ and $Q_t = \{q^{k_t}\}$.
Or, if $q^j \stackrel{\triangledown}{\longrightarrow}_i \{q^\varepsilon_f\}$ exists in
$A_i$, we may have $S_t = \{B^a_{1}\}$ and $Q_t = \{q^\varepsilon_f\}$.

\item
If $o_t = push_w$ then $S_t =
\{(a, push_w, \theta)\}$ and there exists a transition
$q^{k_t} \stackrel{\theta}{\longrightarrow}_i Q_t$ in $A_i$.
\end{enumerate}
Finally, we give the transition relation $\Delta^{i+1}$.
\[ \Delta^{i+1} = \begin{array}{l}
                     \left\{ q \stackrel{B}{\longrightarrow} Q\ |\
                        \begin{array}{l}
                           (q \stackrel{B}{\longrightarrow} Q) \in \Delta^i \\
                                \mathrm{and\ } B \in \mathcal{B}
                        \end{array}\ \right\}\ \cup\
                     \left\{ q
\xrightarrow{\idG^{i+1}_{(q, Q)}}
Q\ |\ \idG^{i+1}_{(q, Q)} \in \idallG^{i+1} \right\}
   \end{array}
\]
We can construct an automaton whose transitions are $1$-store automata by replacing
each set $\idG^{i+1}_{(q, Q)}$ with the automaton $G^{i+1}_{(q, Q)}$ which is $\mathcal{G}^{i+1}$ with
initial state $g^{i+1}_{(q, Q)}$, where $\mathcal{G}^{i+1} =
T_{\idallG^{i+1}}(\mathcal{G}^i)$.  Note that $\mathcal{G}^i$ is assumed by
induction.  In the base case, $\mathcal{G}^0$ is the disjoint union of all
automata in $\mathcal{B}$.
\end{defi}
The above construction is similar to Definition~\ref{TGorder2}.
However, because we do
not change the initial states of the automaton, we do not have to
perform the inheritance step.  Furthermore the set of commands
$\mathcal{D}$ specify how the automata should be updated, rather than
a set $\idallG^i$.  A command $(p^j, a, \{(o_1, p^{k_1}),\ldots,(o_m,
p^{k_m})\})$ takes the place of a set $\{\alpha_1,\ldots,\alpha_m\}$.

The
contents of $S_t$ and $Q_t$ depend on the operation $o_t$.
If $o_t$ is of a lower order than $2$ (that is, a $push_w$ command)
 then $o_t(\gamma w) = o_t(\gamma) w$
for any store $\gamma w$.  Hence we inherit the first transition from
the initial state of the automaton represented $\theta$, but pass the required
constraint (using $S_t = \{(a, o_t, B)\}$) to the lower orders of the
automaton.

Otherwise $o_t$ is a $pop_2$ or $push_2$ operation.  If is a $push_2$ command,
then $push_2(\gamma w') = \gamma\gamma w'$, and hence we use $S_t$ to ensure
that the top store $\gamma$ of $\gamma w'$ is accepted by the first two
transitions from the initial state of the automaton represented by $\theta$ and
we use $Q_t$ to ensure that the tails of the stores match.

In case $o_t$ is a $pop_2$ operation and the new store is simply the old store
with an additional $2$-store on top (that is $pop_2(\gamma w') = w'$).  Thus, $Q_t$ is the initial state
of the automaton represented by $\theta$ and $S_t$ contains the
automaton $B^a_1$, which ensures that the $top_1$ character of the new
store is $a$.
We also need to consider
the undefined store $\triangledown$.  This affects the
processing of $pop_2$ operations since their result is not always
defined.  Hence, when considering which new stores may be accepted by
$A_{i+1}$, we check whether
the required undefined configuration is accepted by $A_i$.
This is witnessed by the presence of a $\triangledown$
transition from $p^j$.  If the result may be undefined, we accept all
stores that do not have an image under the $pop_2$ operation.
That is, all stores of the form $[\gamma]$.

By repeated applications of $T_\mathcal{D}$ we construct the sequence
$A_0, A_1, \ldots$ which is sound and complete with respect to $Pre^*(C_{Init})$.

\begin{property}
For any configuration $\langle p^j, \gamma\rangle$ it is the case that  $\gamma \in
\lang(A_i^{q^j})$ for some $i$ iff $\langle p^j, \gamma\rangle \in
Pre^*(C_{Init})$.
\end{property}
\begin{proof}
From Property~\ref{infseqsound} and Property~\ref{infseqcomp}.
\end{proof}

\subsection{Constructing the Automaton $A_*$}
\label{astarorder2}

We need to
construct a finite representation of the sequence $A_0, A_1, \ldots$ in a finite amount
of time.  To do this we will construct an automaton $A_*$ such that
$\lang(A_*) = \bigcup_{i \geq 0} \lang(A_i)$.
We begin by introducing some notation and a notion of subset modulo $i$ for
the sets $\idG^i_{(q_1, Q_2)}$.

\begin{defi} \
\begin{enumerate}[(1)]
\item
Given $\theta \in \mathcal{B} \cup \idallG^{i'}$ for some $i'$, let
\[
   \theta[j/i] = \left\{\begin{array}{ll}
                            \theta & \mathrm{if\ } \theta \in
\mathcal{B} \\
                            G^j_{(q_1, Q_2)} & \mathrm{if\ } \theta =
G^i_{(q_1, Q_2)} \in \idallG^{i'}
                         \end{array}\right.
\]

\item
For a set $S$ we define $S[j/i]$ such that,
\begin{enumerate}[(a)]
\item
We have $\theta \in S$ iff we have $\theta[j/i] \in
S[j/i]$, and
\item
We have $(a, o, \theta) \in S$ iff we have $(a, o, \theta[j/i]) \in
S[j/i]$.
\end{enumerate}
\item
We extend the notation $[j/i]$ to nested sets of sets structures in a
point-wise fashion.
\end{enumerate}
\end{defi}

\begin{defi} \
\begin{enumerate}[(1)]
\item
We write $\idG^i_{(q_1, Q_2)} \lesssim \idG^j_{(q_1, Q_2)}$ iff for each
$S \in \idG^i_{(q_1, Q_2)}$ we have $S[j-1/i-1] \in \idG^j_{(q_1,
Q_2)}$.
\item
If $\idG^i_{(q_1, Q_2)} \lesssim \idG^j_{(q_1, Q_2)}$ and
$\idG^j_{(q_1,
Q_2)} \lesssim \idG^i_{(q_1, Q_2)}$, then we write  $\idG^i_{(q_1, Q_2)}
\simeq \idG^j_{(q_1, Q_2)}$.

\item
Furthermore, we extend the notation to sets.  That is, $\idallG^i_l
\lesssim \idallG^j_l$ iff for all $\idG^i_{(q_1, Q_2)} \in
\idallG^i_l$ we have $\idG^j_{(q_1, Q_2)} \in \idallG^j_l$ and
$\idG^i_{(q_1, Q_2)} \lesssim \idG^j_{(q_1, Q_2)}$.
\end{enumerate}
\end{defi}

We now show that a fixed point is reached at order-$2$.  That we reach
a fixed point is important, since, when $\idallG^i \simeq
\idallG^{i+1}$ there are two key consequences.  Firstly, for all $q_1$
and $Q_2$, we
have $\idG^i_{(q_1, Q_2)} \in \idallG^{i}$ iff we
also have $\idG^{i+1}_{(q_1, Q_2)} \in \idallG^{i+1}_{(q_1, Q_2)}$.  This
means that, if we ignore the automata labelling the edges of $A_i$ and
$A_{i+1}$, the two automata have the same transition structure.
The second consequence follows from the first: we have $\idG^i_{(q_1, Q_2)} \simeq \idG^{i+1}_{(q_1, Q_2)}$ for all
$q_1$ and $Q_2$.  That is, the automata labelling the edges of $A_{i}$ and
$A_{i+1}$ will be updated in the same manner.  It is this repetition
that allows us to fix the state-set at order-$1$, and thus reach a
final fixed point.

\begin{property}
\label{nfixedpointorder2}
There exists $i_{1} > 0$ such that $\idallG^{i} \simeq
\idallG^{i_{1}}$ for all $i \geq i_{1}$.
\end{property}
\begin{proof}(Sketch)
Since the order-$1$ state-set in $A_i$ remains constant and we add at most one
transition between any state $q_1$ and set of states $Q_2$, there
is some $i_{1}$
where no more transitions are
added at order-$2$.  That  $\idallG^{i} \simeq
\idallG^{i_{1}}$ for all $i \geq i_{1}$ follows since the
contents of $\idG^i_{(q_1, Q_2)}$ and $\idG^{i_{1}}_{(q_1, Q_2)}$ are derived from the
same transition structure.
\end{proof}

Once a fixed point has been reached at order-$2$, we can fix the
state-set at order-$1$.

\begin{lem}
\label{finiteorderorder2}
Suppose we have constructed, as above, a sequence of automata $\mathcal{G}^0, \mathcal{G}^1, \ldots$
with the
associated sets $\idallG^0, \idallG^1, \ldots$.  Further, suppose
there exists an $i_1$ such that for all $i \geq i_1$ we have $\idallG^i \simeq
\idallG^{i_1}$.  We can define a sequence of automata
$\newautG^{i_1},\newautG^{i_1 + 1}, \ldots$ such that the
state-set in $\newautG^i$ remains constant
and there exists $i_0$ such that $\newautG^{i_0}$ characterises the
sequence --- that is, the following are equivalent for all $w$,
\begin{enumerate}[\em(1)]
\item
The run $g^{i_1}_{(q, Q')}
\stackrel{w}{\longrightarrow}_{i} Q_1$ with $Q_1 \subseteq
\mathcal{Q}_f$ exists in $\newautG^i$ for some $i$.
\item
The run $g^{i_1}_{(q, Q')} \stackrel{w}{\longrightarrow}_{i_0} Q_2$
with $Q_2 \subseteq \mathcal{Q}_f$ exists in $\newautG^{i_0}$.
\item
The run $g^{i'}_{(q, Q')} \stackrel{w}{\longrightarrow}_{i'}
Q_3$ with $Q_3 \subseteq \mathcal{Q}_f$ exists in $\mathcal{G}^{i'}$
for some $i'$.
\end{enumerate}
\end{lem}
\begin{proof}
Follows from the definition of $\newautG^{i+1} = T_{\idallG^{i_1}[i_1/i_1
- 1]}(\newautG^i_1)$, Lemma~\ref{lis1fixedpoint},
Lemma~\ref{lis1oneway} and Lemma~\ref{lis1theother}.
\end{proof}

We use $\newautG^{i+1} = T_{\idallG^{i_1}[i_1/i_1-1]}(\newautG^i)$ to construct
the sequence $\newautG^{i_1}, \newautG^{i_1 + 1}, \ldots$ with $\newautG^{i_1} =
\mathcal{G}^{i_1}$.  Intuitively, since the transitions from the states
introduced to define $\mathcal{G}^{i}$ for $i \geq i_1$ are derived from similar
sets, we can compress the subsequent repetition into a single set of new states.
The substitution $\idallG^{i_1}[i_1/i_1 - 1]$ makes the sets in $\idallG^{i_1}$
self-referential.  This generates the loops shown in Figure~\ref{selfloops}.
Since the state-set of this new sequence does not change and the alphabet
$\Sigma$ is finite, the transition structure will become saturated.

We define $\newautG^* = \newautG^{i_0}$ letting $g^*_{(q_1, Q_2)} =
g^{i_1}_{(q_1, Q_2)}$ for each $g^{i_1}_{(q_1, Q_2)}$.  Finally, we show that we
can construct the automaton $A_*$.

\begin{property}
\label{constructAstarorder2}
There exists an automaton $A_*$ which is sound and complete with
respect to $A_0, A_1, \ldots$ and hence computes the set $Pre^*(C_{Init})$.
\end{property}
\begin{proof}
By Property~\ref{nfixedpointorder2} there is some $i_1$
with $\idallG^i \simeq \idallG^{i_{1}}$ for all $i
\geq i_{1}$.
By Lemma~\ref{finiteorderorder2}, we have $\newautG^* = \newautG^{i_0}$.  We then define
$A_*$ from $A_{i_{1}}$ with each transition $q \longrightarrow_* Q'$ in $A_*$
labelled with the automaton $\finitesingleautG^*_{(q, Q')}$ from $\newautG^* =
\newautG^{i_0}$.
\end{proof}

Thus, we have the following algorithm for constructing $A_*$:

\medskip

\begin{center}
\psframebox[framearc=.2]{
\parbox{.95\textwidth}{
  \begin{enumerate}[(1)]
  \item
  Given $A_0$, iterate $A_{i+1} = T_\mathcal{D}(A_i)$ until the fixed
  point $A_{i_{1}}$ is reached.
  \item
  Iterate $\newautG^{i+1} =
  T_{\idallG^{i_1}_l[i_1/i_1-1]}(\newautG^i)$ to generate
  the fixed point $\newautG^{*}$ from $\mathcal{G}^{i_1}$.

  \item
  Construct $A_{*}$ by labelling the transitions of
  $A_{i_1}$ with automata derived from $\newautG^{*}$.
  \end{enumerate}
}}
\end{center}

\subsection{The General Case and Complexity}
\label{fullconstruct}

We may generalise our algorithm to order-$n$ for all $n$ by extending
Definition~\ref{TGorder2} to $n$-store automata using similar techniques to
those used in Definition~\ref{gena0a1order2}.  Termination is reached through a
cascading of fixed points.  As we fixed the state-set at order-$1$ in the
order-$2$ case, we may fix the state-set at order-$(n-1)$ in the order-$n$ case.
We may then generalise Property~\ref{nfixedpointorder2} and
Lemma~\ref{finiteorderorder2} to find a sequence of fixed points
$i_n,\ldots,i_0$, from which $A_*$ can be constructed.  For a complete
description of this procedure, we refer the reader to Hague's forthcoming Ph.D.
thesis~\cite{HThesis}.

We claim our algorithm runs in $n$-EXPTIME.  Intuitively, when the
state-set $\mathcal{Q}$ is fixed at order-$1$ of the store automaton, we add at
most $\mathcal{O}(2^{|\mathcal{Q}|})$ transitions (since we never remove states,
it is this final stage that dominates the complexity).  At orders $l > 1$ we
add at most $\mathcal{O}(2^{|\mathcal{Q}|})$ new transitions, which
exponentially increases the state-set at order-$(l-1)$.  Hence, the
algorithm runs in $n$-EXPTIME.  This algorithm is optimal since reachability
games over higher-order PDSs are $n$-EXPTIME-complete~\cite{CW07}.
An alternative proof of $n$-EXPTIME-hardness --- by reduction from the
non-emptiness of order-$(n+1)$ PDA --- is due to appear in Hague's Ph.D.
thesis~\cite{HThesis}.  It was shown by Engelfriet that the non-emptiness problem
for order-$(n+1)$ PDSs is $n$-EXPTIME-complete~\cite{E83}.

When the higher-order PDS is nondeterministic (rather than alternating), we add
at most $|\mathcal{Q}|^2$ transitions at order-$n$.  Hence, the complexity is
$(n-1)$-EXPTIME, matching the lower-bound of the non-emptiness problem for
higher-order PDA (as acceptors of word languages).

\section{Applications}
\label{applications}

In this section we discuss some of the applications of our algorithm
to decision problems over higher-order PDSs.

\subsection{Model-Checking Linear-Time Temporal Logics}

Bouajjani\etal\ use their backwards reachability algorithm to provide a
model-checking algorithm for linear-time temporal logics over the
configuration graphs of pushdown systems~\cite{BEM97}.  In this
section we show that this work permits a simple generalisation to
higher-order PDSs.

Let $Prop$ be a finite set of  atomic propositions and $(\mathcal{P},
\mathcal{D}, \Sigma)$ be a higher-order PDS with a labelling function $\Lambda : \mathcal{P} \rightarrow
2^{Prop}$ which assigns to each control state
a set of propositions deemed to be \emph{true} at that state.  Given
formula $\phi$ of an $\omega$-regular logic such as LTL or $\mu$TL, we
calculate the set of configurations $C$ of  $(\mathcal{P},
\mathcal{D}, \Sigma)$ such
that every run from each $c \in C$ satisfies $\phi$.

It is well known that any formula of an $\omega$-regular logic has a B\"uchi
automaton representation~\cite{T90,V95,V88} etc..  We form the product of the
higher-order PDS and the B\"uchi automaton corresponding to the negation of
$\phi$.  This gives us a higher-order B\"uchi PDS; that is, a higher-order PDS with a set
$\mathcal{F}$ of accepting control states.  Thus, model-checking
reduces to the non-emptiness problem for higher-order B\"uchi PDSs.  Specifically,
we compute the
set of configurations from which there is an infinite run visiting
configurations with control states in $\mathcal{F}$ infinitely often.
Note that $C$ is the complement of this set.

This problem can be reduced further to a number of applications of the
reachability problem.  We present a generalisation of the reduction of
Bouajjani\etal.  Let $[^1 a ]^1$ denote the order-$1$ stack
consisting of a single character $a$ and $[^l a ]^l$ for $l > 1$
denote the stack consisting of a single order-$(l - 1)$ stack $[^{(l-1)} a
]^{(l-1)}$.
\begin{prop}
\label{buchiprop}
Let $c$ be a configuration of an order-$n$ B\"uchi PDS $BP$.  It is the case
that $BP$ has an
accepting run from $c$ iff there exist distinct configurations $\langle p^j,
[^n a ]^n \rangle$ and $\langle p^j, \gamma_2
\rangle$ with $top_1(\gamma_2) = a$ and a configuration $\langle p^f,
\gamma_1\rangle$ such that $p^f \in \mathcal{F}$ and,
\begin{enumerate}[\em(1)]
\item
$c \pdsreach \langle p^j, \gamma_3 \rangle$ for some $\gamma_3$ with $top_1(\gamma_3)
= a$, and

\item
$\langle p^j, [^n a ]^n \rangle \pdsreach \langle p^f, \gamma_1 \rangle
\pdsreach \langle p^j, \gamma_2 \rangle$
\end{enumerate}
\end{prop}
\begin{proof}
See Appendix~\ref{buchipropproof}.
\end{proof}

We reformulate these conditions as follows, where $C^\Sigma_n$ is
the set of all order-$n$ stacks over the alphabet $\Sigma$.   We remind the reader that $B^a_n$ is the $n$-store
automaton accepting all $n$-stores $\gamma$ such that $top_1(\gamma) =
a$.
\begin{enumerate}[(1)]
\item
$c \in Pre^*(\{p^j\} \times \lang(B^a_n))$,
\item
\label{ltlreach2}
$\langle p^j, [^n a ]^n\rangle \in Pre^*((\mathcal{F} \times
C^\Sigma_n) \cap Pre^+(\{p^j\} \times \lang(B^a_n)))$
\end{enumerate}
We can compute the set of pairs $\langle p^j, [^n a]^n\rangle$
satisfying (\ref{ltlreach2}) in
$n$-EXPTIME by calculating $Pre^*(\{p^j\} \times \lang(B^a_n))$ over
the following higher-order PDS:
\begin{defi}
Given an order-$n$  B\"uchi PDS $BP = (\mathcal{P}, \mathcal{D}, \Sigma, \mathcal{F})$
we define $BP' = (\mathcal{P} \times \{0,1\}, \mathcal{D}', \Sigma)$
where,
\[ \begin{array}{ll}
     \mathcal{D}' = & \{\ ((p, 0), b, o, (p', 0))\ |\ p \in \mathcal{P}
\cap \overline{\mathcal{F}}\ \land\ (p, b, o, p') \in \mathcal{D}\ \}\
\cup
\\
        &\{\ ((p, 0), b, o, (p', 1))\ |\ p \in \mathcal{F}\
\land\ (p, b, o, p') \in \mathcal{D}\ \}\ \cup \\
        &\{\ ((p, 1), b, o, (p', 1))\ |\ (p, b, o, p') \in
\mathcal{D}\ \}
   \end{array}
\]
\end{defi}
\begin{lem}
\label{ltlreachflaglem}
There exists a run $\langle (p, 0), [^n a]^n \rangle \pdsreach \langle (p,
1), w'\rangle$ with $w' \in \lang(B^a_n)$ in $BP'$ iff $\langle p,[^n
a]^n\rangle$ satisfies (\ref{ltlreach2}).
\end{lem}
\begin{proof}
See Appendix~\ref{ltlreachflag}.  Since $BP'$ is twice as large as
$BP$,
$Pre^*(\{p^j\} \times \lang(B^a_n))$  for $BP'$ can be calculated in
$n$-EXPTIME.  This gives the set of configurations satisfying (\ref{ltlreach2}).
\end{proof}

To construct
an $n$-store automaton accepting all configurations from which there is
an accepting run, we calculate the configurations $\langle p^j, [^n
a]^n\rangle$ satisfying the second condition.  Since there are only
finitely many $p^j \in \mathcal{P}$ and $a \in \Sigma$ we can perform
a simple enumeration.  We then construct an $n$-store automaton $A$
corresponding to the $n$-store
automata accepting configurations satisfying (\ref{ltlreach2}) and compute $Pre^*(\lang(A))$.
\begin{thm}
Given an order-$n$ B\"uchi PDS $BP = (\mathcal{P}, \mathcal{D}, \Sigma,
\mathcal{F})$, we can calculate in $n$-EXPTIME the set of configurations $C$ such
that from all $c \in C$ there is an accepting run of $BP$.	
\end{thm}
\begin{proof}
Let $exp_0(x) = x$ and $exp_n(x) = 2^{exp_{n-1}(x)}$.
We appeal to Lemma~\ref{ltlreachflaglem} for each $p^j$ and $a$ (of
which there are polynomially many) to construct an $n$-store automaton
$\mathcal{O}(exp_n(2 \times |\mathcal{P}|))$ in size which accepts $\langle p^j,
[^n a]^n\rangle$ iff it satisfies (\ref{ltlreach2}).  Membership can
be checked in polynomial time (Proposition~\ref{nstoremembership}).

It is straightforward to construct an automaton $A$ polynomial in size
which accepts $\langle p, w\rangle$ iff $\langle p, [^n
top_1(w)]^n\rangle$ satisfies (\ref{ltlreach2}).  We can construct $Pre^*(\lang(A))$
in $n$-EXPTIME.  Thus, the algorithm requires $n$-EXPTIME.
\end{proof}

\begin{cor}
Given an order-$n$ PDS $(\mathcal{P},
\mathcal{D}, \Sigma)$ with a labelling function $\Lambda : \mathcal{P} \rightarrow
2^{Prop}$ and a
formula $\phi$ of an $\omega$-regular logic, we
can calculate in $(n+2)$-EXPTIME the set of configurations $C$ of  $(\mathcal{P},
\mathcal{D}, \Sigma)$ such
that every run from each $c \in C$ satisfies $\phi$.
\end{cor}
\begin{proof}
The construction of $BP$ is exponential in size.  Hence, we construct the $n$-store multi-automaton $A$ that accepts
 the set of configurations from which there is a run
satisfying the negation of $\phi$ as described above in time $\mathcal{O}(exp_n(2^{|\phi|}))$.  To calculate
$C$ we complement $A$ as described in Appendix~\ref{nstoreboolops}.
This may include an exponential blow-up in the transition relation of
$A$, hence we have $(n+2)$-EXPTIME.
\end{proof}

Observe that since we can test $c \in C$ by checking $c \notin
\lang(A)$ where $A$ is defined as above, we may avoid the
complementation step, giving us an $(n+1)$-EXPTIME algorithm.

\subsection{Reachability Games}

Our algorithm may be used to compute the winning region for a player in a
two-player reachability game over higher-order PDSs.  This generalises a result
due to Cachat~\cite{C03}.  We call our players Eloise and Abelard.

\begin{defi}
Given an order-$n$ PDS $(\mathcal{P}, \mathcal{D}, \Sigma)$, an order-$n$
\emph{Pushdown Reachability Game} (PRG) $(\mathcal{P}, \mathcal{D}, \Sigma,
\mathcal{R})$ over the order-$n$ PDS is given by a partition $\mathcal{P} =
\mathcal{P}_A \uplus \mathcal{P}_E$ and a set $\mathcal{R}$ of configurations
considered winning for Eloise.
\end{defi}

We write $\langle p, \gamma\rangle \in \mathcal{C}_E$ iff $p \in \mathcal{P}_E$
and $\langle p, \gamma\rangle \in \mathcal{C}_A$ iff $p \in \mathcal{P}_A$. From
a configuration $\langle p, \gamma \rangle$ play proceeds as follows:
\begin{enumerate}[$\bullet$]
\item
If $\langle p, \gamma\rangle  \in \mathcal{C}_A$, Abelard chooses a move $(p, a, o, p') \in
\mathcal{D}$ with $top_1(\gamma) = a$ and $o(\gamma)$ defined.  Play moves to
the configuration $\langle p', o(\gamma)\rangle$.

\item
If $\langle p, \gamma\rangle \in \mathcal{C}_E$, Eloise chooses a move $(p, a,
o, p') \in \mathcal{D}$ with $top_1(\gamma) = a$ and $o(\gamma)$ defined.  Play
moves to the configuration $\langle p', o(\gamma)\rangle$.
\end{enumerate}
Eloise wins the game iff play reaches a configuration $\langle p, \gamma\rangle$
where $\langle p, \gamma \rangle \in \mathcal{R}$ or $p \in \mathcal{P}_A$ and
Abelard is unable to choose a move.  Abelard wins otherwise.

The \emph{winning region} for a given player is the set of all configurations
from which that player can force a win.  The winning region for Eloise can be
characterised using an \emph{attractor} $Attr_E(\mathcal{R})$ defined as
follows,
\[ \begin{array}{rcl}
     Attr^0_E(\mathcal{R}) & = & \mathcal{R} \\
     Attr^{i+1}_E(\mathcal{R}) & = & Attr^i_E(\mathcal{R}) \cup \{\ c
\in \mathcal{C}_E\ |\ \exists c'. c \hookrightarrow c'\land c' \in
Attr^i_E(\mathcal{R})\ \} \\
      & & \cup\
      \{\ c
\in \mathcal{C}_A\ |\ \forall c'. c \hookrightarrow c'\Rightarrow c' \in
Attr^i_E(\mathcal{R})\ \} \\
     Attr_E(\mathcal{R}) & = & \bigcup_{i \geq 0} Attr^i_E(\mathcal{R})

   \end{array}
\]
Conversely, the winning region for Abelard is $\overline{Attr_E(\mathcal{R})}$.
Intuitively, from a position in $Attr^i_E(\mathcal{R})$, Eloise's winning
strategy is to simply choose a move such that the next configuration is in
$Attr^{i-1}_E(\mathcal{R})$.  Abelard's strategy is to avoid Eloise's winning
region.

We can use backwards-reachability for order-$n$ APDSs to calculate
$Attr_E(\mathcal{R})$, and hence the winning regions of both Abelard and Eloise.
To simplify the reduction, we make a \emph{totality} assumption.  That is, we
assume a bottom-of-the-stack symbol $\perp$ that is never popped nor pushed, and
for all $a \in \Sigma \cup \{\perp\}$ and control states $p \in \mathcal{P}$,
there exists a command $(p, a, o, p') \in \mathcal{D}$.  This can be ensured by
adding sink states $p^E_{lose}$ and $p^A_{lose}$ from which Eloise and Abelard
lose the game.  In particular, for every $p \in \mathcal{P}$ and $a \in \Sigma
\cup \{\perp\}$ we have $(p, a, push_a, p^x_{lose})$ where $x = E$ if $p \in
\mathcal{P}_E$ or $x = A$ otherwise.  Furthermore, the only commands available
from $p^x_{lose}$ are of the form $(p^x_{lose}, a, push_a, p^x_{lose})$ for $x
\in \{A, E\}$.  To ensure that $p^A_{lose}$ is losing for Abelard, we set
$\langle p^A_{lose}, \gamma\rangle \in \mathcal{R}$ for all $\gamma$.
Conversely, $\langle p^E_{lose}, \gamma\rangle \notin \mathcal{R}$ for all
$\gamma$.

\begin{defi}
\label{reachencoding}
Given an order-$n$ PRG $(\mathcal{P}, \mathcal{D}, \Sigma, \mathcal{R})$ we
define an order-$n$ APDS $(\mathcal{P}, \mathcal{D}', \Sigma)$ where,
\[ \begin{array}{rcl}
     \mathcal{D}' & = & \{\ (p, a, \{(o, p')\})\ |\ (p, a, o, p') \in
\mathcal{D}\land p \in P_E\ \} \\
      & & \cup\ \left\{\ \left(p, a, \{\ (o, p')\ |\ (p, a, o, p') \in
\mathcal{D}\ \}\right)\ |\ p \in \mathcal{P}_A\ \right\}
   \end{array}
\]
Furthermore, let $R_{stuck}$ be the set of configurations $\langle
p, \triangledown\rangle$ such that $p \in \mathcal{P}_A$.  The set $\mathcal{R}_{stuck}$ is
regular and represents the configurations reached if Abelard performs an move
with an undefined next stack.
\end{defi}

Let $C^\triangledown_A$ be the set of order-$n$ configurations with an undefined
stack and a control state belonging to Abelard.

\begin{thm}
\label{gamesth}
Given an order-$n$ PRG, where $\mathcal{R}$ is a regular set of configurations,
and an order-$n$ APDS as defined above, $Attr_E(\mathcal{R})$ is regular and
equivalent to $Pre^*(\mathcal{R} \cup \mathcal{R}_{stuck})\setminus
C^\triangledown_A$.  Hence, computing the winning regions in the order-$n$
PRG is $n$-EXPTIME.
\end{thm}

\subsection{Model-Checking Branching-Time Temporal Logics}

Generalising a further result of Bouajjani\etal~\cite{BEM97}, we show that
backwards-reachability for higher-order APDSs may be used to perform
model-checking for the alternation-free (propositional)
$\mu$-calculus over higher-order PDSs.  Common logics such as CTL are sub-logics of
the alternation-free $\mu$-calculus.

\subsubsection{Preliminaries}

Given a set of atomic propositions $Prop$ and a finite set of variables
$\chi$, the propositional $\mu$-calculus is defined by the following
grammar,
\[ \phi\ :=\ \pi \in Prop\ |\ X \in \chi\ |\ \neg\phi\ |\ \phi_1 \cup
\phi_2\ |\ \diamond\phi\ |\ \mu X. \phi \]
with the condition that, for a formula $\mu X. \phi$, $X$ must occur
under an even-number of negations.  This ensures that the logic is
monotonic. As well as the usual abbreviations for $\Rightarrow$ and $\land$, we
may also use, $\Box\phi = \neg\diamond\neg\phi$, $\nu X.\phi(X) =
\neg\mu X.\neg\phi(\neg X)$ and $\sigma$ for either $\mu$ or $\nu$.  A
$\sigma$-formula is of the form $\sigma X.\phi$.

A variable $X$ is bound in $\phi$ if it occurs as part of a sub-formula $\sigma
X. \phi'(X)$.  We call an unbound variable \emph{free} and write
$\phi(X)$ to indicate that $X$ is free in $\phi$.  A \emph{closed}
formula has no variables occurring free, otherwise the formula is
\emph{open}.

Formulae in \emph{positive normal form} are defined by the following
syntax,
\[ \phi\ :=\ \pi \in Prop\ |\ \neg\pi\ |\ X \in \chi\ |\ \phi_1 \cup
\phi_2\ |\ \phi_1 \cap \phi_2\ |\ \diamond\phi\ |\ \Box\phi\ |\ \mu
X. \phi\ |\ \nu X. \phi \]
We can translate any formula into positive normal form by ``pushing
in'' the negations using the abbreviations defined above.

A $\sigma$-sub-formula of $\sigma X.\phi(X)$ is \emph{proper} iff it
does not contain any occurrence of $X$.  We are now ready to define the
alternation-free $\mu$-calculus:
\begin{defi}
The alternation-free $\mu$-calculus is the set of formulae in positive
normal form such that for every $\sigma$-sub-formula $\psi$ of $\phi$
we have,
\begin{enumerate}[$\bullet$]
\item
If $\psi$ is \ $\mu$-formula, then all $\nu$-sub-formulae of $\psi$ are
proper, and
\item
If $\psi$ is a $\nu$-formula, then all $\mu$-sub-formulae of $\psi$ are proper.
\end{enumerate}
\end{defi}
The \emph{closure} $cl(\phi)$ of a formula $\phi$ is the smallest set such
that,
\begin{enumerate}[$\bullet$]
\item
If $\psi_1 \land \psi_2 \in cl(\phi)$ or $\psi \lor \psi \in cl(\phi)$,
then $\psi_1 \in cl(\phi)$ and $\psi_2 \in cl(\phi)$, and
\item
If $\diamond\psi \in cl(\phi)$ or $\Box\psi \in cl(\phi)$, then $\psi
\in cl(\phi)$, and

\item
If $\sigma X.\psi(X) \in cl(\phi)$, then $\psi(\sigma X.\psi(X)) \in cl(\phi)$.
\end{enumerate}
The closure of any formula is a finite set whose size is bounded by
the length of the formula.

Finally, we give the semantics of the $\mu$-calculus over higher-order
PDSs.
Given a formula $\phi$, an order-$n$ PDS $(\mathcal{P}, \mathcal{D}, \Sigma)$, a labelling
function $\Lambda : \mathcal{P} \rightarrow 2^{Prop}$, and a valuation
function $\mathcal{V}$ assigning a set of configurations to each
variable $X \in \chi$, the set of configurations
$\llbracket\phi\rrbracket_\mathcal{V}$ satisfying $\phi$ is defined,
\[ \begin{array}{rcl}
     \llbracket\pi\rrbracket_\mathcal{V} &=& \Lambda^{-1}(\pi)
\times C^\Sigma_n \\
     \llbracket X\rrbracket_\mathcal{V} &=& \mathcal{V}(X) \\
     \llbracket\neg\psi\rrbracket_\mathcal{V} &=& (\mathcal{P}
\times C^\Sigma_n)\setminus\llbracket\psi\rrbracket_{\mathcal{V}} \\
     \llbracket\psi_1 \lor \psi_2\rrbracket_\mathcal{V} &=&
\llbracket\psi_1\rrbracket_\mathcal{V} \cup
\llbracket\psi_2\rrbracket_\mathcal{V} \\
     \llbracket\diamond\psi\rrbracket_\mathcal{V} &=&
Pre(\llbracket\psi\rrbracket_\mathcal{V}) \\
     \llbracket\mu X.\psi\rrbracket_\mathcal{V} &=& \bigcap\{\ C
\subseteq \mathcal{P} \times C^\Sigma_n\ |\
\llbracket\psi\rrbracket_{\mathcal{V}[X \mapsto C]} \subseteq C\ \} \\
   \end{array}
\]
where $\mathcal{V}[X \mapsto C]$ is the valuation mapping all
variables $Y \neq X$ to $\mathcal{V}(Y)$  and $X$ to $C$.

\subsubsection{Model-Checking the Alternation-Free $\mu$-Calculus}

Given an order-$n$ PDS $(\mathcal{P}, \mathcal{D}, \Sigma)$ with a labelling
function $\Lambda : \mathcal{P} \rightarrow 2^{Prop}$, a formula
$\phi$ of the alternation-free $\mu$-calculus, and a valuation
$\mathcal{V}$ we show that we can generalise the construction of
Bouajjani\etal\ to produce an $n$-store multi-automata $A_\phi$
accepting the set $\llbracket \phi \rrbracket_\mathcal{V}$.

Initially, we only consider formulae whose $\sigma$-sub-formulae are
$\mu$-formulae.  We construct a product of the higher-order PDS and the usual
``game'' interpretation of $\phi$~\cite{SE89,S99} as follows:
observing that commands of the form $(\_, a, push_a, \_)$ do not
alter the contents of the stack, we construct the order-$n$ PRG $A =
(\mathcal{P}^{(\mathcal{P}, \phi)},
\mathcal{D}^\phi_\mathcal{P}, \Sigma, \mathcal{R})$ where
$\mathcal{P}^{(\mathcal{P}, \phi)}_A$, $\mathcal{P}^{(\mathcal{P},
\phi)}_E$ and
$\mathcal{D}^\phi_\mathcal{P}$ are the smallest sets such
that for every $(p, \psi) \in \mathcal{P} \times cl(\phi)$ and $a \in \Sigma$,
\begin{enumerate}[$\bullet$]
\item
  If $\psi = \psi_1 \lor \psi_2$, then we have $(p, \psi) \in
\mathcal{P}^{(\mathcal{P}, \phi)}_E$ and $((p,\psi), a, push_a, (p,
\psi_1)) \in \mathcal{D}^\phi_\mathcal{P}$ and $((p,\psi), a, push_a, (p,
\psi_2)) \in \mathcal{D}^\phi_\mathcal{P}$,

\item
  If $\psi = \psi_1 \land \psi_2$, then we have $(p, \psi) \in
\mathcal{P}^{(\mathcal{P}, \phi)}_A$ and $((p,\psi), a, push_a, (p,
\psi_1)) \in \mathcal{D}^\phi_\mathcal{P}$ and $((p,\psi), a, push_a, (p, \psi_2)) \in \mathcal{D}^\phi_\mathcal{P}$,

\item
If $\psi = \mu X.\psi'(X)$, then $(p, \psi) \in
\mathcal{P}^{(\mathcal{P}, \phi)}_A$ and $((p,\psi), a, push_a, (p,
\psi'(\psi))) \in \mathcal{D}^\phi_\mathcal{P}$,

\item
If $\psi = \diamond\psi'$ and $(p, a, o, p') \in \mathcal{D}$, then
$(p, \psi) \in \mathcal{P}^{(\mathcal{P}, \phi)}_E$ and $((p, \psi), a, o, (p', \psi')) \in \mathcal{D}^\phi_\mathcal{P}$,

\item
If $\psi = \Box\psi'$, then $(p, \psi) \in \mathcal{P}^{(\mathcal{P},
\phi)}_A$ and for every $(p, a, o, p') \in \mathcal{D}$ it is the case that
$((p, \psi), a, o, (p', \psi')) \in \mathcal{D}^\phi_\mathcal{P}$.
\end{enumerate}
Finally, we define the set of configurations $\mathcal{R}$ that indicate
that the formula $\phi$ is satisfied by $(\mathcal{P}, \mathcal{D},
\Sigma)$, $\Lambda$ and $\mathcal{V}$.  The set $\mathcal{R}$ contains all configurations of the
form,
\begin{enumerate}[$\bullet$]
\item
$\langle (p, \pi), \gamma \rangle$ where $\pi \in \Lambda(p)$,
\item
$\langle (p, \neg\pi), \gamma \rangle$ where $\pi \notin \Lambda(p)$,
\item
$\langle (p, X), \gamma\rangle$, where $X$ is free in $\phi$ and $\langle
p, w\rangle \in \mathcal{V}(X)$.
\end{enumerate}
If $\mathcal{V}(X)$ is regular for all $X$ free in $\phi$, then
$\mathcal{R}$ is also regular.

Commands of the form $(\_, a, push_a, \_)$ are designed to deconstruct
sub-formulae into literals that can be evaluated immediately.  These
commands require that the top order-one stack is not empty ---
otherwise play would be unable to proceed.  Correctness
of the construction requires the top order-one stack to contain at least one stack symbol.  This condition may be ensured
with a special ``bottom of the stack'' symbol $\perp \in \Sigma$.
This symbol marks the bottom of all order-one stacks and is
never pushed or popped, except in the case of a
command $(\_, \perp, push_\perp, \_)$.  The use of such a symbol is
common throughout the literature~\cite{W96,KNU02,C03} etc..

\begin{prop}
\label{allmu}
Given the order-$n$ PRG $A =
(\mathcal{P}^{(\mathcal{P}, \phi)},
\mathcal{D}^\phi_\mathcal{P}, \Sigma, \mathcal{R})$ constructed from
the order-$n$ PDS $(\mathcal{P}, \mathcal{D}, \Sigma)$, a labelling function
$\Lambda$, a valuation $\mathcal{V}$, and a formula $\phi$ of the
alternation-free $\mu$-calculus such that all
$\sigma$-sub-formulae of $\phi$ are $\mu$-sub-formulae, we have $\langle
p, \gamma \rangle \in \llbracket \phi \rrbracket_\mathcal{V}$ iff $\langle
(p, \phi), \gamma \rangle \in Attr_E(\mathcal{R})$.
\end{prop}
\begin{proof}(Sketch)
The result follows from the fundamental theorem of the propositional
$\mu$-calculus~\cite{SE89,BS01}.  If  $\langle
(p, \phi), \gamma \rangle \in Attr_E(\mathcal{R})$, then there is a winning
strategy for Eloise in $A$.  In the absence of $\nu$-sub-formulae, this winning strategy defines a
well-founded choice function and hence a well-founded pre-model for $(\mathcal{P},
\mathcal{D}, \Sigma)$, $\Lambda$, $\mathcal{V}$ and $\phi$ with initial
state $\langle p, \gamma\rangle$.  Thus, by the fundamental theorem,  $\langle p, \gamma\rangle$ satisfies
$\phi$.

In the opposite direction, if $\langle p, \gamma\rangle$ satisfies $\phi$,
then --- by the fundamental theorem --- there is a well-founded pre-model with choice
function $f$.  Since there are no $\nu X. \psi$ sub-formula in $\phi$,
all paths in the pre-model are finite and all leaves are of a form
accepted by $\mathcal{R}$.  Hence, a winning strategy for Eloise is
defined by $f$ and we have  $\langle
(p, \phi), \gamma \rangle \in Attr_E(\mathcal{R})$.
\end{proof}

In the dual case --- when all $\sigma$-sub-formulae of $\phi$ are
$\nu$-sub-formulae --- we observe that the negation $\bar{\phi}$ of $\phi$
has only $\mu$-sub-formulae.  We construct $Attr_E(\mathcal{R})$ for
$\bar{\phi}$ and complement the resulting $n$-store multi-automaton
(see Appendix~\ref{nstoreboolops}) to
construct the set of configurations satisfying $\phi$.

We are now ready to give a recursive algorithm for model-checking with
the alternation-free $\mu$-calculus.  We write $\Phi =
\{\phi_i\}^m_{i=1}$ to denote a set of sub-formulae such that no
$\phi_i$ is a sub-formula of another.  Furthermore, we write
$\phi[U/\Phi]$ where $U = \{U_i\}^m_{i=1}$ is a set of fresh variables
to denote the simultaneous substitution in $\phi$ of $\phi_i$ with $U_i$ for
all $i \in \{1,\ldots,m\}$.  The following proposition is taken
directly from~\cite{BEM97}:
\begin{prop}
Let $\phi$ be a $\mu$-formula ($\nu$-formula) of the alternation-free
$\mu$-calculus, and let $\Phi = \{\phi_i\}^n_{i=1}$ be the family of
maximal $\nu$-sub-formulae ($\mu$-sub-formulae) of $\phi$ with respect to
the sub-formula relation.  Then,
\[ \llbracket \phi \rrbracket_{\mathcal{V}} =
\llbracket\phi[U/\Phi]\rrbracket_{\mathcal{V}'} \]
where $U = \{U_i\}^n_{i=1}$ is a suitable family of fresh variables,
and $\mathcal{V}'$ is the valuation which extends $\mathcal{V}$ by
assigning to each $U_i$ the set $\llbracket\phi_i\rrbracket_{\mathcal{V}}$.\qed
\end{prop}
Since, given a $\mu$-formula ($\nu$-formula) $\phi$, the formula
$\phi[U/\Phi]$ has only $\mu$-sub-formulae ($\nu$-sub-formulae) we can
calculate $\llbracket \phi_i \rrbracket_{\mathcal{V}}$ for all $\phi_i
\in \Phi$, using the  above propositions to calculate an
automaton recognising $\llbracket \phi\rrbracket_{\mathcal{V}}$.
\begin{thm}
Given an order-$n$ PDS $(\mathcal{P}, \mathcal{D}, \Sigma)$, a labelling function
$\Lambda$, a valuation function $\mathcal{V}$ and a formula $\phi$
of the alternation-free $\mu$-calculus, we can construct an $n$-store
multi-automaton $A$ such that $\lang(A) = \llbracket
\phi\rrbracket_\mathcal{V}$.\qed
\end{thm}

\subsubsection{Complexity}

Let $exp_0(x) = x$ and $exp_n(x) = 2^{exp_{n-1}(x)}$. A
 formula $\phi$ can be described as a tree structure with $\phi$ at
the root.  Each node in the tree is a $\mu$-sub-formula or a
$\nu$-sub-formula $\psi$ of $\phi$.  The children of the node are all maximal
$\nu$-sub-formulae or $\mu$-sub-formulae of $\psi$ respectively.  There
are at most $n_\phi$ nodes in the tree, where $n_\phi$ is the length
of $\phi$.  Let $n_\mathcal{R}$ be the number of states in the
$n$-store automaton recognising $\mathcal{R}$.  The size of this
automata is linear in the size of the automata specifying
$\mathcal{V}$ for each variable $X$.

The $n$-store
multi-automaton recognising $\llbracket \psi \rrbracket_{\mathcal{V}}$
for a leaf node $\psi$ has $\mathcal{O}(exp_n(n_\mathcal{R}))$
states.  Together with a possible complementation step (which does not
increase the state-set) we require
$\mathcal{O}(exp_{n+1}(n_\mathcal{P}\cdot n_\phi))$ time and
$\mathcal{B}$ may be of size $\mathcal{O}(exp_{n+1}(n_\mathcal{V}))$.

Similarly, the $n$-store multi-automaton recognising $\llbracket \psi
\rrbracket_{\mathcal{V}'}$ for an internal node $\psi$ with children
$\phi_1,\ldots,\phi_m$ has $\mathcal{O}(exp_n(\Sigma^m_{i=1}n_i
+ n_\mathcal{R}) \times 2^{b_i})$ states, where $n_i$ is the size of the
automaton recognising $\llbracket \phi_i \rrbracket_{\mathcal{V}_i}$
for $i \in \{1,\ldots,m\}$ and $b_i$ is the size of $\mathcal{B}$ for
that automaton.  Due to the final complementation step,
$|\mathcal{B}|$ may be of size $\mathcal{O}(exp_{n+1}(\Sigma^m_{i=1}n_i
+ n_\mathcal{R}))$, which is also the total time required.

Subsequently, the automaton $A$ recognising $\llbracket \phi
\rrbracket_{\mathcal{V}'}$ has $\mathcal{O}(exp_{n_\phi\cdot
n}(n_\mathcal{R}))$ states and can be constructed in  $\mathcal{O}(exp_{(n_\phi\cdot
n) + 1}(n_\mathcal{R}))$ time.  Since we may test $c \in C$
for any configuration $c$ and set of configurations $C$ by checking $c
\notin \overline{C}$, we may avoid the final complementation step to
give us an  $\mathcal{O}(exp_{n_\phi\cdot
n}(n_\mathcal{R}))$ time algorithm.

\section{Conclusion}
\label{conclusion}

Given an automaton representation of a regular set of higher-order
APDS configurations $C_{Init}$, we have shown that the set
$Pre^*(C_{Init})$ is regular and computable via automata-theoretic
methods.  This builds upon previous work on pushdown
systems~\cite{BEM97} and higher-order context-free
processes~\cite{BM04}.  The main innovation of this generalisation is
the careful management of a complex automaton construction.  This
allows us to identify a sequence of cascading fixed points, resulting
in a terminating algorithm.

Our result has many applications.  We have shown that it can be used
to provide a solution to the model-checking problem for linear-time
temporal logics and the alternation-free $\mu$-calculus.  In
particular we compute the set of configurations of a higher-order PDS
satisfying a given constraint.  We also show that the winning regions
can be computed for a reachability game played over an higher-order
PDS.

There are several possible extensions to this work.  We plan to investigate the
applications of this work to higher-order pushdown games with more general
winning conditions.  In his Ph.D. thesis, Cachat adapts the reachability
algorithm of Bouajjani\etal~\cite{BEM97} to calculate the winning regions in
B\"uchi games over pushdown processes~\cite{C03}.  It is likely that our work
will permit similar extensions. We also intend to generalise this work to
higher-order collapsible pushdown automata, which can be used to study
higher-order recursion schemes~\cite{KNUW05,HMOS06}.  This may provide the first
steps into the study of the global model-checking problem over these structures.
Finally, an alternative definition of higher-order pushdown systems defines the
higher-order pop operation as the inverse of the push operation.  That is, a
stack may only be popped if it matches the stack below.  The results of
Carayol~\cite{C05} show that the set $Pre^*(C_{Init})$ over these structures is
regular, using Carayol's notion of regularity.  However, the complexity of
computing this set is unknown.  We may attempt to adapt our algorithm to this
setting, proving the required complexity bounds.

\section*{Acknowledgments}We thank Olivier Serre and Arnaud
Carayol for helpful discussions.  We also thank the anonymous referees for their
invaluable remarks.

\bibliographystyle{plain}
\bibliography{higherorderreachBib}

\appendix

\section{Notions of Regularity}
\label{ourregularityis}

We show that our notion of a regular set of $n$-stores coincides with
the definition of Bouajjani and Meyer~\cite{BM04}.  Bouajjani and
Meyer show that a set of $n$-stores is regular iff it is accepted by
a \emph{level $n$ nested store automata}.

Because we are considering $n$-stores rather than configurations, we
assume that there is only one control state, and hence, an $n$-store
multi-automaton has only a single initial state.  We also disregard
the undefined store $\triangledown$, since it is not strictly a
store.  Observe that we are left with $n$-store automata.

In the absence of
alternation, the set of $n$-store automata is definitionally
equivalent to
the set of level $n$ nested store
automata in the sense of Bouajjani and Meyer.  Hence, it is the case
that every level $n$ nested store automaton is also an $n$-store
automaton.

We need to prove that every $n$-store automaton has an equivalent
level $n$ nested store automata.
We present the following definition:
\begin{defi}
Given an $n$-store automaton $A = (\mathcal{Q}, \Sigma,
\Delta, q_0, \mathcal{Q}_f)$ we define a level $n$
nested store automaton $\hat{A} = (2^\mathcal{Q}, \Sigma, \hat{\Delta},
\{q_0\}, 2^{\mathcal{Q}_f})$, where, if $n = 1$,
\[
   \hat{\Delta} = \left\{\ (\{q_1,\ldots,q_m\}, a, Q')\ |\ \forall i
\in \{1,\ldots,m\}.\left(\exists(q_i, a, Q_i) \in
\Delta \right) \land Q' = Q_1 \cup \ldots \cup Q_m\ \right\}
\]
and if $n > 1$,
\[
   \hat{\Delta} = \left\{\ (\{q_1,\ldots,q_m\}, \hat{B}, Q')\ |\
     \begin{array}{l}
        \forall i \in \{1,\ldots,m\}.\left(\exists(q_i, B_i, Q_i) \in
           \Delta \right) \land \\
        Q' = Q_1 \cup \ldots \cup Q_m \land B = B_1
\cap \ldots \cap B_m\end{array} \ \right\}
\]
where $\hat{B}$ is defined recursively and the construction of $B_1 \cap \cdots \cap B_m$ is given in section~\ref{nstoreboolops}.

\end{defi}

\begin{property}
For any $w$, the run $\{q_1,\ldots,q_m\} \stackrel{w}{\longrightarrow}
Q'$ exists in the $n$-store automaton $A$ iff the run
$\{q_1,\ldots,q_m\} \stackrel{w}{\longrightarrow} Q'$ exists in $\hat{A}$.
\end{property}
\begin{proof}
The proof is by induction over $n$ and then by a further induction
over the length of $w$.

Suppose $n = 1$.  When $w = \varepsilon$ the proof is immediate.  When
$w = aw'$ we have in one direction,
\[ \{q_1,\ldots,q_m\} \stackrel{a}{\longrightarrow} Q_1
\stackrel{w'}{\longrightarrow} Q' \]
in $A$, and by induction over the length of the run, $Q_1
\stackrel{w'}{\longrightarrow} Q'$ in $\hat{A}$.  By definition of the
runs of $A$ we have $q_i \stackrel{a}{\longrightarrow} Q^i_1$ for each
$i \in \{1,\ldots,m\}$ with $Q_1 = Q^1_1 \cup \ldots \cup Q^m_1$.
Hence, by definition of $\hat{A}$ we have the transition
$\{q_1,\ldots,q_m\} \stackrel{a}{\longrightarrow} Q^1_1 \cup \ldots
\cup Q^m_1 = Q_1$.  Hence we have the run
$\{q_1,\ldots,q_m\}\stackrel{w}{\longrightarrow} Q'$ in $\hat{A}$ as
required.

In the other direction we have a run of the form
\[ \{q_1,\ldots,q_m\} \stackrel{a}{\longrightarrow} Q_1
\stackrel{w'}{\longrightarrow} Q' \]
in $\hat{A}$, and by induction over the length of the run, $Q_1
\stackrel{w'}{\longrightarrow} Q'$ in $A$.  By definition of the
transition relation of $\hat{A}$ we have $q_i
\stackrel{a}{\longrightarrow} Q^i_1$ in $A$ for each
$i \in \{1,\ldots,m\}$ with $Q_1 = Q^1_1 \cup \ldots \cup Q^m_1$.
Hence, we have the transition
$\{q_1,\ldots,q_m\} \stackrel{a}{\longrightarrow} Q^1_1 \cup \ldots
\cup Q^m_1 = Q_1$ in $A$.  Thus, we have the run
$\{q_1,\ldots,q_m\}\stackrel{w}{\longrightarrow} Q'$ in $A$ as
required.

When $n > 1$, when $w = \varepsilon$ the proof is immediate.  When
$w = \gamma w'$ we have in one direction,
\[ \{q_1,\ldots,q_m\} \stackrel{\gamma}{\longrightarrow} Q_1
\stackrel{w'}{\longrightarrow} Q' \]
in $A$, and by induction over the length of the run, $Q_1
\stackrel{w'}{\longrightarrow} Q'$ in $\hat{A}$.  By definition of the
runs of $A$ we have $q_i \stackrel{B_i}{\longrightarrow} Q^i_1$ with
$\gamma \in \lang(B_i)$ for each
$i \in \{1,\ldots,m\}$ with $Q_1 = Q^1_1 \cup \ldots \cup Q^m_1$.
Consequently, we have $\gamma \in \lang(B)$ where $B = B_1 \cap \ldots
\cap B_m$.  By induction over $n$ we have $\gamma \in \lang(\hat{B})$.
Hence, by definition of $\hat{A}$ we have the transition
$\{q_1,\ldots,q_m\} \stackrel{\gamma}{\longrightarrow} Q^1_1 \cup \ldots
\cup Q^m_1 = Q_1$.  Hence we have the run
$\{q_1,\ldots,q_m\}\stackrel{w}{\longrightarrow} Q'$ in $\hat{A}$ as
required.

In the other direction we have a run of the form
\[ \{q_1,\ldots,q_m\} \stackrel{\gamma}{\longrightarrow} Q_1
\stackrel{w'}{\longrightarrow} Q' \]
in $\hat{A}$.  In particular, we have $\{q_1,\ldots,q_m\}
\stackrel{\hat{B}}{\longrightarrow} Q_1$ in $\hat{A}$ with $\gamma \in
\lang(\hat{B})$.  By induction over the length of the run, $Q_1
\stackrel{w'}{\longrightarrow} Q'$ in $A$.  By definition of the
transition relation of $\hat{A}$ we have $q_i
\stackrel{B_i}{\longrightarrow} Q^i_1$ for each $i \in \{1,\ldots,m\}$
with $B = B_1 \cap \ldots \cap
B_m$ and $Q_1 = Q^1_1 \cup \ldots \cup Q^m_1$.  By induction over $n$
we have $\gamma \in \lang(B)$ and hence $\gamma \in \lang(B_i)$ for all
$i \in \{1,\ldots,m\}$.  Therefore, we have $q_i
\stackrel{\gamma}{\longrightarrow} Q^i_1$ in $A$ for all $i \in
\{1,\ldots,m\}$.
Thus, we have the transition
$\{q_1,\ldots,q_m\} \stackrel{\gamma}{\longrightarrow} Q^1_1 \cup \ldots
\cup Q^m_1 = Q_1$ in $A$ and the run
$\{q_1,\ldots,q_m\}\stackrel{w}{\longrightarrow} Q'$ as
required.
\end{proof}

\begin{cor}
A set of $n$-stores is definable by an $n$-store automaton iff it is
definable by a level $n$ nested store automaton.\qed
\end{cor}

\section{Algorithms over $n$-Store (Multi-)Automata}
\label{storeautalg}

In this section we describe several algorithms over $n$-store automata
and $n$-store multi-automata.  Observe that an $n$-store automaton is a
special case of an $n$-store multi-automaton.

\subsection{Enumerating Runs}

\begin{prop}
\label{enumeratingorder1}
Given a $1$-store (multi-)automaton $A = (\mathcal{Q}, \Sigma, \Delta, \_,
\mathcal{Q}_f)$, a set of states $Q$ and word $w$, the set of
all $Q'$ reachable via a run $Q \stackrel{w}{\longrightarrow} Q'$ can be calculated in time
$\mathcal{O}(2^{|\mathcal{Q}|})$.
\end{prop}
\begin{proof}
We define the following procedure, which given a set of sets of states
$Q_1$ computes the set of sets $Q'$ with $Q \in Q_1$ and $Q
\stackrel{a}{\longrightarrow} Q'$.
\begin{quote}
\textsc{Expand}$(a, Q_1)$
  \begin{quote}
    {\bf let} $Q_{next} = \emptyset$ \\
    {\bf for each} $\{q_1,\ldots,q_m\} \in Q_1$
    \begin{quote}
      {\bf let} $ok = (\exists (q_1,a, \_) \in \Delta)$ \\
      {\bf let} $Q = \Delta(q_1, a)$ \\
      {\bf for} $i = 2$ {\bf to} $m$
      \begin{quote}
        $ok = ok \land (\exists (q_i, a, \_) \in \Delta)$ \\
        $Q = \{\ Q' \cup Q''\ |\ Q' \in Q \land (q_i, a, Q'') \in
\Delta\ \}$
      \end{quote}
      {\bf if} $ok$
      {\bf then} $Q_{next} = Q_{next} \cup Q$
    \end{quote}
  {\bf return} $Q_{next}$
  \end{quote}
\end{quote}
The outer loop repeats $\mathcal{O}(2^{|\mathcal{Q}|})$ times and the
inner loop $\mathcal{O}(|\mathcal{Q}|)$.  Since the number of $Q'
\in Q$ is
$\mathcal{O}(2^{|\mathcal{Q}|})$ and the number of $(q_i, a, Q'') \in
\Delta$ is also $\mathcal{O}(2^{|\mathcal{Q}|})$, construction of $Q$ takes time
$\mathcal{O}(2^{|\mathcal{Q}|})$.  Hence the procedure takes time
$\mathcal{O}(2^{|\mathcal{Q}|} \times |\mathcal{Q}| \times 2^{|\mathcal{Q}|})$, that is $\mathcal{O}(2^{|\mathcal{Q}|})$.

\textsc{Expand} is correct since $Q \in Q_{next}$ at the end of the
procedure iff we have $\{q_1,\ldots,q_m\} \in Q_1$ and some $(q_i, a,
Q^i_{next}) \in \Delta$ for each $i \in \{1,\ldots, m\}$ with
$Q_{next} = Q^1_{next} \cup \ldots \cup Q^m_{next}$.

Over a word $w = a_1\ldots a_m$ we define the following procedure,
\begin{quote}
  \textsc{ExpandWord}$(a_1\ldots a_m, Q)$
  \begin{quote}
    {\bf let} $Q_1 = \{Q\}$ \\
    {\bf for} $i = 1$ {\bf to} $m$
    \begin{quote}
      $Q_1 =\ $\textsc{Expand}$(a_i, Q_1)$
    \end{quote}
    {\bf return} $Q_1$
  \end{quote}
\end{quote}
This procedure requires $m$ runs of \textsc{Expand} and consequently
runs in time $\mathcal{O}(2^{|\mathcal{Q}|})$.

We prove the correctness of \textsc{ExpandWord} by induction over the
length of the word.  When $w = a_1$ correctness follows from the
correctness of \textsc{Expand}.  In the inductive case $w = a_1\ldots
a_m$.  We have all
runs of the form $Q \stackrel{a_1}{\longrightarrow} Q_1$ as before, and
all runs over $a_2\ldots a_m$ from all $Q_1$ by induction.  We have all
runs of the form $Q \stackrel{w}{\longrightarrow} Q'$ therefrom.
\end{proof}

\begin{prop}
\label{enumeratingorderl}
Given an $l$-store (multi-)automaton $A = (\mathcal{Q}, \Sigma, \Delta, \_,
\mathcal{Q}_f)$ with $l > 1$, and a set of states $Q$, the set of
all $Q'$ reachable via a run $Q \stackrel{\widetilde{B}_1}{\longrightarrow} Q'
\stackrel{\widetilde{B}_2}{\longrightarrow} Q''$ can be calculated in time
$\mathcal{O}(2^{|\Delta| + |\mathcal{Q}|})$.
\end{prop}
\begin{proof}
We define the following procedure, which given a set of states
$Q_1$ computes the set of sets $Q'$ and set of $(l-1)$-store automata $\widetilde{B}$ with $Q \in Q_1$ and $Q
\stackrel{\widetilde{B}}{\longrightarrow} Q'$.
\begin{quote}
\textsc{Expand}$(Q_1)$
  \begin{quote}
    {\bf let} $Q_{next} = \emptyset$ \\
    {\bf for each} $\{q_1,\ldots,q_m\} \in Q_1$
    \begin{quote}
      {\bf for each} set
$\{(q_1,B_1,Q^1_{next}),\ldots,(q_m,B_m,Q^m_{next})\} \subseteq
\Delta$
      \begin{quote}
        $\begin{array}{rcl} Q_{next} &=& Q_{next}\ \cup \\
& & \{(\{B_1,\ldots,B_m\},Q^1_{next}\cup\ldots\cup Q^m_{next})\}\end{array}$
      \end{quote}
    \end{quote}
  {\bf return} $Q_{next}$
  \end{quote}
\end{quote}
The outer loop repeats at most $\mathcal{O}(2^{|\mathcal{Q}|})$
times.  At most $\mathcal{O}(2^{|\Delta|})$ sets need to be enumerated
during the inner loop.   Hence, \textsc{Expand} runs in time
$\mathcal{O}(2^{|\Delta| + |\mathcal{Q}|})$.  The correctness of
\textsc{Expand} is immediate.

To complete the algorithm, we define the following procedure,
\begin{quote}
  \textsc{ExpandETimes}$(e, Q)$
  \begin{quote}
    {\bf let} $Q_1 =\ $\textsc{Expand}$(\{Q\})$ \\
    {\bf for} h = 1 {\bf to} e
      \begin{quote}
      {\bf for each} $(\widetilde{B}_1,\ldots,\widetilde{B}_h, Q') \in Q_1$
        \begin{quote}
        $Q_1 = Q_1 \cup (\{(\widetilde{B}_1,\ldots,\widetilde{B}_h)\} \times$\textsc{Expand}$(\{Q'\}))$
        \end{quote}
     \end{quote}
    {\bf return} $Q_1 \cap ((\mathcal{B}_l)^e \times 2^\mathcal{Q})$
  \end{quote}
\end{quote}
This procedure requires $\mathcal{O}(e \times (e \times 2^{|\Delta|}) \times 2^{|\mathcal{Q}|})$
iterations of the loop.  Each iteration requires time
$\mathcal{O}(2^{|\Delta| + |\mathcal{Q}|})$ and consequently
the procedure runs in time $\mathcal{O}(2^{|\Delta| +
|\mathcal{Q}|})$.

By the correctness of \textsc{Expand} we have $(\widetilde{B}, Q') \in
Q_1$ iff we have the path $Q \stackrel{\widetilde{B}}{\longrightarrow} Q'$
in $A$.  After execution of the loop we have, by correctness of
\textsc{Expand}, $(\widetilde{B}_1,\ldots,\widetilde{B}_e, Q') \in Q_1$ iff we have
the following path in $A$: $Q \stackrel{\widetilde{B}_1}{\longrightarrow} \ldots \stackrel{\widetilde{B}_e}{\longrightarrow} Q'$.
\end{proof}

\subsection{Membership}

\begin{prop}
\label{nstoremembership}
Given an $n$-store (multi-)automaton $A = (\mathcal{Q}, \Sigma, \Delta, _,
\mathcal{Q}_f)$ and an $n$-store $w$ we can determine whether there is an
accepting run over $w$ in $A$ from a given state $q \in \mathcal{Q}$ in time
$\mathcal{O}(|w||\Delta||\mathcal{Q}|)$.
\end{prop}
\begin{proof}
When $w = \triangledown$ we can check membership immediately.  Otherwise
the algorithm is recursive.  In the base case, when $n = 1$ and $w =
a_1\ldots a_m$, we
present the following well-known algorithm,
\begin{quote}
  {\bf let} $Q = \mathcal{Q}_f$ \\
  {\bf for} $i = m$ {\bf downto} $1$
  \begin{quote}
    $Q = \{\ q'\ |\ (q', a_i, Q') \in \Delta\land Q' \subseteq Q\ \}$
  \end{quote}
  {\bf return} $(q \in Q)$
\end{quote}
This algorithm requires time $\mathcal{O}(m|\Delta||\mathcal{Q}|)$.  We prove
that this algorithm is correct at order-$1$ by induction over $m$.  When $m =
1$, we have $q \in Q$ at the end of the algorithm iff there exists a transition
$(q, a_1, Q') \in \Delta$ where $Q' \subseteq \mathcal{Q}_f$.  When $w =
a_1a_2\ldots a_m$ we have $q \in Q$ at the end of the algorithm iff there exists
a transition $(q, a_1, Q')$ where, by induction if $q' \in Q'$ then the word
$a_2\ldots a_m$ is accepted from $q'$.  Hence, we have $q \in Q$ iff there is an
accepting run over $w$ from $q$.

When $n > 1$ we generalise the algorithm given above.  Let $w = \gamma_1\ldots\gamma_m$,
\begin{quote}
  {\bf let} $Q = \mathcal{Q}_f$ \\
  {\bf for} $i = m$ {\bf downto} $1$
  \begin{quote}
    $Q = \{\ q'\ |\ (q', B, Q') \in \Delta\land \gamma \in \lang(B)\land Q' \subseteq Q\ \}$
  \end{quote}
  {\bf return} $(q \in Q)$
\end{quote}
The outer loop of the program repeats $m$ times, there are $|\Delta|$
transitions to be checked. By considering all labelling automata as a single
automaton with an initial state for each (as in the backwards reachability
construction), we make a single recursive call (for each $\gamma$ in $w$),
obtaining all states accepting $\gamma$.  Checking $\gamma
\in \lang(B)$ then requires checking whether the appropriate initial state is in
the result of the recursive call.  We have $|w| = |\gamma_1| + \cdots +
|\gamma_m|$, hence the algorithm requires
$\mathcal{O}(|\gamma_1||\Delta_1||\mathcal{Q}| + \cdots +
|\gamma_m||\Delta_1||\mathcal{Q}|) = \mathcal{O}(|w||\Delta_1||\mathcal{Q}|)$
time for the pre-computation, then $\mathcal{O}(m|\Delta_2||\mathcal{Q}|)$ time
for the body of the algorithm, where $\Delta = \Delta_1 \cup \Delta_2$ is the
partition of $\Delta$ into the order-$n$ and lower-order parts.  Hence, we
require $\mathcal{O}(|w||\Delta||\mathcal{Q}|)$ time.

We prove that this algorithm is correct at order $n > 1$ by induction over $m$.
When $m = 1$, we have $q \in Q$ at the end of the algorithm iff there exists a
transition $(q, B, Q') \in \Delta$ with $\gamma \in \lang(B)$ and $Q' \subseteq
\mathcal{Q}_f$.  When $w = \gamma_1\gamma_2\ldots \gamma_m$ we have $q \in Q$ at
the end of the algorithm iff there exists a transition $(q, B, Q')$ where
$\gamma \in \lang(B)$ and, by induction, if $q' \in Q'$ then the word $a_2\ldots
a_m$ is accepted from $q'$.  Hence, we have $q \in Q$ iff there is an accepting
run over $w$ from $q$.
\end{proof}

\subsection{Boolean Operations}
\label{nstoreboolops}

We can form the intersection, union and complement of $n$-store automata.
Intersection and union are straightforward.  We omit the details.
To complement $n$-store multi-automata, we begin by
defining an operation on sets of sets.

\begin{defi}
\label{inversesets}
Given a set of sets $\{Q_1,\ldots,Q_m\}$ we define,
\[
invert(\{Q_1,\ldots,Q_m\}) = \{\ \{q_1,\ldots,q_m\}\ |\ q_i \in
Q_i \land 1 \leq i \leq m\ \}
\]
\end{defi}

\begin{defi}
Given an $n$-store multi-automaton $A = (\mathcal{Q}, \Sigma, \Delta, \{q^1,\ldots,q^z\},
\mathcal{Q}_f)$, we define $\bar{A}$ as follows.
\begin{enumerate}[$\bullet$]
\item
When $n = 1$ we assume $A$ is \emph{total} (this is a standard assumption
that can easily be satisfied by adding a sink state).
We define $\bar{A} = (\mathcal{Q}, \Sigma, \Delta', \{q^1,\ldots,q^z\},
\mathcal{Q}\setminus\mathcal{Q}_f)$
where $\Delta'$ is the smallest set
such that for each $q \in \mathcal{Q}$ and $a \in \Sigma$ we have,
\begin{enumerate}[(1)]
\item
The transitions from $q$
in $\Delta$ over $a$ are  $(q, a,
Q_1),\ldots,(q, a, Q_m)$, and
\item
$Q_{a} = invert\left(\bigcup_{1 \in
\{1,\ldots,m\}} \{Q_i\}\right)$, and
\item
$\Delta'(q,
a) = Q_{a}$.
\end{enumerate}
Since $Q_a$ may be exponential in size, the construction
runs in exponential time when $n = 1$.

\item
When $n > 1$ we define $\bar{A} = (\mathcal{Q} \cup \{q^*_f, q^\varepsilon_f\}, \Sigma, \Delta', \{q^1,\ldots,q^z\},
(\mathcal{Q}\cup\{q^*_f,q^\varepsilon_f\})\setminus\mathcal{Q}_f)$
where $q^*_f, q^\varepsilon_f \notin \mathcal{Q}$, all $n$-stores are
accepted from $q^*_f$ and $q^\varepsilon_f$ has no outgoing transitions.

Furthermore $\Delta'$ is the smallest set
such that for each $q \in \mathcal{Q}$ we have,
\begin{enumerate}[(1)]
\item
The non-$\triangledown$ transitions from $q$
in $\Delta$ are  $(q, B_1,
Q_1),\ldots,(q, B_m, Q_m)$ (we assume $m \geq 1$), and
\item
For all $\widetilde{B} \in 2^{\{B_1,\ldots,B_m\}}$ we
have,
\[
   \begin{array}{rcl}
Q_{\widetilde{B}} &=& \left\{\begin{array}{ll}
                     \{q^*_f\} & \mathrm{if\ } \widetilde{B} = \emptyset
\\
                    invert\left(\bigcup_{B_i \in \widetilde{B}}
\{Q_i\}\right) & \mathrm{otherwise}
                   \end{array}\right. \\
   & & \\
  B_{\widetilde{B}} &=& \bigcap_{B_i
\in \widetilde{B}}B_i \cap \bigcap_{B_i \notin \widetilde{B}}\bar{B}_i
   \end{array}
\]
Note we
have $\bar{B}_i$ recursively; and
\item
$\Delta'(q,
B_{\widetilde{B}}) = Q_{\widetilde{B}}$, and
\item
For all $j \in
\{1,\ldots,z\}$ we have
$(q^j, \triangledown, \{q^\varepsilon_f\}) \in \Delta'$ iff there is no
$\triangledown$-transition from $q^j$ in $A$.
\end{enumerate}
Overall, when $n > 1$ there may be an exponential blow up
in the number of transitions and the construction of each
$B_{\widetilde{B}}$ may take exponential time.  The construction is therefore
exponential.
\end{enumerate}
\end{defi}

We now show that the above definition is correct.

\begin{property}
Given an $n$-store multi-automaton $A$, we have $\lang(\bar{A}^{q^j}) =
\overline{\lang(A^{q^j})}$ for all $q^j \in \{q^1,\ldots,q^z\}$.
\end{property}
\begin{proof}
We propose the following induction hypothesis: an accepting run $q
\stackrel{w}{\longrightarrow} Q$ exists in $\bar{A}$ iff there is no
accepting run $q \stackrel{w}{\longrightarrow} Q'$ in $A$.
We proceed first by induction over $n$ and then by induction over
the length of the run.

When $n = 1$, and the length of the run is zero, the induction
hypothesis follows since $\mathcal{Q}_f \cap
(\mathcal{Q}\setminus\mathcal{Q}_f) = \emptyset$.
When the length of the run is larger than zero, we begin by showing the
if direction.  Assume we have an accepting run,
\[ q \stackrel{a}{\longrightarrow} Q^1
\stackrel{w}{\longrightarrow} Q \]
in $\bar{A}$ for some $a$ and $w$.  Suppose for contradiction we have a run,
\[ q \stackrel{a}{\longrightarrow} Q^2
\stackrel{w}{\longrightarrow} Q' \]
in $A$ with $Q' \subset \mathcal{Q}_f$.  Then, by induction over the
length of the run, there are no accepting runs over $w$ in $\bar{A}$
from any state in $Q^2$.  In $\Delta$ we have the transition $(q, a,
Q^2)$.  By definition
there is some $q' \in Q^2$ with $q' \in Q^1$ and consequently the
accepting run
$Q^1 \stackrel{w}{\longrightarrow} Q$ cannot exist in $\bar{A}$.  We have a
contradiction.

In the only-if direction, assume there is no run,
\[ q \stackrel{a}{\longrightarrow} Q^1
\stackrel{w}{\longrightarrow} Q' \]
with $Q' \subseteq \mathcal{Q}_f$ in $A$.
For all transitions of the form $q
\stackrel{a}{\longrightarrow} Q^1$  (guaranteed to exist since $A$ is total)
there is no accepting
run $Q^1
\stackrel{w}{\longrightarrow} Q'$.  Hence, there is some $q' \in Q^1$
with no accepting run over $w$, and by induction over the length of
the run, there is an accepting run from $q'$ over $w$ in $\bar{A}$.

Let $\{(q, a, Q^\top_1),\ldots,(q, a, Q^\top_e)\}$ be the
set of all transitions in $\Delta$ from $q$ over $a$.  For each $i \in
\{1,\ldots,e\}$, let $q^\top_i
\in Q^\top_i$ be the
state from which there is no accepting run over $w$ in $A$ and
hence an accepting run over $w$ in $\bar{A}$.  By
definition of $\Delta'$ the transition $q \stackrel{a}{\longrightarrow}
\{q^\top_1,\ldots,q^\top_e\}$ exists in $\bar{A}$.
Hence we have the accepting run,
\[ q \stackrel{a}{\longrightarrow} \{q^\top_1,\ldots,q^\top_e\}
\stackrel{w}{\longrightarrow} Q' \]
in $\bar{A}$ as required.

We now consider the inductive case $n > 1$.  If $q = q^*_f$ or $q^\varepsilon_f$ the result is
immediate.  Similarly, when the length
of the run is zero, then the property follows since $\mathcal{Q}_f
\cap (\mathcal{Q}\cup\{q^\varepsilon_f,q^*_f\})\setminus\mathcal{Q}_f = \emptyset$.
Furthermore, since we have an (accepting) $\triangledown$-transition from
$q^j$ for all $j \in \{1,\ldots,z\}$ in $A$ iff there is no (accepting)
$\triangledown$-transition from $q^j$ in $\bar{A}$ the result is also
straightforward in this case.

Otherwise, in the if direction, assume we have an accepting run,
\[ q \stackrel{\gamma}{\longrightarrow} Q^1
\stackrel{w}{\longrightarrow} Q \]
in $\bar{A}$ for some $\gamma$ and $w$.  Suppose for contradiction we have a run,
\[ q \stackrel{\gamma}{\longrightarrow} Q^2
\stackrel{w}{\longrightarrow} Q' \]
in $A$ with $Q' \subset \mathcal{Q}_f$.  Then, by induction over the
length of the run, there are no accepting runs over $w$ in $\bar{A}$
from any state in $Q^2$.  In $\Delta$ we have the transition $(q, B,
Q^2)$ with $\gamma \in \lang(B)$, hence $B$ must appear positively on
the transition in $\Delta'$ from $q$ to $Q^1$ (else $\bar{B}$ appears, and by induction
over $n$, $\gamma \notin \lang(\bar{B})$).  By definition
there is some $q' \in Q^2$ with $q' \in Q^1$ and consequently the run
$Q^1 \stackrel{w}{\longrightarrow} Q$ cannot exist in $\bar{A}$.  We have a
contradiction.

In the only-if direction, assume there is no run,
\[ q \stackrel{\gamma}{\longrightarrow} Q^1
\stackrel{w}{\longrightarrow} Q' \]
with $Q' \subseteq \mathcal{Q}_f$ in $A$.  There are two cases.
\begin{enumerate}[$\bullet$]
\item
If there are no transitions $q \stackrel{\gamma}{\longrightarrow} Q^1$
in $A$
then for all $q \stackrel{B}{\longrightarrow} Q^1$ we have $\gamma
\in \bar{B}$ by induction over $n$.  Hence, in $\bar{A}$ we have a
run,
\[ q \stackrel{\gamma}{\longrightarrow} q^*_f
\stackrel{w}{\longrightarrow} Q^* \]
which is an accepting run as required.

\item
If there are transitions of the form $q
\stackrel{\gamma}{\longrightarrow} Q^1$ in $A$ then for each of these runs
there is no accepting
run $Q^1
\stackrel{w}{\longrightarrow} Q'$.  Hence, there is some $q' \in Q^1$
with no accepting run over $w$, and by induction over the length of
the run, there is an accepting run from $q'$ over $w$ in $\bar{A}$.

Let $\{(q, B^t_1, Q^t_1),\ldots,(q, B^t_e, Q^t_e),(q,
B^f_1,Q^f),\ldots,(q,B^f_h,Q^f_h)\}$ be the
set of all transitions in $\Delta$ from $q$ such that $\gamma \in
B^t_i$ for all $i \in \{1,\ldots,e\}$ and $\gamma \notin
B^f_i$ for all $i \in \{1,\ldots,h\}$ (and consequently $\gamma
\in \bar{B}^f_i$).  For each $i \in \{1,\ldots,e\}$ let $q^t_i \in Q^t_i$ be the
state from which $\bar{A}$ has no accepting run over $w$ in $A$ and
hence has an accepting run over $w$ in $\bar{A}$.  By
definition of $\Delta'$ the transition $q \stackrel{B}{\longrightarrow}
\{q^t_1,\ldots,q^t_e\}$ with $B = B^t_1 \cap \ldots \cap
B^t_e \cap \bar{B}^f_1 \cap \ldots \cap \bar{B}^f_h$ exists in $\bar{A}$.
Hence we have the accepting run,
\[ q \stackrel{\gamma}{\longrightarrow} \{q^t_1,\ldots,q^t_e\}
\stackrel{w}{\longrightarrow} Q' \]
in $\bar{A}$ as required.
\end{enumerate}
We have shown that $\bar{A}$ has an accepting run from any state
iff there is no accepting run from that state in $A$ as required.
\end{proof}

\section{Soundness and Completeness for $A_0, A_1, \ldots$}
\label{soundnessandcompleteness}

In this section we show that the sequence $A_0, A_1, \ldots$ is sound
and complete with respect to $Pre^*(C_{Init})$, where $C_{Init} = \lang(A_0)$.

\subsection{Preliminaries}

We begin by proving some useful properties of the automaton construction.  These
properties assert that the automata constructed from the sets in $\idallG^i_l$
are well-behaved.  Once this has been established, we need only consider
order-$n$ of the automata $A_0, A_1, \ldots$ to show soundness and completeness.
Note that since no $g^i_{(q_1, Q_2)}$ is accepting, any store accepted by some
$G^i_{(q_1, Q_2)}$ has a $top_1$ element.

In order to reason about a particular transition, we need to know its origin.
Hence we introduce the notion of an \emph{inherited} and a \emph{derived}
transition.  The remaining lemmata fall into four categories:
\begin{enumerate}[(1)]
\item
Lemma~\ref{inheritedis} shows that inherited runs are sound.
\item
Lemma~\ref{runspersist} shows the completeness of inherited runs.
\item
Lemma~\ref{thetaisinS} and
Lemma~\ref{othetaisinS} show that derived runs are sound.
\item
Lemma~\ref{thetaandothetaSbackwards} shows the completeness of derived runs.
\end{enumerate}

\begin{defi}
A non-empty run $g^{i}_{(q_1, Q_2)} \stackrel{w}{\longrightarrow}_i Q$ of
$\mathcal{G}^i$ or $q^j \stackrel{w}{\longrightarrow}_i Q$ of $A_i$ can be characterised by
its initial transition.
There are two cases,
\begin{enumerate}[$\bullet$]
\item
A run of $\mathcal{G}^i$:
Then $w = aw'$ and we have $g^i_{(q_1, Q_2)} \stackrel{a}{\longrightarrow}_i Q'$.
If the transition was inherited from $g^{i-1}_{(q_1, Q_2)}$ then we
say that the run is an {\bf inherited} run.  Otherwise the transition
was introduced by some $S \in \idG^i_{(q_1, Q_2)}$.  We say that the
run was {\bf derived} from $S$.

\item
A run of $A_i$:
We have $w = \gamma w'$ and $q^j \stackrel{G}{\longrightarrow}_i Q'$ with
$\gamma \in \lang(G)$.

If the accepting run of $G$ was inherited, then the run is {\bf inherited}.
If the accepting run of $G$ is derived from some $S' \in \idG$ and $S'$
was added to $\idG$ by $T_\mathcal{D}$ and the command $d$, then the run $g^{i}_{(q_1, Q_2)} \stackrel{w}{\longrightarrow}_i Q$ is {\bf
derived} from $d$.
\end{enumerate}
\end{defi}


The language accepted by
the sequence $A_0, A_1,\ldots$ or
$\mathcal{G}^0,\mathcal{G}^1,\ldots$ is increasing.  In
particular, if $q \stackrel{w}{\longrightarrow}_i Q$ exists in $A_i$, then $q
\stackrel{w}{\longrightarrow}_{i+1} Q$ exists in $A_{i+1}$.

\vbox{\begin{lem}
\label{runspersist}
$\quad$
\begin{enumerate}[\em(1)]
\item
If $g^{i}_{(q_1, Q_2)} \stackrel{w}{\longrightarrow}_i Q$ is a run of
$G^i_{(q_1, Q_2)}$ for some $i$ (and $w \neq \varepsilon$), then $g^{i+1}_{(q_1, Q_2)}
\stackrel{w}{\longrightarrow}_{i+1} Q$ is a run of $G^{i+1}_{(q_1, Q_2)}$.

\item
For all transitions $q \stackrel{\gamma}{\longrightarrow}_i Q'$ in $A_i$
for some $i$, we have the transition $q
\stackrel{\gamma}{\longrightarrow}_{i+1} Q'$ in $A_{i+1}$.

\item
For all runs $q \stackrel{w}{\longrightarrow}_i Q'$ of $A_i$ for some $i$,
we have the run $q \stackrel{w}{\longrightarrow}_{i+1} Q'$ in $A_{i+1}$.
\end{enumerate}
\end{lem}}
\begin{proof}
To prove (2) we observe that there are two cases.  In the first case,
the transition from $q$ to $Q'$ is labelled by an
automaton $B \in \mathcal{B}$ or $\triangledown$.  Because this transition will remain
unchanged by $T_\mathcal{D}$, the lemma follows immediately. In the
second case, the transition is labelled by $G^i_{(q, Q')}$ and the
property follows directly from (1) and the run $g^i_{(q, Q')}
\stackrel{w_\gamma}{\longrightarrow}_i Q$ with $Q \subseteq
\mathcal{Q}_f$ for $[w_\gamma] = \gamma$.  Since $g^i_{(q, Q')}$ is
not an accepting state, it is the case that $w_\gamma \neq \varepsilon$.

We note that (3) can be shown by repeated applications of (2).

Finally, we show (1).  The automaton $G^i_{(q_1, Q_2)}$ has the run,
\[ g^i_{(q_1, Q_2)} \stackrel{a}{\longrightarrow}_i Q^1
\stackrel{w'}{\longrightarrow}_i Q
\]
where $w = aw'$.

By definition the
automaton $G^{i+1}_{(q_1, Q_2)}$ has the transition
$g^{i+1}_{(q_1,Q_2)} \stackrel{a}{\longrightarrow}_i Q^2$ for every
transition $g^{i}_{(q_1, Q_2)} \stackrel{a}{\longrightarrow}_i
Q^2$.  Hence we have the run,
\[ g^{i+1}_{(q_1, Q_2)} \stackrel{a}{\longrightarrow}_{i+1} Q^1
\stackrel{w'}{\longrightarrow}_i Q \]
as required.
\end{proof}

\begin{lem}
\label{inheritedis}
If a run $g^{i+1}_{(q_1, Q_2)} \stackrel{w}{\longrightarrow}_{i+1}
Q$ in $\mathcal{G}^{i+1}$ is inherited, then the run $g^{i}_{(q_1, Q_2)}
\stackrel{w}{\longrightarrow}_i Q$ exists in $\mathcal{G}^i$.
\end{lem}
\begin{proof}
Observe that an inherited run
cannot be empty.
We have $w = aw'$ and,
\[ g^{i+1}_{(q_1, Q_2)} \stackrel{a}{\longrightarrow}_{i+1} Q'
\stackrel{w'}{\longrightarrow}_i Q \]
Since the run is an inherited run, we have $g^i_{(q_1, Q_2)}
\stackrel{a}{\longrightarrow}_i Q'$ in $\mathcal{G}^i_1$ and hence,
\[ g^{i}_{(q_1, Q_2)} \stackrel{a}{\longrightarrow}_{i} Q'
\stackrel{w'}{\longrightarrow}_i Q \]
in $\mathcal{G}^i$ as required.
\end{proof}

\begin{lem}
\label{thetaisinS}
Suppose the run $g^{i+1}_{(q_1, Q_2)}
\stackrel{w}{\longrightarrow}_{i+1} Q$ derived from $S$ exists
in $\mathcal{G}^{i+1}$ and $\theta_1 \in S$. We have
$q^{\theta_1} \stackrel{w}{\longrightarrow}_i Q'$
in $\mathcal{G}^i$, where $Q' \subseteq Q$.
\end{lem}
\begin{proof}
 Observe that, since
the run is derived, we have $w \neq \varepsilon$.
 Let $w = aw'$.  We have the
following run in $\mathcal{G}^{i+1}$,
\[ g^{i+1}_{(q_1, Q_2)} \stackrel{a}{\longrightarrow}_{i+1} Q^1
\stackrel{w'}{\longrightarrow}_i Q \]
and by definition, since the run is derived from $S$ and $\theta_1 \in S$, we have $q^{\theta_1} \stackrel{a}{\longrightarrow}_i Q^2$
in $\mathcal{G}^i$ where $Q^2 \subseteq Q^1$, and hence,
\[ q^{\theta_1} \stackrel{a}{\longrightarrow}_i Q^2
\stackrel{w'}{\longrightarrow}_i Q' \]
with $Q' \subseteq Q$ as required.
\end{proof}

\begin{lem}
\label{othetaisinS}
Suppose the run $g^{i+1}_{(q_1, Q_2)}
\stackrel{w}{\longrightarrow}_{i+1} Q$ derived from $S$ exists in
$\mathcal{G}^{i+1}$ and there is some $(a, o,\theta_1) \in S$.
If $[w'] = o([w])$, we have
$q^{\theta_1} \stackrel{w'}{\longrightarrow}_i Q'$
in $\mathcal{G}^i_l$, where $Q' \subseteq Q$.
\end{lem}
\begin{proof}
Since the run is derived, we have $w \neq \varepsilon$.
We have $w = aw''$.  There is only one
value of $o$,
$o = push_{w_p}$ and $[w'] = o([w]) = [w_pw'']$.  We have the
following run in $\mathcal{G}^{i+1}_l$,
\[ g^{i+1}_{(q_1, Q_2)} \stackrel{a}{\longrightarrow}_{i+1} Q^1
\stackrel{w''}{\longrightarrow}_i Q \]
and by definition, since the run is derived from $S$ and $(a, o,
\theta_1) \in S$, we have $q^{\theta_1} \stackrel{w_p}{\longrightarrow}_i Q^2$
in $\mathcal{G}^i_l$ where $Q^2 \subseteq Q^1$, and hence,
\[ q^{\theta_1} \stackrel{w_p}{\longrightarrow}_i Q^2
\stackrel{w''}{\longrightarrow}_i Q' \]
with $Q'_f \subseteq Q$ as required.
\end{proof}

\begin{lem}
\label{thetaandothetaSbackwards}
Let $S = \{\alpha_1,\ldots,\alpha_m\} \in \idG^{i+1}_{(q, Q)}$.
Given some $\gamma$ with $top_1(\gamma) = a$ such that for each $e \in \{1,\ldots,m\}$ we have,
\begin{enumerate}[$\bullet$]
\item
If $\alpha_e = \theta_e$ then $\gamma_e = \gamma$ and $\gamma_e \in
\lang(\theta_e)$
\item
If $\alpha_e = (b, o_e, \theta_e)$ then $b = a$, $o_e(\gamma) = \gamma_e$ and $\gamma_e
\in \lang(\theta_e)$
\end{enumerate}
we have $\gamma \in \lang(G^{i+1}_{(q, Q)})$.
\end{lem}
\begin{proof}
Let $\gamma = [aw]$.  We have $\alpha_e = \theta_e$ or
$\alpha_e = (a, push_{w_e},
\theta_e)$.  We have,
\begin{enumerate}[$\bullet$]
\item
When $\alpha_e = \theta_e$, the run,
\[ q^{\theta_e} \stackrel{a}{\longrightarrow}_i Q_e
\stackrel{w}{\longrightarrow}_i Q^e_f \]
with $Q^e_f \subseteq \mathcal{Q}_f$ in $\mathcal{G}^i$.
Furthermore, $\gamma_e = \gamma$.
\item
When $\alpha_e = (a, push_{w_e}, \theta_e)$, the run,
\[ q^{\theta_e} \stackrel{w_e}{\longrightarrow}_i Q_e
\stackrel{w}{\longrightarrow}_i Q^e_f \]
with $Q^e_f \subseteq \mathcal{Q}_f$ in $\mathcal{G}^i_l$.
Furthermore, we have $\gamma_e = [w_ew]$.
\end{enumerate}
Hence, since $S \in \idG^{i+1}_{(q, Q)}$, we have from the
definition of $G^{i+1}_{(q, Q)}$ the run,
\[ g^{i+1}_{(q, Q)} \stackrel{a}{\longrightarrow}_{i+1} Q_1 \cup
\ldots \cup Q_m \stackrel{w}{\longrightarrow}_i Q^1_f \cup \ldots \cup
Q^m_f \]
with $Q^1_f \cup \ldots \cup Q^m_f \subseteq \mathcal{Q}_f$.  Hence
$\gamma \in \lang(G^{i+1}_{(q, Q)})$ as required.
\end{proof}

\subsection{Soundness}

We show that for any configuration $\langle p^j, \gamma\rangle$ such that $\gamma
\in \lang(A^{q^j}_i)$, for some $i$, we have $\langle p^j, \gamma\rangle
\pdsreach C$ with $C \subseteq C_{Init}$.  Let $I = \{q^1,\ldots,q^z\}$.  The
following lemma describes the relationship between added transitions and the
evolution of the order-$2$ PDS.

In the following lemma, the restrictions on $w'$ are technical requirements in
the case of $pop_2$ operations.  They may be justified by observing that only
the empty store is accepted from the state $q^\varepsilon_f$, and that, since
initial states are never accepting, the empty store cannot be accepted from an
initial state.

\begin{lem}
\label{wholerun}
For a given run $q^j \stackrel{w}{\longrightarrow}_{i} Q$ of $A_i$ there exists
for any $w'$ satisfying the conditions below, some $C$ such that $\langle p^j,
[w w']\rangle \pdsreach C$, where $C$ contains configurations of the form
$\langle p^k, w''w'\rangle$ with $q^{k}\stackrel{w''}{\longrightarrow}_0 Q'$ or $\langle
p^j, \triangledown\rangle$ with $q^j \stackrel{\triangledown}{\longrightarrow}_0
Q'$.  Furthermore, the union of all such $Q'$ is $Q$.  We require $w' \neq
\triangledown$ and,
\begin{enumerate}[(1)]
\item
 If $q^\varepsilon_f \in Q$ then $w' = \varepsilon$,
\item
 If $q^k \in Q$ for some $q^k$ then $w' \neq \varepsilon$.
\end{enumerate}
\end{lem}
\begin{proof}
The proof proceeds by induction on $i$.  In the base case $i = 0$ and
the property holds trivially.  We now consider the case for $i + 1$.
Since $T_\mathcal{D}$ does not add any
$\triangledown$-transitions, we can assume $w \neq \triangledown$.

We perform a further induction over the length of the run.  In the
base case we have $w = \gamma$ (the case $w = \varepsilon$ is immediate with $C
= \{\langle p^j, [w']\rangle\}$)
and consider the single transition $q^j
\stackrel{\gamma}{\longrightarrow}_{i+1} Q$.  We assume that the
transition is not inherited, else the
property holds by Lemma~\ref{inheritedis} and induction over $i$.  If the transition is
not inherited, then the run is derived from some $d$ and we have
$\gamma \in \lang(G^{i+1}_{(q^j, Q)})$
 and the accepting run of $G^{i+1}_{(q^j, Q)}$ is derived from
some $S \in \idG^{i+1}_{(q^j, Q)}$ introduced by during the
processing of $d$.

Let $d = (p^j, a, \{(o_1, p^{k_1}),\ldots,(o_m, p^{k_m})\})$.  We have
$\langle p^j, [\gamma w']\rangle \hookrightarrow C'$ where,
\[ \begin{array}{rcl}
    C' &=& \{\ \langle p^{k_t},
\gamma'\rangle\ |\ t \in\{1,\ldots,m\} \land\
\gamma' = o_t([\gamma w'])\
\}  \\
    & & \cup\ \{\ \langle p^j, \triangledown\rangle\ |\ \mathrm{if\ }
o_t([\gamma w'])\
\mathrm{with\ } t \in \{1,\ldots,m\}
\mathrm{\ is\ not\ defined}\ \}
    \end{array}
\]

We can decompose the new
transition as per the definition of $T_\mathcal{D}$.  That is $Q =
Q'_1\cup\ldots\cup Q'_m$.
  There are several
cases:
\begin{enumerate}[$\bullet$]
\item
$o_t = push_2$.

By definition of $T_\mathcal{D}$, we have the run,
\[ q^{k_t} \stackrel{\widetilde{\theta}_1}{\longrightarrow}_i Q'
\stackrel{\widetilde{\theta}_2}{\longrightarrow}_i Q'_t \]
with $\{B^a_1\} \cup \widetilde{\theta}_1 \cup \widetilde{\theta}_2
\subseteq S$. By
Lemma~\ref{thetaisinS} we have $\gamma \in
\lang(\{B^a_1\} \cup \widetilde{\theta}_1 \cup \widetilde{\theta}_2)$.  Hence we have,
\[ q^{k_t} \stackrel{\gamma}{\longrightarrow}_i
Q' \stackrel{\gamma}{\longrightarrow}_i Q'_t \]

We have $push_2[\gamma w'] = [\gamma\gamma w']$ and $\langle p^{k_t},
[\gamma\gamma w']\rangle \in C'$.
Via induction over $i$ we have the set $C_t$ with $\langle p^{k_t}, o_t[\gamma w']\rangle
\pdsreach C_t$ which satisfies the lemma.

\item
$o_t = pop_2$.

We have  $B^a_1
\in S$.  We have, by
Lemma~\ref{thetaisinS}, $\gamma \in \lang(B^a_1)$.

If $Q_t
= \{q^{k_t}\}$ then $pop_2[\gamma w'] = [w']$ since $w'$ is non-empty
and $C_t = \{\langle p^{k_t}, [w']\rangle\}$.
Note $q^{k_t} \stackrel{\varepsilon}{\longrightarrow}_0 \{q^{k_t}\}$.

If $Q_t = \{q^\varepsilon_f\}$ then $w' = \varepsilon$ and $pop_2[\gamma w']$ is
undefined.  By definition of $T_\mathcal{D}$ we have
$q^j \stackrel{\triangledown}{\longrightarrow}_0 \{q^\varepsilon_f\}$.  Let $C_t
= \{\langle p^j, \triangledown\rangle\}$.

\item
$o_t = push_w$.

By definition, we have $q^{k_t} \stackrel{\theta}{\longrightarrow}_i
Q_t$ in $A_i$ and $(a, o_t, \theta) \in S$.  Hence, by
Lemma~\ref{othetaisinS}, we have $o_t[\gamma] \in \lang(\theta)$ and
the run $q^{k_t} \stackrel{o_t[\gamma]}{\longrightarrow}_i Q_t$ in
$A_i$.
Furthermore, it is the case that $\langle p^{k_t}, o_t[\gamma w']\rangle \in C'$ and
via induction over $i$ we have a set $C'$ with $\langle p^{k_t}, o_t[\gamma w']\rangle
\pdsreach C_t$ which satisfies the lemma.
\end{enumerate}
Hence, we have $\langle p^j, [ww']\rangle \hookrightarrow C' \pdsreach C_1 \cup
\ldots \cup C_m = C$ where $C$ satisfies the lemma.

This completes the proof of the single transition case.  Let $w =
\gamma_1\ldots\gamma_m$ and (for any $Q$) let $Q = Q^I \cup Q^{\setminus I}$ where
$Q^I$ contains all initial states in $Q$ and $Q^{\setminus I} = Q \setminus Q^I$.
We have the run,
\[ q^j \stackrel{\gamma_1}{\longrightarrow}_{i+1} Q_1
\stackrel{\gamma_2}{\longrightarrow}_{i+1} \ldots
\stackrel{\gamma_m}{\longrightarrow}_{i+1} Q_m \]
For each $q^k \in Q^I_1$ we have a run,
\[ q^k \stackrel{\gamma_2}{\longrightarrow}_{i+1} Q^k_2
\stackrel{\gamma_3}{\longrightarrow}_{i+1} \ldots
\stackrel{\gamma_m}{\longrightarrow}_{i+1} Q^k_m\]
and by induction on the length of the run we have $C_k$ such that $\langle p^k,
[\gamma_2\ldots\gamma_m w']\rangle \pdsreach C_k$ and $C_k$ satisfies
the lemma. Furthermore, since we only
add new transitions to initial states, we have,
\[ Q^{\setminus I}_1 \stackrel{\gamma_2}{\longrightarrow}_0 \ldots
\stackrel{\gamma_m}{\longrightarrow}_0 Q'_m \]
and $Q_m = Q'_m \cup \bigcup_{q^k \in Q^I_1}Q^k_m$.

From $q^j \stackrel{\gamma_1}{\longrightarrow}_{i+1} Q_1$
we have $C_1$ with $\langle p^j, [\gamma_1\ldots\gamma_mw']\rangle
\pdsreach C_1$ satisfying the lemma.
Let $C^I_1$ be the set of all $\langle p^k,
\gamma_2\ldots\gamma_mw'\rangle \in C_1$ and $C'_1 =
C_1\setminus C^I_1$.
For each $q^k \in Q^I_1$ we have $\langle p^k,
[\gamma_2\ldots\gamma_m w']\rangle \in C_1$ since there are no transitions to
initial states in $A_0$ (and hence we must have $q^k
\stackrel{\varepsilon}{\longrightarrow}_0 \{q^k\}$ to satisfy the conditions of
the lemma for $C_1$).  From $\langle p^k,
[\gamma_2\ldots\gamma_m w']\rangle \pdsreach C_k$ and since we have
$Q^{\setminus I}_1
\xrightarrow{\gamma_2\ldots\gamma_m}_0 Q'_m$, it is the
case that the set $C = C'_1 \cup \bigcup_{q^k \in
Q^I_1}C_k$ which has $\langle p^j, [\gamma_1\ldots\gamma_mw']\rangle
\pdsreach C_1 \pdsreach C$ and satisfies the lemma as required.
\end{proof}

\begin{property}[Soundness]
\label{infseqsound}
For any configuration $\langle p^j, \gamma\rangle$ such that $\gamma \in
\lang(A_i^{q^j})$ for some $i$, we have $\langle p^j, \gamma \rangle
\pdsreach C$ such that $C \subseteq C_{Init}$.
That is, $\langle p^j, \gamma\rangle \in Pre^*(C_{Init})$.
\end{property}
\begin{proof}
Let $\gamma = [w_\gamma]$.  Since $\gamma \in \lang(A_i^{q^j})$ we have a run $q^j
\stackrel{w_\gamma}{\longrightarrow}_i Q_f$ with $Q_f \subseteq
\mathcal{Q}_f$.  Since $\mathcal{Q}_f$ contains no initial states, we
apply Lemma~\ref{wholerun} with $w' = \varepsilon$.  Therefore, we
have $\langle p^j, \gamma\rangle \pdsreach C \subseteq \lang(A_0^{q^k})$.  Since $A_0$ is
defined to represent $C_{Init}$, soundness follows.
\end{proof}

\subsection{Completeness}
\label{completeness}

\begin{property}[Completeness]
\label{infseqcomp}
For all $\langle p^j, \gamma\rangle \in Pre^*(C_{Init})$ there is some $i$ such
that $\gamma \in \lang(A^{q^j}_i)$.
\end{property}
\begin{proof}
We take $\langle p^j, \gamma\rangle \in Pre^*(C_{Init})$ and reason by induction
over the length of the shortest path $\langle p^j, \gamma\rangle \pdsreach C$ with $C \subseteq C_{Init}$.

In the base case the path length is zero and we have $\langle p^j, \gamma
\rangle \in C_{Init}$ and hence $\gamma \in \lang(A^{q^j}_0)$.

For the inductive step we have $\langle p^j, \gamma\rangle \hookrightarrow
C_1 \pdsreach C_2$ with
$C_2 \subseteq C_{Init}$ and some $i$ such that $C_1 \subseteq \lang(A_{i})$ by induction.  We show $\gamma \in
\lang(A^{q^j}_{i+1})$ by analysis of the higher-order APDS command $d$ used in
the transition $\langle p^j, \gamma\rangle \hookrightarrow
C_1$.

Let $d = (p^j, a, \{(o_1, p^{k_1}),\ldots,(o_m, p^{k_m})\})$.  We have
\[ \begin{array}{rcl}
    C_1 &=& \{\ \langle p^{k_t},
\gamma'\rangle\ |\  t \in\{1,\ldots,m\} \land\ \gamma' = o_t(\gamma)\
\}  \\
    & & \cup\ \{\ \langle p^j, \triangledown\rangle\ |\ \mathrm{if\ } o_t(\gamma)\
\mathrm{with\ } t \in \{1,\ldots,m\}
\mathrm{\ is\ not\ defined}\ \}
    \end{array}
\]
By
induction we have for each $e \in \{1,\ldots,m\}$ that $q^{k_e} \xrightarrow{w_{o_e(\gamma)}}_i
Q^e_f$ with $Q^e_f \subseteq \mathcal{Q}_f$ in $A_i$ if $o_e(\gamma) =
[w_{o_e(\gamma)}]$
is defined. Otherwise we have $q^j \stackrel{\triangledown}{\longrightarrow}_i
\{q^\varepsilon_f\}$ in $A_i$.

Let $\gamma = [\gamma'w]$. We have
  $S' = S'_1 \cup \ldots \cup S'_m$ and $Q' = Q_1 \cup \ldots \cup
Q_m$ where,  for each $e \in \{1,\ldots,m\}$,
\begin{enumerate}[$\bullet$]
\item
When $o_e = push_2$, $o_e(\gamma) = [\gamma'\gamma'w]$.
Additionally, we have the transitions,
\[ q^{k_e} \stackrel{\theta^1_e}{\longrightarrow}_i Q'
\stackrel{\widetilde{\theta}^2_e}{\longrightarrow}_i Q_e \]
in $A_i$ where $\gamma' \in
\lang(\{B^a_1, \theta^1_e\} \cup \widetilde{\theta}^2_e)$.  Furthermore, we
have the run $Q_e \stackrel{w}{\longrightarrow}_i Q^e_f$ with $Q^e_f
\subseteq \mathcal{Q}_f$ and $S'_e = \{B^a_1, \theta^1_e\} \cup \widetilde{\theta}^2_e$.

\item
When $o_e = pop_2$.  If $o_e(\gamma) = [w]$, we have the
run,
\[ q^{k_e} \stackrel{w}{\longrightarrow}_i Q^e_f \]
in $A_i$ with $Q^e_f \subseteq \mathcal{Q}_f$, $S'_e = \{B^a_{1}\}$, $\gamma'
\in \lang(B^a_1)$ and $Q_e = \{q^{k_e}\}$.

If $o_e(\gamma)$ is undefined we have $w = \varepsilon$ and the run,
\[ q^j \stackrel{\triangledown}{\longrightarrow}_i \{q^\varepsilon_f\} \]
if $A_i$.  Hence we have $S'_e = \{B^a_1\}$, $\gamma' \in
\lang(B^a_1)$ and $Q_e = Q^e_f =  \{q^\varepsilon_f\}$.

\item
When $o_e = push_w$, and we have
$o_e(\gamma) = [o_e(\gamma')w]$, and the transition $q^{k_e}
\stackrel{\theta'_e}{\longrightarrow}_i Q_e$ and run $Q_e
\stackrel{w}{\longrightarrow}_i Q^e_f$ with $Q^e_f \subseteq
\mathcal{Q}_f$ in $A_i$.  Additionally, $o_e(\gamma') \in
\lang(\theta'_e)$ and $S'_e = \{(a, o_e, \theta'_e)\}$.
\end{enumerate}
Hence, by definition of $A_{i+1}$, we have the transition,
\[ q^j \stackrel{\idG}{\longrightarrow}_{i+1} Q_1 \cup
\ldots \cup Q_m \]
with $S' \in \idG$ and by
Lemma~\ref{thetaandothetaSbackwards} $\gamma' \in \lang(G)$.  Hence we have the run,
\[ q^j \stackrel{\gamma'}{\longrightarrow}_{i+1} Q_1
\cup \ldots \cup Q_m \stackrel{w}{\longrightarrow}_i Q^1_f \cup \ldots
\cup Q^m_f \]
with $Q^1_f \cup \ldots \cup Q^m_f \subseteq \mathcal{Q}_f$ in
$A_{i+1}$.  That is, $\gamma \in \lang(A^{q^j}_{i+1})$ as required.
\end{proof}

\section{Proofs for $A_*$}
\label{termination}

In this section we provide a proof of Lemma~\ref{finiteorderorder2}.
The main idea of the proof is that the loops in $\newautG^i$ can simulate,
correctly, the prefix of any run in $\mathcal{G}^{i'}$ and vice-versa.  That is, a run in
$\newautG^i$ begins by traversing it's initial loops before progressing to its
accepting states.  If we unroll this looping we will construct a run of
$\mathcal{G}^{i'}$ for a sufficiently large $i'$.  In the other direction, the
prefix of a run in $\mathcal{G}^{i'}$ can be simulated by the initial looping
behaviour of $\newautG^i$.

We begin by proving a small lemma that will ease the remaining proofs.

\begin{lem}
\label{imax}
Given $g^{i_y}_{(q_y, Q_y)} \stackrel{w}{\longrightarrow}_{i_y} Q_y$ for
all $y \in \{1,\ldots,h\}$ for some $h$, let $i_{max}$ be the maximum $i_y$.  We
have $\{g^{i_{max}}_{(q_1, Q_1)},\ldots,g^{i_{max}}_{(q_h, Q_h)}\}
\stackrel{w}{\longrightarrow} \bigcup_{y \in \{1,\ldots,h\}} Q_y$.
\end{lem}
\begin{proof}
By Lemma~\ref{runspersist} we have $g^{i_{max}}_{(q_y, Q_y)}
\stackrel{w}{\longrightarrow}_{i_{max}} Q_y$ for each
$y \in \{1,\ldots,h\}$.  Hence we have the run as required.
\end{proof}

\subsection{Proofs of Lemma~\ref{finiteorderorder2}}

\begin{lem}
\label{lis1fixedpoint}
There exists some $i_0$ such that $\newautG^i = \newautG^{i_0}$ for
all $i > i_0$.  Furthermore, we have the run $g^{i_1}_{(q, Q')}
\stackrel{w}{\longrightarrow}_i Q_f$ with $Q_f \subseteq
\mathcal{Q}_f$ for some $i$ iff we have  $g^{i_1}_{(q, Q')}
\stackrel{w}{\longrightarrow}_{i_0} Q_f$ in $\newautG^{i_0}$.
\end{lem}
\begin{proof}
This is a simple consequence of the finiteness of $\Sigma$ and that
$T_{\idallG^{i_1}[i_1/i_1 - 1]}$ only adds transitions and never states.  The
automaton will eventually become saturated and no new transitions will
be added.
\end{proof}

\begin{lem}
\label{lis1oneway}
For all $w$, if $g^i_{(q, Q')} \stackrel{w}{\longrightarrow}_i Q_1$
with $Q_1 \subseteq \mathcal{Q}_f$
is a run in $\mathcal{G}^i$ for some $i$, then we have
$g^{i_1}_{(q, Q')} \stackrel{w}{\longrightarrow}_{i_0} Q_2$ with $Q_2
\subseteq \mathcal{Q}_f$ in $\newautG^{i_0}$.
\end{lem}
\begin{proof}
We prove the following property.    For any path $g^{i}_{(q, Q')}
\stackrel{w}{\longrightarrow}_{i} \{q_1,\ldots,q_h\}$ in
$\mathcal{G}^i$,
we have a path $g^{i_1}_{(q, Q')}
\stackrel{w}{\longrightarrow}_{i_0} \{q^!_1,\ldots,q^!_h\}$ in
$\newautG^{i_0}$ with,
\[  q^!_y = \left\{\begin{array}{ll}
                g^{i_1}_{(q', Q'')} & \mathrm{if\ } q_y =
g^{i'}_{(q', Q'')}\mathrm{\ and\ } i' \geq
i_1 \\
                q_y & \mathrm{otherwise}
                   \end{array}\right.
\]
for all $y \in \{1,\ldots,h\}$.  Since $q^!_f = q_f$ for all $q_f \in \mathcal{Q}_f$, the lemma follows.
When $Q = \{q_1,\ldots,q_h\}$ we write $Q^!$ to denote the set $\{q^!_1,\ldots,q^!_h\}$.

There are two cases.  When $i \leq i_1$, then using
 that we have only added transitions to
$\mathcal{G}^{i_1}$ to define $\newautG^{i_0}$ and that $q^!_y =
q_y$ for all $y$, we have
 $g^{i_1}_{(q, Q')}
\stackrel{w'}{\longrightarrow}_{i_0} \{q^!_1,\ldots,q^!_h\}$ in $\newautG^{i_0}$.

We now consider the case $i > i_1$.  We begin by proving that for a single transition,
\[ g^i_{(q, Q')} \stackrel{b}{\longrightarrow}_i \{q_1,\ldots,q_{h}\} \]
in $\mathcal{G}^i$ with $b \in \Sigma$, we have the
following transition in $G^{i_0}_{(q, Q')}$,
\[ g^{i_1}_{(q, Q')} \stackrel{b}{\longrightarrow}_{i_0} \{q^!_1,\ldots,q^!_h\}
\]
We consider the source $S = \{\alpha_1,\ldots,\alpha_m\} \in
\idG^i_{(q, Q')}$ of the transition from $g^i_{(q, Q')}$.  Since
 $\idG^i_{(q, Q')} \simeq \idG^{i_1}_{(q, Q')}$ we have
$S[i_1/i-1] \in \idG^{i_1}_{(q,
Q')}[i_1/i_1-1]$.  Furthermore, we have $\{q_1,\ldots,q_h\} = Q_1 \cup \ldots
\cup Q_m$.
For $e \in \{1,\ldots,m\}$ there are two cases,
\begin{enumerate}[$\bullet$]
\item
If $\alpha_e = \theta$, then let $g = q^{\theta}$.  We have $g
\stackrel{b}{\longrightarrow}_{i-1} Q_e$ exists in
$\mathcal{G}^{i-1}_1$.  By induction over $i$ we have $g^!
\stackrel{b}{\longrightarrow}_{i_0} Q^!_e$ in $\newautG^{i_0}_1$.

\item
$\alpha_e = (a, push_{w_p}, \theta)$.  Then $b = a$. Let $g = q^{\theta}$.
By definition of $T_{\idallG^{i_1}[i_1/i_1 - 1]}$, we have the path $g
\stackrel{w_p}{\longrightarrow}_{i-1} Q_e$ in $\mathcal{G}^i$.  By induction on $i$ we have
the path $g^! \stackrel{w_p}{\longrightarrow}_{i_0} Q^!_e$ in
$\newautG^{i_0}$.
\end{enumerate}
We have $Q^!_1 \cup \ldots \cup Q^!_m = \{q^!_1,\ldots,q^!_h\}$.
Since $\idG^i_{(q, Q')} \simeq \idG^{i_1}_{(q, Q')}$ and
$S[i_1/i - 1] \in \idG^{i_1}_{(q,
Q')}[i_1/i_1-1]$, by definition of $\newautG^{i_0}$, we have,
\[ g^{i_1}_{(q, Q')} \stackrel{b}{\longrightarrow}_{i_0}
\{q^!_1,\ldots,q^!_h\} \]
in $\newautG^{i_0}_1$ as required.

We now prove the result for a run of more than one step by induction over the
length of the run.  In the base case we have a run of a single
transition.  The result in this case has already been shown.

In the inductive case we have a run of the form,
\[ g^i_{(q, Q')} \stackrel{a_0}{\longrightarrow}_i \{q^1_1,\ldots,q^1_{h_1}\}
\stackrel{a_1}{\longrightarrow}_i \ldots \stackrel{a_m}{\longrightarrow}_i
\{q^m_1,\ldots,q^m_{h_m}\} \]
in $\mathcal{G}^i$.  For each $y \in \{1,\ldots,h_1\}$ we have a run $q^1_y
\stackrel{a_1\ldots a_m}{\longrightarrow}_i Q_y$ such that $\bigcup_{y \in
\{1,\ldots,h_1\}} Q_y = \{q^m_1,\ldots,q^m_{h_m}\}$.  By induction
over the length of the run we have
$q^{!1}_y \stackrel{a_1\ldots a_m}{\longrightarrow}_{i_0} Q^!_y$ for each $y$.  Hence, since we have
$g^{i_1}_{(q, Q')} \stackrel{a_0}{\longrightarrow}_{i_0}
\{q^{!1}_1,\ldots,q^{!1}_{h_1}\}$ from the above proof for one
transition, we have a run of the form,
\[ g^{i_1}_{(q, Q')} \stackrel{a_0}{\longrightarrow}_{i_0}
\{q^{!1}_1,\ldots,q^{!1}_{h_1}\} \stackrel{a_1}{\longrightarrow}_{i_0} \ldots \stackrel{a_m}{\longrightarrow}_{i_0}
\{q^{!m}_1,\ldots,q^{!m}_{h_m}\} \]
in $\newautG^{i_0}$ as required.
 \end{proof}

\begin{lem}
\label{lis1theother}
For all $w$, if we have $g^{i_1}_{(q, Q')}
\stackrel{w}{\longrightarrow}_{i} Q_f$ with $Q_f \subseteq \mathcal{Q}_f$ in $\newautG^i$ for some $i$,
then there is some $i'$
such that the run $g^{i'}_{(q, Q')} \stackrel{w}{\longrightarrow}_{i'}
Q_f$ exists in $\mathcal{G}^{i'}$.
\end{lem}
\begin{proof}

We take a run of $\newautG^{i}_{(q, Q')}$,
\[ g^{i_1}_{(q, Q')} \stackrel{w}{\longrightarrow}_{i} \{q_1,\ldots,q_h\} \]
We show that for all $i^1 \geq i_1$, there is some $i^2 > i^1$ such that,
\[ g^{i^2}_{(q, Q')} \stackrel{w}{\longrightarrow}_{i^2} \{q^?_1,\ldots,q^?_h\}
\]
in $G^{i^2}_{(q, Q')}$ where, for $y \in \{1,\ldots,h\}$,
\[ q^?_y = \left\{\begin{array}{ll}
                      g^{i^1}_{(q', Q'')} & \mathrm{if\ } q_1 =
g^{i_1}_{(q', Q'')} \\
                      q_y & \mathrm{otherwise}
                  \end{array}\right.
\]
Since $q^?_f = q_f$ for all $q_f \in \mathcal{Q}_f$, the lemma follows.
For a set $Q = \{q_1,\ldots,q_h\}$ we write $Q^? = \{q^?_1,\ldots,q^?_h\}$.

The proof proceeds by induction over $i$.  In the base case $i \leq i_1$
and the property holds by Lemma~\ref{runspersist} and since
$\newautG^{i_1} = \mathcal{G}^{i_1}_1$ and there are no incoming
transitions to any $g^{i_1}_{(q', Q'')}$ in $\mathcal{G}^{i_1}$.

In the inductive case, we begin by showing for a single transition,
\[ g^{i_1}_{(q, Q')} \stackrel{b}{\longrightarrow}_{i}
\{q_1,\ldots,q_{h}\} \]
in $\newsingleautG^i_{(q,Q')}$ with $b \in \Sigma$, we have,  for all $i^1 \geq i_1$, there is some $i^2 > i^1$ such that,
\[ g^{i^2}_{(q, Q')} \stackrel{b}{\longrightarrow}_{i^2} \{q^?_1,\ldots,q^?_h\} \]
in $G^{i^2}_{(q, Q')}$.  We analyse the $S \in \idG^{i_1}_{(q, Q')}[i_1/i_1-1]$
that spawned the transition from $g^{i_1}_{(q, Q')}$ (we assume the
transition is new, else the property holds by induction).

Let $S = \{\alpha_1,\ldots,\alpha_m\}$.
We have $\{q_1,\ldots,q_h\} = Q_1 \cup \ldots \cup Q_m$.
For each $e \in \{1,\ldots, m\}$, there are several cases,
\begin{enumerate}[$\bullet$]
\item
$\alpha_e = \theta$.

Let $g_e = q^{\theta}$.
By definition of
$\newautG^i$ we have the transition $g_e
\stackrel{b}{\longrightarrow}_{i-1} Q_e$ in $\newautG^{i-1}$.

If  $\theta = \idG^{i_1}_{(q', Q'')}$ then by induction we have
$i^2_e > i^1$ such that $g^{i^2_e}_{(q', Q'')}
\stackrel{b}{\longrightarrow}_{i^2_e} Q^?_e$ in
$\mathcal{G}^{i^2_e}$.
 Otherwise $g_e$ is initial in some $B \in \mathcal{B}$ and the transition $g_e \stackrel{b}{\longrightarrow}_{i-1}
Q_e$ also exists in $\mathcal{G}^0$ and is the same as $g_e
\stackrel{b}{\longrightarrow}_0 Q^?_e$.  Let $w_e =
b$.

\item
$\alpha_e = (a, push_{w_p}, \theta)$.  Then $b=a$.

Let $g_e = q^{\theta}$. By definition of
$\newautG^i$ we have the run $g_e \stackrel{w_p}{\longrightarrow}_{i-1}
Q_e$ in $\newautG^{i-1}$.

If  $\theta = \idG^{i_1}_{(q', Q'')}$ then by induction we have
$i^2_e > i^1$ such that $g^{i^2_e}_{(q', Q'')}
\stackrel{w_p}{\longrightarrow}_{i^2_e} Q^?_e$ in
$\mathcal{G}^{i^2_e}$.
 Otherwise $g_e$ is initial in some $B \in \mathcal{B}$ and the transition $g_e \stackrel{w_p}{\longrightarrow}_{i-1}
Q_e$ also exists in $\mathcal{G}^0$ and is the same as $g_e
\stackrel{w_p}{\longrightarrow}_0 Q^?_e$.
Let $w_e = w_p$.
\end{enumerate}
Let $i_{max}$ be the maximum $i^2_e$.  If $g_e = g^{i_1}_{(q', Q'')}$,
we have, by Lemma~\ref{runspersist}, $g^{i_{max}}_{(q', Q'')}
\stackrel{w_e}{\longrightarrow}_{i_{max}} Q^?_e$.  Also, by
Lemma~\ref{runspersist} we have $g_e
\stackrel{w_e}{\longrightarrow}_{i_{max}} Q^?_e$ when $g_e$ is not of the form
$g^{i_1}_{(q', Q'')}$.  Since we have
$\idG^{i_{max}+1}_{(q, Q')} \simeq \idG^{i_1}_{(q, Q')}$ we have
$S[i_{max}/i_1 - 1] \in \idG^{i_{max} + 1}_{(q, Q')}$ and since $Q^?_1
\cup \ldots \cup Q^?_m = \{q^?_1,\ldots,q^?_h\}$ we have,
\[ g^{i_{max}+1}_{(q, Q')}
\stackrel{b}{\longrightarrow}_{i_{max}+1} \{q^?_1,\ldots,q^?_h\} \]
in $G^{i_{max}+1}_{(q, Q')}$.  Let $i^2 = i_{max}+1$ and we are
done in the case of a single transition.

We now expand the result to a complete run by induction over the
length of the run.  That is, we take a run of $\newautG^{i}_{(q, Q')}$,
\[ g^{i_1}_{(q, Q')} \stackrel{w}{\longrightarrow}_{i} \{q_1,\ldots,q_h\} \]
and show that for all $i^1 \geq i_1$ there is some $i^2 > i^1$ such that,
\[ g^{i^2}_{(q, Q')} \stackrel{w}{\longrightarrow}_{i^2} \{q^?_1,\ldots,q^?_h\}
\]
in $G^{i^2}_{(q,Q')}$.

The base case has already been shown.  We now
consider the run,
\[  g^{i_1}_{(q, Q')} \stackrel{a_0}{\longrightarrow}_{i}
\{q^1_1,\ldots,q^1_{h_1}\} \stackrel{a_1}{\longrightarrow}_{i} \ldots
\stackrel{a_m}{\longrightarrow}_{i} \{q^m_1,\ldots,q^m_{h_m}\} \]
We have
$q^1_y \stackrel{a_1\ldots a_m}{\longrightarrow}_{i} Q_y$ for each $y
\in \{1,\ldots,h_1\}$ and $\bigcup_{y \in
\{1,\ldots,h_1\}} Q_y = \{q^m_1,\ldots,q^m_{h_m}\}$.  Then
for all $y \in \{1,\ldots,h_1\}$ via induction and Lemma~\ref{imax} we
have for all $i^1 > i_1$ an $i_{max}$ with
\[ \{q^{?1}_1,\ldots,q^{?1}_{h_1}\}
\xrightarrow{a_1\ldots a_m}_{i_{max}} \{q^{?m}_1,\ldots,q^{?m}_{h_m}\}
\]
We then use the result for a single transition to obtain the result
for the complete run.  That is, we have for $i_{max}$ an $i^2 > i_{max}$
such that,
\[ g^{i^2}_{(q, Q')} \stackrel{a_0}{\longrightarrow}_{i^2} \{q^{?1}_1,\ldots,q^{?1}_{h_1}\}
\xrightarrow{a_1\ldots a_m}_{i^2} \{q^{?m}_1,\ldots,q^{?m}_{h_m}\}
\]
exists in $\mathcal{G}^{i^2}$ as required.
\end{proof}

\section{Applications: Proofs and Definitions}
\label{buchipropproof}

\subsection{Proof of Proposition~\ref{buchiprop}}

\begin{proof}
We show a higher-order B\"uchi PDS has an accepting run iff the following
condition holds: let $c$ be a configuration of an order-$n$ B\"uchi PDS $BP$.
There is an
accepting run in $BP$ from $c$ iff there exist distinct configurations $\langle p^j,
[^n a ]^n \rangle$ and $\langle p^j, \gamma_2
\rangle$ with $top_1(\gamma_2) = a$ and configuration $\langle p^f,
\gamma_1\rangle$ such that $p^f \in \mathcal{F}$ and,
\begin{enumerate}[(1)]
\item
$c \pdsreach \langle p^j, \gamma_3 \rangle$ for some $\gamma_3$ with $top_1(\gamma_3)
= a$, and

\item
$\langle p^j, [^n a ]^n \rangle \pdsreach \langle p^f, \gamma_1 \rangle
\pdsreach \langle p^j, \gamma_2 \rangle$
\end{enumerate}

$\Rightarrow$: Every higher-order stack may be flattened into a well bracketed
string, as per Definition~\ref{nstoredef}.  Given a suffix of an
$n$-store $w$, let $comp(w)$ be a number of symbols ``$[$'' added to the
beginning of $w$ to form an $n$-store proper.

Given an accepting run of BP $\rho = c_0c_1\ldots$,
there exists a sequence of suffixes $w_1,w_2,\ldots$
such that there exists an increasing sequence of natural numbers $i_1,
i_2,\ldots$ and for all $j > 0$ and $i \geq i_j$
$c_{i}$ has a stack with the suffix $w_j$.  Additionally $c_{i_j}$ has
the $n$-store $comp(w_j)$ and $w_i$ is a
suffix of $w_j$ for all $i \leq j$ (it may be the case that $w_i =
w_j$).  Take the sequence $c_{i_1}c_{i_2}\ldots$.  Due to the
finiteness of $\mathcal{P}$ and $\Sigma$ there must be $p, a$ with an
infinite number of $c_{i_j}$ with control state $p$ and a stack whose
$top_1$ element is $a$.  Furthermore, since $\rho$ is accepting, we
must have distinct $c_{i_a}$ and $c_{i_b}$ with $p$ as their control states and
$a$ as the $top_1$ element, with a $c_f$ whose control state is
$p^f \in \mathcal{F}$, and,
\[ c_0 \pdsreach c_{i_a} \pdsreach c_f \pdsreach c_{i_b} \]
We have $(1)$ from $c_0 \pdsreach c_{i_a}$.  By definition of
$c_{i_1},c_{i_2}\ldots$ we have $c_{i_a} = \langle p,
comp(w_{i_a})\rangle$ and all configurations between $c_{i_a}$ and
$c_{i_b}$ have the suffix $w_{i_a}$.  This implies,
\[ \langle p, [^na]^n \rangle \pdsreach \langle p^f, u\rangle
\pdsreach \langle p, v\rangle \]
with $top_1(v) = a$.  Hence, $(2)$ holds as required.

$\Leftarrow$: From $(1)$ we have $c \pdsreach \langle p, \gamma_1 \rangle$
with $top_1(\gamma_1) = a$.  From $(2)$ we can construct a path,
\[ \langle p,
\gamma_2 \rangle \pdsreach \langle p^f, \gamma_3 \rangle \pdsreach \langle p,
\gamma_4 \rangle \]
with $p^f \in \mathcal{F}$ and $top_1(\gamma_4) = a$ for any $\gamma_2$ with
$top_1(\gamma_2) = a$.  Thus, through infinite applications of $(2)$, we can construct an accepting run of $BP$.
\end{proof}

\subsection{Proof of Lemma~\ref{ltlreachflaglem}}
\label{ltlreachflag}

\begin{proof}
We begin by showing that if $\langle p, [^n a]^n\rangle$ satisfies
(\ref{ltlreach2}), then a run $\langle (p, 0), [^n a]^n\rangle \pdsreach
\langle (p, 1), \gamma\rangle$ with $\gamma \in \lang(B^a_n)$ exists in $BP'$.
The run over $BP$ satisfying (\ref{ltlreach2}) can be split into two
parts,
\[ \langle p, [^n a]^n\rangle \pdsreach \langle p^f, \gamma_f \rangle
\pdsreach \langle p, \gamma\rangle \]
with $\gamma \in \lang(B^a_n)$ and $p^f$ is the first accepting state seen
in the run.  We consider each part separately.
\begin{enumerate}[$\bullet$]
\item
Suppose we have a run,
\[ \langle p_0, \gamma_0\rangle \hookrightarrow \ldots \hookrightarrow \langle p_m, \gamma_m
\rangle \]
such that $p_m$ is the only accepting control state in the run.  This
run is derived from a sequence of commands $d_1,\ldots,d_m$.  Let $d_i
= (p_{i-1}, a_i, o_i, p_i)$ for all $i \in \{1,\ldots,m\}$.  We show
the run,
\[ \langle (p_0, 0), \gamma_0\rangle \hookrightarrow \ldots
\hookrightarrow \langle (p_m, 0), \gamma_m
\rangle \]
exists in $BP'$ by induction over $m$.  In the base case $m = 0$ and the result is
trivial.  Suppose we have,
\[ \langle (p_1, 0), \gamma_1\rangle \hookrightarrow \ldots
\hookrightarrow \langle (p_m, 0), \gamma_m
\rangle \]
by the induction hypothesis.  Since $d_1 = (p_0, a_1, o_1, p_1)$ and
$p_0 \notin \mathcal{F}$, we have that $((p_0, 0), a_1, o_1, (p_1, 0))$ is in
$\mathcal{D}'$.  Hence we have the run,
\[ \langle (p_0, 0), \gamma_0\rangle \hookrightarrow \ldots
\hookrightarrow \langle (p_m, 0), \gamma_m
\rangle \]
as required.

\item
We have $\langle p^f, \gamma_f\rangle \in (\mathcal{F} \times C^\Sigma_n)
\cap Pre^+(\{p\} \times \lang(B^a_n)))$, we show there exists the run $\langle (p^f, 0), \gamma_f
\rangle \pdsreach \langle (p, 1), \gamma\rangle$ in BP' with $\gamma \in
\lang(B^a_n)$.

We have the run $\langle p^f, \gamma_f\rangle \pdsreach \langle p,
\gamma\rangle$ in BP with $\gamma \in \lang(B^a_n)$.  This run is of the form,
\[ \langle p_0, \gamma_0 \rangle \hookrightarrow \langle p_1, \gamma_1\rangle
\hookrightarrow \ldots \hookrightarrow \langle p_m, \gamma_m\rangle \]
with $m \geq 1$, $p_0 = p^f$, $\gamma_0 = \gamma_f$, $p_m = p$ and $\gamma_m = \gamma$.  The run is the consequence of a
sequence of commands $d_1,\ldots,d_m$.  Let $d_i = (p_{i-1}, a_{i},
o_{i}, p_i)$.  Since $p_0 \in \mathcal{F}$ we have $((p_0, 0),
a_1, o_1, (p_1, 1))$ in $\mathcal{D}'$ by definition.  Furthermore,
for $i \in \{2,\ldots,m\}$ we have  $((p_{i-1}, 1),
a_i, o_i, (p_i, 1))$ in $\mathcal{D}'$. We have the run
\[ \langle (p_0, 0), \gamma_0 \rangle \hookrightarrow \langle (p_1, 1), \gamma_1\rangle
\hookrightarrow \ldots \hookrightarrow \langle (p_m, 1), \gamma_m\rangle \]
in BP' therefrom.
\end{enumerate}
The proof of this direction follows immediately.

We now consider the proof in the opposite direction.  Suppose we have
$\langle (p, 0), [^n a]^n\rangle \pdsreach \langle (p, 1), \gamma\rangle$
with $\gamma \in \lang(B^a_n)$.  From the definition of $\mathcal{D}'$ it
follows that the run is of the form,
\[ \langle (p, 0), [^n a]^n\rangle \hookrightarrow \ldots
\hookrightarrow \langle (p^f, 0), \gamma_f\rangle \hookrightarrow \langle
(p', 1), \gamma'\rangle \hookrightarrow \ldots \hookrightarrow \langle (p,
1), \gamma\rangle \]
where the second element of each control state/flag pair changes only
in the position shown.  Furthermore, $p^f$ is the first occurrence of
an accepting control state in $BP$.  This run is the result of a
sequence of commands $d_1,\ldots,d_m$ where $m \geq 1$.  From a simple
projection on the first element of each control state/flag pair, we
immediately derive a sequence commands $d'_1,\ldots,d'_m$ in
$\mathcal{D}$ and the following run of $BP$,
\[ \langle p, [^n a]^n\rangle \hookrightarrow \ldots
\hookrightarrow \langle p^f, \gamma_f\rangle \hookrightarrow \langle
p', \gamma'\rangle \hookrightarrow \ldots \hookrightarrow \langle p,
\gamma\rangle \]
Since $\langle p^f, \gamma_f\rangle$ and $\langle p', \gamma'\rangle$ must be
distinct, the existence of this run implies $\langle p, [^n a]^n\rangle$
satisfies $(\ref{ltlreach2})$.
\end{proof}

\subsection{Proof of $Attr_E(\mathcal{R}) = Pre^*(\mathcal{R}')\setminus\mathcal{C}^\triangledown_A$}
\label{attrispre}
\begin{proof}
We show $Attr_E(\mathcal{R}) = Pre^*(\mathcal{R}')\setminus\mathcal{C}^\triangledown_A$.  We begin by
proving $Attr_E(\mathcal{R}) \supseteq Pre^*(\mathcal{R}')\setminus\mathcal{C}^\triangledown_A$.

Take a configuration $\langle p, \gamma \rangle \in
Pre^*(\mathcal{R}')\setminus\mathcal{C}^\triangledown_A$.  We show $\langle p,
\gamma \rangle \in Attr_E(\mathcal{R})$ by induction over the shortest path
$\langle p, \gamma \rangle \pdsreach C$ of the order-$n$ APDS with $C \subseteq
\mathcal{R}'$.

For the base case, we have $\langle p, \gamma \rangle \in
\mathcal{R}'\setminus\mathcal{C}^\triangledown_A$.  Hence, $\langle p, \gamma \rangle \in
Attr_E(\mathcal{R})$ since $\mathcal{R} \subseteq Attr_E(\mathcal{R})$.

Now, suppose we have $\langle p, \gamma \rangle \hookrightarrow C$ via the
command $d = (p, a, OP)$ in the higher-order APDS with $C \in
Pre^*(\mathcal{R})\setminus\mathcal{C}^\triangledown_A$ and by induction $C
\subseteq Attr^i_E(\mathcal{R})$ for some $i$.  There are two cases,
\begin{enumerate}[$\bullet$]
\item
If $p \in \mathcal{P}_A$ then for each $(o, p') \in OP$ and hence each
move $(p, a, o, p')$ in the higher-order PDS we have a corresponding $\langle p', \gamma'\rangle \in C$.
We have either $\langle p', \gamma'\rangle \in
Pre^*(\mathcal{R}')\setminus\mathcal{C}^\triangledown_A$ or we
have $\langle p', \gamma'\rangle = \langle p, \triangledown\rangle$.

If we have $\langle p', \gamma'\rangle \in
Pre^*(\mathcal{R}')\setminus\mathcal{C}^\triangledown_A$ then $\langle p',
\gamma'\rangle \in Attr^i_E(\mathcal{R})$ for some $i$ by induction.

If we have $\langle p', \gamma'\rangle = \langle p, \triangledown\rangle$ then
$o(\gamma)$ is undefined.  Hence $(p, a, o, p')$ is not a valid move for
Abelard.

Hence we have $\langle p, \gamma\rangle \in \mathcal{C}_A$ and $\forall c'.
\langle p, \gamma\rangle \hookrightarrow c' \Rightarrow c' \in
Attr^i_E(\mathcal{R})$ which implies $\langle p, \gamma \rangle \in
Attr^{i+1}_E(\mathcal{R}) \subseteq Attr_E(\mathcal{R})$.

\item
If $p \in \mathcal{P}_E$ then $C = \{\langle p', o(\gamma)\rangle\}$ and
$(p, a, o, p') \in \mathcal{D}$.  Thus, we have $\exists c'. \langle
p, \gamma\rangle
\hookrightarrow c' \land c' \in Attr^{i}_E(\mathcal{R})$ and $\langle
p, \gamma \rangle \in \mathcal{C}_E$.  Therefore $\langle p, \gamma  \rangle \in
Attr^{i+1}_E(\mathcal{R}) \subseteq Attr_E(\mathcal{R})$.
\end{enumerate}
Thus, we have $Attr_E(\mathcal{R}) \supseteq
Pre^*(\mathcal{R}')\setminus\mathcal{C}^\triangledown_A$ as required.

To show $Attr_E(\mathcal{R}) \subseteq
Pre^*(\mathcal{R}')\setminus\mathcal{C}^\triangledown_A$ we induct over 
$i$ in $Attr_E(\mathcal{R}) = \bigcup_{i \leq 0} Attr^i_E(\mathcal{R})$.  When
$i = 0$ we have $Attr^0_E(\mathcal{R}) = \mathcal{R} \subseteq
\mathcal{R}'\setminus\mathcal{C}^\triangledown_A \subseteq
Pre^*(\mathcal{R}')\setminus\mathcal{C}^\triangledown_A$.  For $i > 1$ there are
two cases for all $c$ such that $c \notin Attr^{i-1}_E(\mathcal{R})$ and $c \in
Attr^i_E(\mathcal{R})$,
\begin{enumerate}[$\bullet$]
\item
$c \in \{\ c \in \mathcal{C}_E\ |\ \exists c'. c \hookrightarrow c'
\land c' \in Attr^{i-1}_E(\mathcal{R})\ \}$.

Hence there is some command $d = (p, a, o, p')$ in the higher-order PDS and
command $(p, a, \{(o, p')\})$ in the higher-order APDS.  By induction $c' \in
Pre^*(\mathcal{R}')\setminus\mathcal{C}^\triangledown_A$ and $c = \langle p,
\gamma\rangle$ and $c' = \langle p', o(\gamma)\rangle$.  Hence $c \in
Pre^*(\mathcal{R}')\setminus\mathcal{C}^\triangledown_A$.

\item
$c \in \{\ c \in \mathcal{C}_A\ |\ \forall c'. c \hookrightarrow c'
\Rightarrow c' \in Attr^{i-1}_E(\mathcal{R})\ \}$.

Let $c = \langle p, \gamma\rangle$.  We have $d = (p, a, OP)$ in the
higher-order APDS such that for all moves $(p, a, o, p')$ we have $(o, p') \in
OP$. If $o(\gamma)$ is defined, we have $\langle p, \gamma \rangle
\hookrightarrow \langle p', o(\gamma)\rangle$ and $\langle p', o(\gamma) \rangle
\in Pre^*(\mathcal{R}')$ by induction.  If $o(\gamma)$ is undefined, then since
we have $\langle p, \triangledown\rangle \in \mathcal{R}'$ we have $\langle p,
\triangledown\rangle \in Pre^*(\mathcal{R}')$.

Thus, we have $\langle p, \gamma\rangle \hookrightarrow C$ via an application of
the command $d$ such that $C \subseteq Pre^*(\mathcal{R}')$.  Hence $\langle p,
\gamma\rangle \in Pre^*(\mathcal{R}')$ and since $\gamma \neq \triangledown$, we
have $\langle p, \gamma\rangle \in
Pre^*(\mathcal{R}')\setminus\mathcal{C}^\triangledown_A$ as required.
\end{enumerate}
Thus, we have  $Attr_E(\mathcal{R}) = Pre^*(\mathcal{R}')\setminus\mathcal{C}^\triangledown_A$.
\end{proof}

\end{document}